\documentclass[master,twoside,BCOR=12mm,DIV=10,ngerman,english]{GAUBMutf8}
\pdfoutput=1

  \usepackage[utf8]{inputenc}
  \usepackage[T1]{fontenc}
  \usepackage{uniinput}
  \usepackage{fourier}

\usepackage{setspace}  
\usepackage{babel}     
\usepackage{calc}      
\usepackage{csquotes}
\usepackage{amsmath,mathtools} 
\usepackage{amsfonts}
\usepackage[amsmath,hyperref,thmmarks]{ntheorem}

\usepackage[backend=bibtex,style=numeric-comp,sorting=nyvt,natbib,isbn=false,url=false,doi=false,eprint=true,firstinits=true,clearlang=true]{biblatex}
\AtEveryBibitem{\clearfield{month}}
\AtEveryBibitem{\clearfield{day}}

\usepackage{booktabs}                      
\usepackage{longtable}                     
\usepackage{array}                         
\usepackage{float}

\usepackage{siunitx} 

\usepackage{graphicx}
\usepackage{color}
\usepackage{caption}
\usepackage{subcaption}
\usepackage{wrapfig}

\usepackage[pdfstartview=FitH,      
            breaklinks=true,        
            bookmarksopen=true,     
            bookmarksnumbered=true,  
	    colorlinks=false,
	    pdfborder={0 0 0},
            ]{hyperref}

\usepackage{xspace}
\usepackage{microtype}
\newcommand{\iflanggerman}[2]{
 \iflanguage{german}{#1}{
  \iflanguage{ngerman}{#1}{#2}
 }
}

\iflanggerman{

}{

}



\newcommand*{\transpose}{\ensuremath{^{\mathrm{T}}}}

\newcommand*{\abs}[1]{\lvert#1\rvert}
\newcommand*{\norm}[1]{\lVert#1\rVert}

\DeclareMathOperator{\imag}{im}

\DeclareMathOperator{\nbh}{nbh}
\DeclareMathOperator{\out}{out}
\DeclareMathOperator{\id}{id}

\DeclareMathOperator{\sgn}{sgn}
\DeclareMathOperator{\diag}{diag}

\DeclareMathOperator{\Asinh}{Asinh}

\renewcommand{\vec}[1]{\boldsymbol{#1}}

\newcommand{\D}{\mathrm{d}}
\newcommand{\I}{\mathrm{i}}
\newcommand{\del}{\partial}
\newcommand{\ddel}{\partial^{*}\!}

\newcommand{\e}[1]{\,\mathrm{e}^{#1}}

\newcommand{\subsc}[1]{_{\text{#1}}}
\newcommand{\supsc}[1]{^{\text{#1}}}

\newcommand{\ie}{\textit{i.\,e.\@}\xspace}
\newcommand{\cf}{\textit{cf.\@}\xspace}
\newcommand{\eg}{\textit{e.\,g.\@}\xspace}

\newcommand{\asure}{\textup{a.\,s.\@\xspace}}

\newcommand{\coloneqq}{\ensuremath{\mathrel{\mathop:}=}} 	
\newcommand{\eqqcolon}{\ensuremath{=\mathrel{\mathop:}}} 	

\newcommand{\Wl}{\ensuremath{w^{-}}}
\newcommand{\Wr}{\ensuremath{w^{+}}}

\theoremstyle{margin}
\theoremheaderfont{\normalfont\bfseries}
\theorembodyfont{\slshape}
\theoremseparator{}
\theoremsymbol{}
\newtheorem{Theorem}{Theorem}[section]
\newtheorem{Proposition}[Theorem]{Proposition}

\theoremstyle{nonumberplain}
\newtheorem{PropositionPlain}{Proposition}

\theoremstyle{margin}
\theoremheaderfont{\normalfont\bfseries}
\theorembodyfont{\rmfamily}
\theoremseparator{}
\newtheorem{Example}[Theorem]{Example}

\theoremheaderfont{\scshape}
\theorembodyfont{\upshape}
\theoremstyle{nonumberplain}
\theoremseparator{}
\theoremsymbol{\ensuremath{\square}}
\newtheorem{Proof}{Proof}

\theoremstyle{margin}
\theoremheaderfont{\normalfont\bfseries}\theorembodyfont{\upshape}
\theoremseparator{}
\theoremsymbol{}
\newtheorem{Definition}[Theorem]{Definition}
\newtheorem{Remark}[Theorem]{Remark}

\usepackage{anysize}
\marginsize{3cm}{4cm}{1cm}{2cm}

\makeatletter
\defbibenvironment{bibliography}
{\list{\printtext[labelnumberwidth]{%
    \printfield{prefixnumber}%
    \printfield{labelnumber}}}
    {\setlength{\bibhang}{0pt}%
    \setlength{\leftmargin}{\bibhang}%
    \setlength{\itemindent}{-\leftmargin}%
    \setlength{\itemsep}{\bibitemsep}%
    \setlength{\parsep}{\bibparsep}}}
{\endlist}
{\item}
\makeatother

\graphicspath{{./figures/}}

\begin{document}
\pagenumbering{roman}
\ThesisAuthor{Artur}{Wachtel}
\PlaceOfBirth{Kara-Balta, Kyrgyzstan}
\ThesisTitle{Fluktuationsspektren und Grobkörnung in Stochastischer Dynamik}{Fluctuation Spectra and Coarse Graining in Stochastic Dynamics}
\FirstReferee{Prof.~Dr.~Jürgen Vollmer}
\Institute{Institute for Nonlinear Dynamics}
\SecondReferee{Prof.~Dr.~Marc Timme}
\ThesisBegin{1}{5}{2013}
\ThesisEnd{1}{11}{2013}
\thispagestyle{empty}
\frontmatter
\maketitle
\cleardoublepage

\cleardoublepage
\thispagestyle{empty}

\vspace*{0.35\textheight}

\begin{center}

  \includegraphics[width=0.8cm]{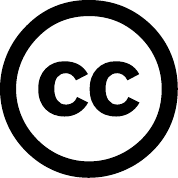}
  \,
  \includegraphics[width=0.8cm]{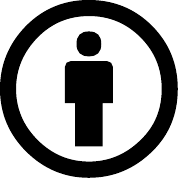}

  This Master’s Thesis by Artur Wachtel is licensed under a \linebreak\href{http://creativecommons.org/licenses/by/4.0/deed.en}{Creative Commons Attribution 4.0 International License}\footnote{\url{http://creativecommons.org/licenses/by/4.0/deed.en}}.
\end{center}

\hspace{1cm}
\thispagestyle{empty}

\begin{center}
  \today\\[2em]
  \begin{slshape}
    The text at hand differs from the thesis officially submitted to Georg-August-Universität Göttingen as part of a master’s degree:\linebreak
    This text is expanded, includes more figures, and contains\linebreak{}fewer typographic errors.
  \end{slshape}
\end{center}

\onehalfspacing
\begin{otherlanguage}{ngerman}
\chapter*{Danksagung}
Die Fertigstellung dieser Abschlussarbeit wäre ohne das Zutun außergewöhnlicher Menschen nicht möglich gewesen.

Zu aller est möchte ich Fabian Telschow danken.
Unsere gemeinsamen, regelmäßig unregelmäßigen, zeremoniösen Teeverköstigungen haben mir sehr häufig über die Frustrationen im Zwist mit der Mathematik hinweggeholfen.
Unsere ausschweifende Lesereise in die mathematischen Untiefen der geometrisierten Mechanik wird mir auf Ewig im Gedächtnis bleiben.

Besonderer Dank gilt Bernhard Altaner.
Seine sehr motivierende Art und Weise, über die Phänomäne der physikalischen Welt zu erzählen, hat meine Faszination für die Wissenschaft immer wieder aufs Neue bestärkt.
Ich hoffe ich war kein allzu anstrengender erster Schützling.
Mir ist durchaus bewusst, dass jede Konfusion die ich verursacht habe, irgendwann auf mich zurückfallen wird.
Ich werde dafür gewappnet sein.

Auch Johannes Blaschke hat viel zu meiner geistigen Gesundheit beigetragen: Seine bürokratische Beharrung auf Rigorosität war ein fester Anker im Meer der unvorhergesehenen Unwägbarkeiten.
Ich wünsche seinem Dominion viele Stückmünzchen und jeden erdenklichen Erfolg beim Kampf gegen die Temperatur\texttrademark.

Meinem hauptamtlichen Betreuer Jürgen Vollmer bin ich zu äußerstem Dank verpflichet.
Seine organisiert-chaotische Aura beeinflusste mich mehrfach, auch über Ländergrenzen hinweg, und lenkte meine Bahn in sichere Gefilde.
Ich schätze besonders sein intuitives Gespür für die komplexen Zusammenhänge, die in der gesamten Arbeitsgruppe bearbeitet werden.
Er ist immer wieder in der Lage, Parallelen zwischen völlig verschiedenen Dingen aufzuzeigen, die meinem Auge verborgen geblieben wären.
Ich hoffe, ich habe andererseits auch ihm noch neues Wissen eröffnet.

Letzten Endes ist selbst die größte Motivation für die  Wissenschaft nicht viel Wert, ohne die beständige Unterstützung aus der „echten Welt“:
Die studentische Lateinformation hat mich oft aufgeheitert und für meine körperliche Ertüchtigung gesorgt.
Für die unschätzbare Hilfe in allen Lebenslagen möchte ich schlussendlich vor allem meinen Eltern und meinen beiden Brüdern danken.

\end{otherlanguage}


\cleardoublepage
\phantomsection
\chapter*{}
\thispagestyle{empty}
\begin{center}
  \vspace*{5cm}
  \itshape{Erkenntnis wächst aus Konfusion}\\
  \begin{align*}
    \rightarrow\leftarrow
  \end{align*}
\end{center}

\cleardoublepage
\onehalfspacing
\tableofcontents


\cleardoublepage
\pagenumbering{arabic}
\chapter{Introduction}
\section*{Thermodynamics}

Thermodynamics is an old and very powerful theory.
It describes the properties of systems that can exchange \emph{extensive} quantities like energy, volume or particles of different types\,\citep{Callen1985}.

A fundamental principle of thermodynamics is \emph{equilibrium}: Bringing two systems in contact and allowing them to exchange arbitrary amounts of a quantity will eventually lead to a steady state.
In this state the systems on average do not exchange any amounts of the quantity anymore.
Depending on which quantity was exchanged one speaks of \emph{thermal} (exchange of energy), \emph{thermo-mechanical} (energy and volume), and \emph{thermo-chemical} (energy and particles) equilibrium.
These equilibria define \emph{intensive} quantities: For every extensive quantity there is an intensive quantity that determines the behavior at contact. \emph{Temperature}, \emph{pressure} and \emph{chemical potential} give equivalence classes of systems that exchange energy, volume or particles, respectively.
If one of the two systems is so large\footnote{Formally a reservoir carries infinite amounts of the extensive quantities.} that it effectively does not feel the exchange due to the contact it is a \emph{reservoir}.
Reservoirs are characterized by their intensive variables.

These exchange processes can be \emph{reversible} or \emph{irreversible}.
If a process is reversible a quantity called \emph{entropy} will not change, irreversible changes cause the entropy to rise.
The \emph{Second Law} of thermodynamics states that the overall entropy cannot decrease.

A small system can be coupled to two reservoirs that differ in their intensive quantity. This will lead to a \emph{non-equilibrium steady state}: In this state the extensive quantity will flow from one reservoir to the other for all times.
Such situations are intrinsically irreversible and can be treated by thermodynamics only with the assumption of \emph{local equilibrium}: At the locations of contact the small system is in equilibrium with the reservoirs and there is a gradient of the corresponding intensive quantity throughout the rest of the small system.

\section*{Statistical Physics}

The development of \emph{Statistical Physics} allowed a mechanistic understanding of thermodynamics: The intensive quantities are emergent phenomena that arise from the interactions of many small particles.
In the \emph{thermodynamic limit} of infinitely many particles the laws of equilibrium thermodynamics can be recovered from statistical physics.
The central quantities in statistical physics are the \emph{partition functions} and their \emph{thermodynamic potentials}.
This approach allows both a generalization of thermodynamics and reveals a tight connection to information theory of ergodic dynamical systems:
Assuming equilibrium, the Shannon entropy of the probability density in phase space turns out converge to the thermodynamic entropy in the thermodynamic limit.
However, a consistent formulation of statistical physics for non-equilibrium systems is still not found.
Neither are the microscopic conditions for local equilibrium fully understood.

\section*{Stochastic Thermodynamics}

In recent years experimental techniques have progressed.
Now it is possible to manipulate single macromolecules and colloidal particles on a sub-micrometer scale: These objects and their environments are neither in the thermodynamic limit, nor necessarily close to equilibrium\,\citep{Speck2007}.
For these cases the second law of thermodynamics no longer holds in its original formulation\,\citep{Evans1993}: The entropy can measurably \emph{decrease}.
However, the ensemble \emph{average} change in entropy remains positive.
The consistent formulation of thermodynamic concepts in these heavily fluctuating systems is called \emph{Stochastic Thermodynamics}.
In terms of statistical physics, the phase space is partitioned into experimentally observable states.
Under the conditions of the chaotic hypothesis, the dynamics on these \emph{mesoscopic} states appears stochastic.

The interesting observables are not only defined on the states: Fluxes and currents are important as well.
So in general, thermodynamic observables are defined on single random trajectories, as
\citet{Seifert2012} describes in a very nice and comprehensive review.

I assume that the partition of phase space is finite, \ie only finitely many different mesoscopic states can be measured experimentally.
Thus, the mesoscopic phase space is a finite set with a structure describing neighborhood relations.
Mathematically, this abstract idea is formalized as a so called \emph{graph}.
Graph theory deals with the properties of these structures. 

\subsubsection*{Fluctuation Relations}

Stochastic thermodynamics allows us to investigate the fluctuations around expectation values more thoroughly: In small biological systems these fluctuations can provide crucial function.
Arguably, the most astonishing insights from stochastic thermodynamics are the so called \emph{Fluctuation Relations}.
For different fluctuating thermodynamic quantities like exchanged work, heat or entropy production, these relations describe symmetries relating the positive and the negative part of the probability distribution.
They generalize fluctuation-dissipation relations to regimes arbitrarily far from equilibrium\,\citep{Gallavotti1996,Ruelle1999} and contain the second law as a special case\thinspace{}—\thinspace{}now as a statement about (ensemble) averages.
A first proof was due to \citet{Gallavotti1995} and later this work was generalized to systems with Markovian dynamics on a continuous\,\citep{Kurchan1998} and a discrete\,\citep{Lebowitz1999} state space.
A very abstract formulation for deterministic dynamics was proven by \citet{Wojtkowski2009}.
In view of these proofs, the fluctuation relations are sometimes called fluctuation \emph{theorems}, even in contexts where a proof is still missing.
For many very different and apparently unrelated systems the fluctuation relations were experimentally and numerically tested: granular matter\,\citep{Joubaud2012,Naert2012}, turbulent flow\,\citep{Bandi2008,Ciliberto2004}, shear flow\,\citep{Schumacher2004,Bonetto1998}, chemical oscillatory waves\,\citep{Sasa2000}, electrical circuits\,\citep{Garnier2005}, colloidal particles\,\citep{Speck2007}, and even macroscopic mechanical oscillators\,\citep{Douarche2005}.
Some of the conceptual problems of the attempts to experimentally or numerically verify the fluctuation relations have been discussed in reference\,\citep{Zamponi2007}.

\section*{This Thesis}

This thesis is divided into six chapters.
The first two chapters form the mathematical basis of the thesis.
They cover the theories of graphs and probability with various aspects.
Their common denominator are Markovian jump processes that are covered in the third chapter.
Additionally, this chapter targets the fluctuations of current-like observables on Markovian jump processes.
The fourth chapter is a mathematical formulation of stochastic thermodynamics for these kinds of stochastic processes.
The fifth chapter presents a local coarse graining of given stochastic systems.
The last chapter analyzes a concrete model system with the tool set introduced in all the other chapters.

\clearpage
\section{General Notation}

I expect the reader is familiar with the basic concepts of sets, groups, fields, vector spaces, and functions on these structures.
So the following might be considered redundant or superfluous, but unluckily there are different conflicting notations and conventions in use.
In order to avoid confusion, I restrict myself to the following notation for this thesis:

\begin{Definition}
  \label{def:sets-tuples}
  A \emph{set} is an unordered collection of elements. Sets are written with curly braces, so $\left\{ 1,2 \right\} = \left\{ 2, 1 \right\} = \left\{ 2, 1, 1 \right\}$ are three descriptions of the same set.
  A \emph{tuple} is an ordered collection of elements. Tuples are written with parentheses, so $\left( 1, 2 \right)\neq\left( 2, 1 \right)\neq\left( 2, 1, 1 \right)$ are three different tuples.
  A tuple with $d$ elements is also called \emph{$d$-tuple}.
\end{Definition}

\begin{Definition}
  \label{def:subset}
  Let $\otherOmega$ be a set.
  Another set $U$ is a \emph{subset} of $\otherOmega$ if every element of $U$ also is an element of $\otherOmega$.
  I will write $U\subset\otherOmega$ to denote a subset relation.
\end{Definition}

\begin{Example}
  \label{exa:subset}
  The empty set $\emptyset\subset\otherOmega$ is a subset of every set $\otherOmega$.
  A set $\otherOmega\subset\otherOmega$ is always a subset of itself.
\end{Example}

\begin{Definition}
  \label{def:power-set}
  For a set $\otherOmega$, the \emph{set of subsets} or the \emph{power set} is
   $ \mathfrak{P}(\otherOmega)\coloneqq \left\{ U\subset\otherOmega \right\}$.
\end{Definition}

\begin{Definition}
  \label{def:cardinality}
  The \emph{cardinality} of a set $\otherOmega$ is the number of elements in $\otherOmega$, written $\abs{\otherOmega}$.
  A set is \emph{finite} if its cardinality is finite.
\end{Definition}

\begin{Example}
  \label{exa:cardinalty}
  The cardinality of $\left\{ 0,1,2,3 \right\}$ is $\abs{ \left\{ 0,1,2,3 \right\} }=4$.
  For any finite set $\otherOmega$ the power set has cardinality $\abs{\mathfrak{P}(\otherOmega)}=2^{\abs{\otherOmega}}$.
\end{Example}

\begin{Definition}
  \label{def:number-fields}
  The symbol $\mathbb{N}=\left\{ 1, 2, 3, \dots \right\}$ denotes the natural numbers, excluding zero.
  The integers, the reals, and the complex numbers are represented by $\mathbb{Z}$, $\mathbb{R}$, and $\mathbb{C}$, respectively.
\end{Definition}

\begin{Definition}
  \label{def:functions-over-sets}
  Let $\otherOmega$ be a set and $\mathbb{F}$ be a field.
  The functions on $\otherOmega$ taking values in $\mathbb{F}$ are denoted by
  \begin{align*}
    \mathbb{F}^{\otherOmega} \coloneqq \left\{ f\colon \otherOmega \to \mathbb{F}\middle|\text{$f$ is a mapping}  \right\}\,.
  \end{align*}
\end{Definition}

\begin{Proposition}
  \label{thm:functions-over-sets}
  The space $\mathbb{F}^{\otherOmega}$ is an $\mathbb{F}$-vector space with the induced operations
  \begin{align*}
    f_1 + f_2\colon x&\mapsto f_1(x) + f_2(x)\,,\\
    \alpha\,f_1\colon x&\mapsto \alpha\,f_1(x)
  \end{align*}
  for any $f_1,f_2\in\mathbb{F}^{\otherOmega}$, $x\in\otherOmega,$ and any $\alpha\in\mathbb{F}$.
\end{Proposition}

\begin{Definition}
  \label{def:isomorphism}
  Let two spaces $\otherOmega_1$, $\otherOmega_2$ be isomorphic.
  I indicate this by $\otherOmega_1\simeq\otherOmega_2$.
\end{Definition}

\begin{Example}
  \label{exa:isomorphism}
  The groups $\left( \left\{ 0,1 \right\}, + \right)\simeq\left( \left\{ 1, -1 \right\},\, \cdot\, \right)$ are isomorphic, just like the $\mathbb{R}$-vector spaces $\mathbb{R}^{ \left\{ 0, 1 \right\}}\simeq \mathbb{R}^2\simeq \mathbb{C}$.
\end{Example}

\begin{Definition}
  \label{def:standard-scalar-product}
  The \emph{standard scalar product} of two vectors $x,y\in\mathbb{R}^d$ is written
  \begin{align*}
    x\cdot\! y = \sum_{i=1}^{d} x_i y_i\,,
  \end{align*}
  where $x=\left( x_1, \dots, x_d \right)$, $y=\left( y_1, \dots, y_d \right)$.
\end{Definition}

\begin{Definition}
  \label{def:iverson-bracket}
  Let $X$ be a logical statement that can be either true or false. Then the \emph{Iverson bracket} is defined as
  \begin{align*}
    \llbracket X \rrbracket \coloneqq
    \begin{cases} 1\,, & \text{if $X$ is true}\,, \\
      0\,, & \text{if $X$ is false}\,.
    \end{cases}
  \end{align*}
\end{Definition}

\begin{Example}
  \label{exa:iverson-bracket}
  There are several important special cases of the Iverson bracket:
  \begin{itemize}
    \item The Kronecker symbol can be written as $\delta^{i}_{j}=\llbracket i=j\rrbracket$.
    \item The Heaviside step function is represented by $\Theta(x)=\llbracket x \geq 0 \rrbracket$.
    \item The signum function is given by $\sgn(x) = \llbracket x>0\rrbracket -\llbracket x<0\rrbracket$.
    \item The indicator function of a set $\otherOmega$ is $\mathbb{1}_{\otherOmega}(x)=\llbracket x \in \otherOmega\rrbracket$.
  \end{itemize}
\end{Example}

\chapter{Graph Theory}
This thesis addresses stochastic dynamics on finite spaces.
Thus, it is important to understand the structure of these spaces.
Graph theory provides a concise and convenient tool set addressing exactly that need.

Just as in other fields of mathematics, there are different conventions in use.
In this chapter I present a consistent set of definitions, notations and results.
My convention is a mix based on different sources\,\citep{Cvetkovic1980,Knauer2011,Gross1987,Diestel2010} but taylored for the later use in stochastic dynamics.
I will treat directed graphs as undirected when it comes to topology, \cf definitions below.
On the other hand, directed graphs are more convenient to analyze with algebraic methods.
Consequently, the difference between directed and undirected is not as important as it might seem initially.

\section{Fundamentals of Graph Theory}
\label{sec:fundamental-graph-theory}

\begin{Definition}
  \label{def:directed-graph}
  Given two countable sets $\mathcal{V}$ and $\mathcal{E}$ together with a map $\iota\colon\mathcal{E}\to \mathcal{V}\times\mathcal{V}$.
  This triple $\mathcal{G}=(\mathcal{V},\mathcal{E},\iota)$ is called a \emph{directed Graph}.
  The set $\mathcal{V}$ is called \emph{vertex set}, its elements are \emph{vertices}.
  The set $\mathcal{E}$ is called \emph{edge set}, its elements are \emph{edges}.
  The map $\iota$ is called \emph{incidence map}.
\end{Definition}

\begin{Definition}
  \label{def:order-of-graph}
  Let $\mathcal{G}=\left( \mathcal{V},\mathcal{E},\iota \right)$ be a graph.
  The cardinality of the vertex set is called the \emph{order} of the graph $\mathcal{G}$.
  The order is written as $\abs{\mathcal{G}}\coloneqq\abs{\mathcal{V}}$ and it can be finite or infinite.
  A graph is \emph{finite} if its order is finite.
  The cardinality of the edge set will be abbreviated as $\norm{\mathcal{G}}\coloneqq\abs{\mathcal{E}}$. 
\end{Definition}

\begin{Remark}
  \label{rmk:graph-notation}
  I will sometimes refer to the vertex and edge sets of a graph $\mathcal{G}$ as $\mathcal{V}(\mathcal{G})$ and $\mathcal{E}(\mathcal{G})$, respectively.
\end{Remark}

\begin{Definition}
  \label{def:origin-end}
  The incidence map defines two more maps $\iota=(o,t)$.
  For a given edge $e\in\mathcal{E}$ the vertex $o(e)\in\mathcal{V}$ is called \emph{origin} or \emph{source}, the vertex $t(e)\in\mathcal{V}$ is called \emph{tail} or \emph{end} of $e$.
  If a vertex is either origin or tail of an edge, this vertex and edge are said to be \emph{incident}.
  Two edges are \emph{incident} if they have a vertex in common.
  The origin and end of an edge are \emph{connected} by $e$ or simply \emph{adjacent}.
  An edge is called \emph{loop} if its origin and tail coincide.
\end{Definition}

\begin{Definition}
  \label{def:undirected-graph}
  Given a vertex set $\mathcal{V}$ and an edge set $\mathcal{E}$.
  Let $\mathcal{U}=\left\{ U\subset \mathcal{V}\colon 1\leq \abs{U}\leq 2 \right\}$ be the set of non-empty subsets with at most two elements.
  An incidence map $\iota\colon\mathcal{E}\to\mathcal{U}$ makes the triple $(\mathcal{V},\mathcal{E},\iota)$ an \emph{undirected graph}.
  The concepts introduced in definition~\ref{def:origin-end} are used for undirected graphs as well, except the fact that the incidence map does not distinguish between origin and tail.
\end{Definition}

\begin{Remark}
  \label{rmk:multiple-edges}
  The above definitions include the case, where for two different edges $e$ and $e'$ we have both $o(e)=o(e')$ and $t(e)=t(e')$.
  Some authors explicitly exclude these so called \emph{multiple edges} from the definition of a graph.
  Sometimes also loops are explicitly forbidden.
  Unless stated otherwise, I will always allow both.
\end{Remark}

\begin{figure}[htbp]
  \centering
  \begin{subfigure}[b]{0.48\columnwidth}
    \centering
    \includegraphics{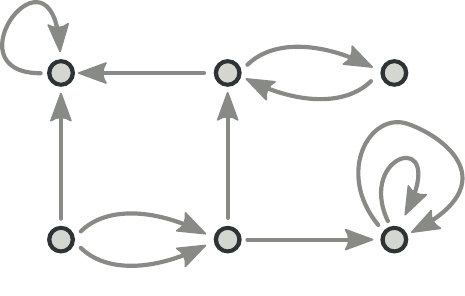}
    \caption{A directed graph $\mathcal{G}_a$.}
    \label{fig:directed-graph-example}
  \end{subfigure}
  \hfill
  \begin{subfigure}[b]{0.48\columnwidth}
    \centering
    \includegraphics{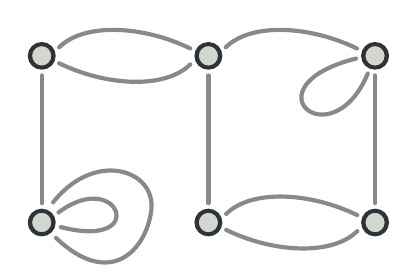}
    \caption{An undirected graph $\mathcal{G}_b$.}
    \label{fig:undirected-graph-example}
  \end{subfigure}
  \hfill
  \caption{
  Two examples for graphs with multiple edges.
  The structures are different although in both cases $\abs{\mathcal{G}_a}=\abs{\mathcal{G}_b}=6$ and $\norm{\mathcal{G}_a}=\norm{\mathcal{G}_b}=11$.
  }
  \label{fig:graph-examples}
\end{figure}

  Graphs can be represented in a very intuitive way: For a vertex you draw a dot.
  In directed graphs you depict edges as arrows from origin to end.
  For undirected graphs you draw edges as lines connecting the incident vertices, \cf figure~\ref{fig:graph-examples}. 
  Note that neither the angles between edges nor the curvature of the edges encode structural properties of a graph.

\begin{Definition}
  \label{def:simple-graph}
  Given a graph $\mathcal{G}$.
  If there is at most one edge connecting any two vertices, we call $\mathcal{G}$ a \emph{simple} graph.
\end{Definition}

\begin{figure}[htbp]
    \centering
    \def\svgwidth{0.195\textwidth}
    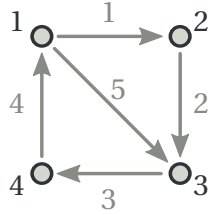
    \caption{
    A simple directed graph $\mathcal{G}_c$ with labelled vertices and edges.
    The edges can be written as $e_1=\left( v_1,v_2 \right)$, $e_5=\left( v_1, v_3 \right)$, $e_4=\left( v_4, v_1 \right)$, etc.
    }
    \label{fig:labeled-simple-graph-example}
  \end{figure}

\begin{figure}[htbp]
    \centering
    \includegraphics{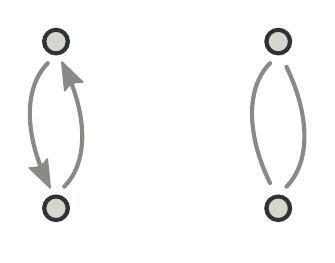}
    \caption{These two graphs are not simple, \cf definition~\ref{def:simple-graph}.}
    \label{fig:not-simple-graphs-example}
  \end{figure}

  For undirected graphs, this requirement is equivalent to the incidence map $\iota$ being injective\thinspace{}---\thinspace{}for directed graphs this requirement is a little bit stronger.
  A vertex with a loop is a simple graph.
  Figure~\ref{fig:labeled-simple-graph-example} shows an example of a nontrivial simple graph.
  The graphs in figure~\ref{fig:not-simple-graphs-example} are not simple.
  The description of a simple graph is much easier, since the edge set $\mathcal{E}$ can be identified with $\iota(\mathcal{E})$.
  Thus, a directed or undirected edge connecting the vertices $v_1$ and $v_2$ can uniquely be written as $\left( v_1,v_2 \right)$ or $\left\{ v_1,v_2 \right\}$, respectively.


The structure of a graph can be described in a very convenient way: If the graph is finite, all the relevant information can be gathered in matrices.

\begin{Definition}
  \label{def:degree-of-vertex}
  Let $v\in\mathcal{V}(\mathcal{G})=\left\{ v_1, v_2, \dots, v_N \right\}$ be a vertex in a finite graph.
  Its \emph{degree} $d_{v}$ is given by the number of incident edges, counting loops twice.
  The \emph{degree matrix} of a graph is the diagonal matrix $\mathbb{D}=\diag\left( d_{v_1}, d_{v_2}, \dots, d_{v_N} \right)$.
\end{Definition}

\begin{Example}
  \label{exa:degree-of-vertex}
  The degree matrix of the graph $\mathcal{G}_c$ in figure~\ref{fig:labeled-simple-graph-example} is $\mathbb{D}=\diag\left( 3,2,3,2 \right)$.
\end{Example}

\begin{Definition}
  \label{def:adjacency-matrix}
  Let $\mathcal{G}$ be a graph.
  Let $\mathcal{V}(\mathcal{G})=\left\{ v_1, v_2, \dots, v_N \right\}$ be its vertex set and $\mathcal{E}(\mathcal{G})$ be its edge set.
  The matrix $\mathbb{A}\in\mathbb{R}^{N\times N}$ with entries 
  \begin{align*}
    a^i_j \coloneqq &\left|\left\{e\in\mathcal{E}(\mathcal{G})\middle| \iota(e)=\left( v_i, v_j \right) \vee \iota(e)=\left( v_j, v_i \right)\right\}\right| \\
    = &\sum_{e\in\mathcal{E}}\left\llbracket \iota(e)=\left( v_i, v_j \right) \vee \iota(e)=\left( v_j, v_i \right)\right\rrbracket
  \end{align*}
  is the \emph{adjacency matrix} of the graph $\mathcal{G}$.
  For undirected graphs, the tuples $\left( v_i, v_j \right)$ have to replaced by sets $\left\{ v_i, v_j \right\}$.
  Thus, the adjacency matrix is always symmetric.
\end{Definition}

This definition of the adjacency matrix faithfully encodes the structure of undirected finite graphs.
For directed graphs, the \emph{incidence matrix} is a more powerful tool.
I will give a definition in the next section, together with further applications and properties.

\begin{Example}
  \label{exa:adjacency-matrix}
  The adjacency matrix of $\mathcal{G}_c$ is
  \begin{align*}
    \mathbb{A}&=
    \begin{pmatrix}
      0 & 1 & 1 & 1\\
      1 & 0 & 1 & 0\\
      1 & 1 & 0 & 1\\
      1 & 0 & 1 & 0
    \end{pmatrix}\,.
  \end{align*}
\end{Example}

Some concepts can only be applied to directed graphs. Thus, we need a tool to transform an undirected graph into a directed one.

\begin{Definition}
  \label{def:orientation-on-graph}
  Let $\mathcal{G}=\left( \mathcal{V},\mathcal{E},\iota \right)$ be an undirected graph and let $e\in\mathcal{E}$ be an edge.
  Let $\left\{ v_1,v_2 \right\}=\iota(e)$ denote the vertices incident to $e$.
  An assignment $e\mapsto \left( v_1, v_2 \right)\in\mathcal{V}\times\mathcal{V}$ is called an \emph{orientation} of $e$.
  The oriented edge $e$ is denoted $e^{+}$.
  Assigning an orientation to every edge makes $\mathcal{G}$ a directed graph.
\end{Definition}

\begin{Remark}
  \label{rmk:natural-orientation-of-directed-edges}
  In a directed graph, every edge has a natural orientation given by $\iota(e)$.
  We write $e^{+}$ to denote an edge with its natural orientation.
\end{Remark}

\begin{Definition}
  \label{def:inverse-orientation}
  Let $\mathcal{G}$ be a directed graph and $e\in\mathcal{E}(\mathcal{G})$ be an edge.
  Then $\left(t(e),o(e)\right)$ is the \emph{inverted orientation of $e$}.
  The edge with inverted orientation is denoted $e^{-}$.
  Formally, it is not an element of the edge set $\mathcal{E}(\mathcal{G})$.
  Nonetheless, the definition $(e^{-})^{-}\coloneqq e^{+}$ makes sense.
\end{Definition}

In the following, directed edges without superscript have no specified orientation. Inversion will be understood as a relative operation.

\begin{Definition}
  \label{def:walk-on-a-graph}
  Let $\mathcal{G}=\left( \mathcal{V},\mathcal{E},\iota \right)$ be a graph.
  A tuple of consecutively incident vertices and edges $\gamma=(v_0, e_1, v_1, e_2,\dots,v_{n-1}, e_n, v_{n})$ is called a \emph{semiwalk} from $v_0$ to $v_n$.
  If $\mathcal{G}$ is directed, the edges need not occur with their natural orientation.
  If the orientations of all edges are aligned, \ie $t(e_i)=v_i= o(e_{i+1})$ for $1\leq i\leq n-1$, then the semiwalk is a \emph{walk} from $v_0$ to $v_n$.
  Choosing a suitable orientation for the edges turns any semiwalk into a walk.
  For undirected graphs, there is no distinction between walks and semiwalks.
  The number $n$ of edges is referred to as \emph{length} of the semiwalk.
  The ordered set of vertices $(v_0,v_1,\dots,v_n)$ is called \emph{trail} of the semiwalk.
  The first vertex $v_0\eqqcolon o(\gamma)$ is the \emph{origin} of the semiwalk, the last vertex $v_n\eqqcolon t(\gamma)$ is its \emph{end}.
\end{Definition}

\begin{Remark}
  \label{rmk:walk-on-a-graph}
  In the following I will denote a semiwalk by its edges, \ie $\gamma = \left( e_1, \dots, e_n \right)$.
  That should not be a source of confusion.
  For simple graphs, the trail determines a walk, so that sometimes it will be convenient to write $\gamma=\left( v_0, v_1, \dots, v_n \right)$.
\end{Remark}

\begin{Example}
  \label{exa:walk-on-a-graph}
  Let us reconsider the graph $\mathcal{G}_c$ depicted in figure~\ref{fig:labeled-simple-graph-example}.
  The tuple of edges $\gamma=\left( e_1, e_2, e_5, e_4, e_3, e_3, e_4, e_5 \right)$ is a semiwalk from $v_1$ to $v_3$.
  With the orientations $\left( e_1^+, e_2^+, e_5^-, e_4^-, e_3^-, e_3^+, e_4^+, e_5^+ \right)$ it is a walk from $v_1$ to $v_3$.
  Its trail is given by $\left( v_1, v_2, v_3, v_1, v_4, v_3, v_4, v_1, v_3 \right)$. 
\end{Example}

\begin{Definition}
  \label{def:path}
  If all the vertices in the trail of a (semi)walk are distinct, we call the (semi)walk a \emph{(semi)path} from $v_0$ to $v_n$.
  If $o(\gamma)=t(\gamma)$ the (semi)walk or the (semi)path is \emph{closed}.
  A path of length $0$, \ie the trail consists only of one vertex, is called \emph{trivial}.
  A trivial path is always closed.
\end{Definition}

\begin{Definition}
  \label{def:connected-graph}
  A graph is called \emph{connected} if for every two vertices $v$ and $v'$ there is a semipath from $v$ to $v'$.
  Note that on a directed graph every semipath can be turned into a path by aligning the orientations of the edges.
\end{Definition}

\begin{Remark}
  \label{rmk:connected-graph}
  For directed graphs one typically defines more subtle concepts of connectedness, but I will not deal with them in this thesis.
  As already mentioned, I treat directed graphs as undirected when it comes to topology, after all connectedness is a topological concept.
  The rationale is the following:
  With the above definition of connectedness, choosing orientations on a \emph{connected undirected} graph results in a \emph{connected directed} graph.
\end{Remark}

\begin{Definition}
  \label{def:circuit}
  A connected graph with constant vertex degree 2 is a \emph{semicircuit}.
  A semicircuit is a \emph{circuit} if all edges (with their natural orientation) form a path.
  I will denote a simple semicircuit of order $N$ by $\mathcal{C}^{N}$.
  Note that also $\norm{\mathcal{C}^N}=N$.
  Furthermore, for $N\neq 2$ a semicircuit is a simple graph.
\end{Definition}

\begin{figure}[htbp]
  \centering
  \begin{subfigure}[b]{0.48\columnwidth}
    \centering
    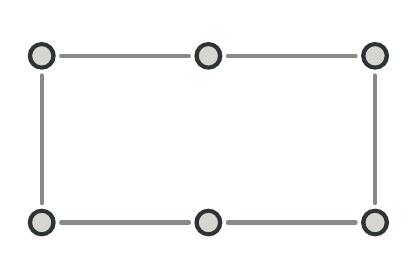
    \caption{undirected circuit}
    \label{fig:undirected-circuit-example}
  \end{subfigure}
  \begin{subfigure}[b]{0.48\columnwidth}
    \centering
    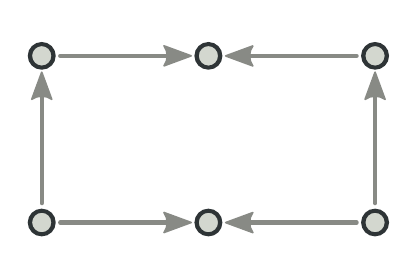
    \caption{directed semicircuit}
    \label{fig:oriented-circuit-example}
  \end{subfigure}
  \caption{Two semicircuits of type $\mathcal{C}^6$.}
  \label{fig:circuits-examples}
\end{figure}

\begin{Example}
  \label{exa:circuit}
  A semicircuit $\mathcal{C}^1$ is a vertex with a loop and therefore always is a circuit.
  Two vertices with two connecting edges are a semicircuit $\mathcal{C}^2$, \cf figure~\ref{fig:not-simple-graphs-example}.
  Two semicircuits $\mathcal{C}^6$ are depicted in figure~\ref{fig:circuits-examples}, one undirected and one directed.
\end{Example}

\begin{Definition}
  \label{def:subgraph}
  Let $\mathcal{G}_1=\left( \mathcal{V}_1,\mathcal{E}_1,\iota_1 \right)$ be a graph.
  Let $\mathcal{V}_2\subset \mathcal{V}_1$ and $\mathcal{E}_2\subset\mathcal{E}_1$ be subsets, such that $\iota_2\coloneqq \left.\iota_1\right|_{\mathcal{E}_2}$ maps to $\mathcal{V}_2\times\mathcal{V}_2$.
  Then $\mathcal{G}_2=\left( \mathcal{V}_2, \mathcal{E}_2,\iota_2 \right)$ is called a \emph{subgraph} of $\mathcal{G}_1$.
  The subgraph $\mathcal{G}_2$ is called \emph{spanning subgraph} if $\mathcal{V}_2=\mathcal{V}_1$.
\end{Definition}

\begin{Definition}
  \label{def:graph-homomorphism}
  Let $\mathcal{G}_1=\left( \mathcal{V}_1,\mathcal{E}_1,\iota_1 \right)$ and $\mathcal{G}_2=\left( \mathcal{V}_2, \mathcal{E}_2,\iota_2 \right)$ be two graphs with incidence maps $\iota_1=(o_1,t_1)$ and $\iota_2=(o_2,t_2)$.
  A mapping $\theta \colon \mathcal{G}_1 \to \mathcal{G}_2$ that is compatible with the incidence maps is called \emph{graph homomorphism}.
  That means it consists of two parts $\theta_\mathcal{V}\colon\mathcal{V}_1\to\mathcal{V}_2$ and $\theta_\mathcal{E}\colon \mathcal{E}_1 \to \mathcal{E}_2$ such that 
  $o_2\circ\theta_\mathcal{E} = \theta_\mathcal{V}\circ o_1$ and
  $t_2\circ\theta_\mathcal{E} = \theta_\mathcal{V}\circ t_1$.
  In case of undirected graphs, a map $\theta$ is a homomorphism if orientations on the graphs exist, such that the above relations are satisfied.
\end{Definition}

\begin{Example}
  \label{exa:subgraph-as-homomorphism}
  Let $\mathcal{G}_2$ be a subgraph of $\mathcal{G}_1$.
  Then the natural inclusion $\mathcal{G}_2\hookrightarrow\mathcal{G}_1$ is a graph homomorphism.
  Any injective graph homomorphism can be understood as a subgraph relation.
\end{Example}

\begin{Definition}
  \label{def:graph-isomorphism}
  A bijective graph homomorphism is called \emph{graph isomorphism}.
  If a graph isomorphism from $\mathcal{G}_1$, onto $\mathcal{G}_2$ exists, the graphs said to be \emph{isomorphic}, written $\mathcal{G}_1\simeq\mathcal{G}_2$.
\end{Definition}

\begin{Definition}
  \label{def:cycle-in-a-graph}
  Let $\mathcal{G}$ be a graph and $\mathcal{C}^N$ be a (semi)circuit.
  Let $\theta\colon\mathcal{C}^N\to\mathcal{G}$ a graph homomorphism and let $\zeta=\theta(\mathcal{C}^N)$ be the image.
  If $\theta$ is injective, then $\zeta\simeq\mathcal{C}^N$ is a \emph{(semi)circuit in $\mathcal{G}$}.
  If $\theta$ is not injective, then the image $\zeta=\theta(\mathcal{C}^N)$ corresponds to a class of closed (semi)walks of length $N$ whose sequences of edges are identical up to cyclic permutations.
  In this case $\zeta$ is called \emph{(semi)cycle in $\mathcal{G}$}.
\end{Definition}

Obviously, not all graphs have semicircuits as subgraphs.

\begin{figure}[htbp]
  \centering
  \includegraphics{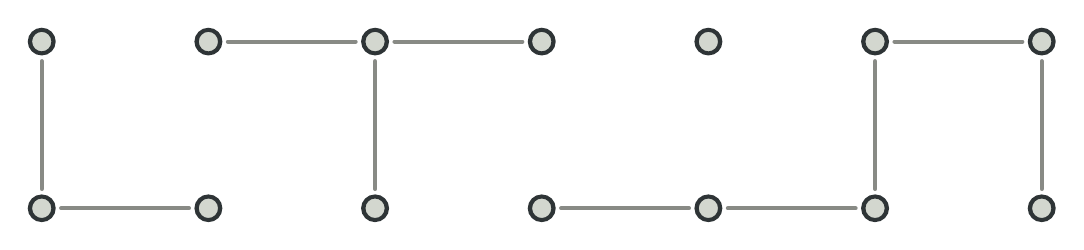}
  \caption{
    An undirected forest $\mathcal{F}$ with $\abs{\mathcal{F}}=14$ and $\norm{\mathcal{F}}=10$.
    It consists of 4 connected components, each of which is a tree.
  }
  \label{fig:undirected-forest-example}
\end{figure}

\begin{Definition}
  \label{def:forest-tree}
  A graph that does not contain a semicircuit as a subgraph is called \emph{forest}.
  A connected forest is a \emph{tree}.
  An example is depicted in figure~\ref{fig:undirected-forest-example}.
\end{Definition}

\begin{Proposition}
  \label{thm:edges-in-a-forest}
  Every connected component of a forest is a tree.
  For every finite forest $\mathcal{F}$ with $k$ connected components we have $\norm{\mathcal{F}}= \abs{\mathcal{F}} -k$.
\end{Proposition}

\begin{Proposition}
  \label{thm:paths-in-trees}
  Let $\mathcal{T}$ be a tree and $v_1, v_2\in\mathcal{V}(\mathcal{T})$ be two vertices. Then there is a unique path from $v_1$ to $v_2$.
\end{Proposition}

\section{Algebraic Structures on Graphs}
\label{sec:algebra-on-graphs}

In this section I construct vector spaces over a field $\mathbb{F}$ for a given finite graph $\mathcal{G}$.
Algebraic methods provide a nice language and make many proofs very short and intuitive.

For undirected graphs I always assume the two-element field, $\mathbb{F}_2$, for directed graphs I assume the field of real numbers, $\mathbb{R}$.
This difference allows to basically use the same definitions, as will become clear later on.
For further reading on algebraic graph theory I recommend the book of \citet{Knauer2011} in which the omitted proofs of this section can be found.

\begin{Definition}
  \label{def:vertex-edge-space}
  Let $\mathcal{G}=\left( \mathcal{V},\mathcal{E},\iota \right)$ be a graph.
  The space $C_{0}(\mathcal{G})\coloneqq \mathbb{F}^{\mathcal{V}}$ is called \emph{vertex space}, the space $C_{1}(\mathcal{G})\coloneqq \mathbb{F}^{\mathcal{E}}$ is the \emph{edge space}.
  The elements of $C_0(\mathcal{G})$ are called \emph{0-chains}, the elements of $C_1(\mathcal{G})$ are \emph{1-chains}.
\end{Definition}

\begin{Definition}
  \label{def:vertex-edge-indicator}
  Given a vertex $x\in\mathcal{V}$ or an edge $x\in\mathcal{E}$.
  Its indicator function is
  \begin{align*}
    \mathbb{1}_{x}\colon y \mapsto \left\llbracket x=y \right\rrbracket
  \end{align*}
  and an element of $C_0(\mathcal{G})$ or $C_1(\mathcal{G})$, respectively.
  Thus $\mathbb{1}_{\mathcal{V}} \subset C_0(\mathcal{G})$ and $\mathbb{1}_{\mathcal{E}}\subset C_{1}(\mathcal{G})$.
\end{Definition}

\begin{Proposition}
  \label{thm:indicator-basis}
  The two sets $\mathbb{1}_{\mathcal{V}}$ and $\mathbb{1}_{\mathcal{E}}$ form bases of the corresponding $\mathbb{F}$-vector spaces.
\end{Proposition}

\begin{Definition}
  \label{def:standard-bases}
  The basis in the preceding proposition are called \emph{standard vertex basis} and \emph{standard edge basis}.
\end{Definition}

\begin{Proposition}
  \label{thm:vertex-edge-space-dimensions}
  Obviously we have $\dim C_0(\mathcal{G}) = \abs{\mathcal{G}}$ and $\dim C_1(\mathcal{G}) = \norm{\mathcal{G}}$.
\end{Proposition}

\begin{Remark}
  \label{rmk:vertex-space-span-of-set}
  In the following I will always identify a vertex or an edge with its indicator function and regard the vertex and edge spaces as linear spans of the vertex and edge sets.
  The formal difference is that the basis $\mathbb{1}_{\mathcal{V}}$ is a tuple of vectors.
  Thus, ordering of the elements is important, whereas in the set $\mathcal{V}$ the order is irrelevant.
\end{Remark}

\begin{Remark}
  \label{rmk:inverted-edges-in-edge-space}
  Let $\mathcal{G}$ be a directed graph and $e\in\mathcal{E}(\mathcal{G})$ be an edge.
  As already noted, the edge with inverted orientation $e^{-}$ formally is not an element of $\mathcal{E}(\mathcal{G})$. However, it can be identified with the 1-chain $(-1)\cdot e\in C_1(\mathcal{G})$.
\end{Remark}

\begin{Definition}
  \label{def:scalar-product}
  For $\ell\in\left\{ 0,1 \right\}$ let $\left( x_1, \dots, x_N \right)$ be the standard basis of $C_\ell\left( \mathcal{G} \right)$.
  The \emph{standard scalar product} on $C_\ell\left( \mathcal{G} \right)$ is given by the linear extension of
  \begin{align*}
    \left\langle x_i, x_j \right\rangle \coloneqq \otherdelta^{i}_{j}\equiv\left\llbracket i=j \right\rrbracket\,.
  \end{align*}
\end{Definition}

\begin{Remark}
  \label{rmk:dual-space}
  The edge space $C_1(\mathcal{G})$ is a finite dimensional vector space. Therefore, the scalar product gives a tool to identify $C_1(\mathcal{G})$  with its dual space $C_1^{*}(\mathcal{G})$, \ie the space of linear functionals on $C_1(\mathcal{G})$.
  So in view of remark~\ref{rmk:inverted-edges-in-edge-space}, it makes sense to define $f(e^{-})\coloneqq f(-e)\coloneqq - f(e)$.
  Thus, in a sense, 1-chains are “linear”.
  Functions on oriented edges not satisfying anti-symmetry are not considered elements of the edge space $C_1(\mathcal{G})$.
\end{Remark}

\begin{Definition}
  \label{def:components-of-vector}
  The numbers $f^{j}\coloneqq \left\langle f, x_j\right\rangle\in\mathbb{F}$ are the \emph{components} of $f\in C_\ell(\mathcal{G})$ with respect to the basis $\left( x_1, \dots, x_N \right)$.
  Thus every $\ell$-chain can uniquely be written as 
  \begin{align*}
    f &= \sum_{j=1}^{N} f^{j} x_j\eqqcolon \left( f^{1}, \dots, f^{N} \right)\,.
  \end{align*}
\end{Definition}

There are two important maps taking 0-chains into 1-chains and vice versa:

\begin{Definition}
  \label{def:boundary}
  Let $e\in\mathcal{E}(\mathcal{G})\subset C_1(\mathcal{G})$ be an edge.
  Then the 0-chain
  \begin{align*}
    \del e \coloneqq o(e) - t(e) \in C_0(\mathcal{G})
  \end{align*}
  is called the \emph{boundary} of $e$.
\end{Definition}

\begin{Definition}
  \label{def:coboundary}
  Let $v\in\mathcal{V}(\mathcal{G})\subset C_0(\mathcal{G})$ be a vertex and $\mathcal{E}=\mathcal{E}(\mathcal{G})\subset C_1(\mathcal{G})$ be the edge set.
  Then the 1-chain
  \begin{align*}
    \ddel{} v \coloneqq \sum_{e\in\mathcal{E}}\left( \left\llbracket v=o(e) \right\rrbracket - \left\llbracket v=t(e) \right\rrbracket \right) e \in C_1(\mathcal{G})
  \end{align*}
  is the \emph{coboundary} of $v$.
\end{Definition}

\begin{Remark}
  \label{rmk:undirected-boundary}
  Note that, if $\mathcal{G}$ is undirected, the above definitions still make sense:
  In $\mathbb{F}_2$ addition and subtraction are the exact same operation and $o(e)$ and $t(e)$ are just the two vertices incident to the edge $e$, in arbitrary order.
\end{Remark}

\begin{Definition}
  \label{def:boundary-coboundary-operators}
  The two operations $\del$ and $\ddel{}$ can be linearly extended to the spaces $C_1(\mathcal{G})$ and $C_0(\mathcal{G})$, respectively. They are the \emph{boundary} and \emph{coboundary operators}.
\end{Definition}

\begin{Remark}
  \label{rmk:coboundary-operator-is-derivative}
  Both the coboundary and the boundary operator can be thought of as some form of discrete derivative.
\end{Remark}

\begin{Example}
  \label{exa:boundary-coboundary-operators}
  Let the vertices and edges of the graph $\mathcal{G}_c$ be labelled as in figure~\ref{fig:labeled-simple-graph-example}.
  Then we have 
  \begin{align*}
    \del\left( e_1 + e_2 \right) &= v_1 - v_2 + v_2 - v_3 = v_1 - v_3 \\
    \ddel\left( v_1 + v_2 + v_3 \right) &= - e_4 + e_5 + e_1 -e_1 + e_2 -e_2 -e_5 +e_3 = e_3 - e_4 
  \end{align*}
\end{Example}

\begin{Proposition}
  \label{thm:boundary-coboundary-dual}
  The boundary and coboundary operators are dual with respect to the standard scalar products on $C_0(\mathcal{G})$ and $C_1(\mathcal{G})$.
  That means for any $f_1 \in C_1(\mathcal{G})$ and $f_0 \in C_0(\mathcal{G})$ we have
  \begin{align*}
    \left\langle \del f_1, f_0 \right\rangle = \left\langle f_1, \ddel{} f_0 \right\rangle\,.
  \end{align*}
  \begin{Proof}
    Let $(v_1, \dots, v_N)$ and $(e_1, \dots, e_M)$ denote the standard bases.
    Then for any $v_j, e_i$ we have
    \begin{align*}
      \left\langle \ddel v_j, e_i\right\rangle
      &= \sum_{\ell=1}^{M}\left( \left\llbracket v_j=o(e_\ell) \right\rrbracket- \left\llbracket v_j=t(e_{\ell}) \right\rrbracket \right) \left\langle e_\ell, e_i\right\rangle \\
      &= \left\llbracket v_j=o(e_i) \right\rrbracket- \left\llbracket v_j=t(e_{i}) \right\rrbracket
      = \left\langle v_j, o(e_i)\right\rangle - \left\langle v_j, t(e_i)\right\rangle\\
      &= \left\langle v_j, \del e_i\right\rangle\,.
    \end{align*}
    Linearity completes the proof.
  \end{Proof}
\end{Proposition}

\begin{Proposition}
  \label{thm:cycles-have-no-boundary}
  Let $\mathcal{G}$ be a graph. A cycle $\zeta=\left( e_1, \dots, e_n \right)$ can be identified with the element $e_1 + \dots + e_n\in C_1(\mathcal{G})$. With this identification we can say that $\zeta$ does not have a boundary or $\del\zeta=0$.
\end{Proposition}

This observation justifies the following 

\begin{Definition}
  \label{def:cycle-space}
  The kernel of the boundary operator is the \emph{cycle space},
  \begin{align*}
    Z(\mathcal{G}) \coloneqq \ker \del \subset C_1(\mathcal{G})\,,
  \end{align*}
  its elements are \emph{cycles}. This generalizes definition~\ref{def:cycle-in-a-graph}.
  The image of the coboundary operator is called \emph{cocycle space},
  \begin{align*}
    Z^{\perp}(\mathcal{G}) \coloneqq \imag \ddel{} \subset C_1(\mathcal{G})\,,
  \end{align*}
  its elements are \emph{cocycles}.
\end{Definition}

\begin{Proposition}
  \label{thm:cycle-orthogonal-cocycle}
  With the standard scalar product on $C_1(\mathcal{G})$, the cocycle space is the orthogonal complement of the cycle space, and therefore $C_1(\mathcal{G}) = Z(\mathcal{G}) \oplus Z^{\perp}(\mathcal{G})$.
  \begin{Proof}
	  Let $z\in Z(\mathcal{G})$ be a cycle and let $\ddel f \in Z^{\perp}(\mathcal{G})$ be a cocycle, where $f\in C_0(\mathcal{G})$ is a 0-chain.
	  Then
	  \begin{align*}
		  \left\langle \ddel f, z \right\rangle = \left\langle f , \del z \right\rangle = 0\,.
	  \end{align*}
  \end{Proof}
\end{Proposition}

\begin{Example}[Electrical networks]
  \label{exa:cycles-cocycles-in-electric-networks}
  Cycles and cocycles are a mathematical generalization of what was known in physics for a long time:
  Electric potentials $u$ are functions on the vertices, \ie elements of $C_0(\mathcal{G})$.
  Their edgewise differences $\ddel u\in Z^{\perp}(\mathcal{G})\subset C_1(\mathcal{G})$ are voltages.
  Thus proposition~\ref{thm:cycle-orthogonal-cocycle} is Kirchhoff’s mesh law.
  Currents are functions on the edges as well.
  On every vertex in the network the currents add up to zero, according to Kirchhoff’s current law.
  In other words, currents have no boundary and thus are elements of the cycle space $Z(\mathcal{G})$.
\end{Example}

\begin{Definition}
  \label{def:incidence-matrix}
  The matrix representation of the boundary operator $\del$ in the standard bases of $C_0(\mathcal{G})$ and $C_1(\mathcal{G})$ is called \emph{incidence matrix} and denoted as $\mathbb{B}$.
  Its transpose, $\mathbb{B}\transpose$, is the matrix representation of $\ddel$ in the standard bases.
\end{Definition}

  The incidence matrix encodes the structure of a directed graph.
  However, this definition of the incidence matrix does not encode loops.
  They have no boundary and do not lie in the image of the coboundary operator.

\begin{Example}
  \label{exa:incidence-matrix}
  Let the edges and vertices of the graph $\mathcal{G}_c$ be labeled as in figure~\ref{fig:labeled-simple-graph-example}.
  Then its incidence matrix is given by
  \begin{align*}
    \mathbb{B}=
    \begin{pmatrix}
      1 & 0 & 0 & -1 & 1\\
      -1 & 1 & 0 & 0 & 0\\
      0 & -1 & 1 & 0 & -1\\
      0 & 0 & -1 & 1 & 0
    \end{pmatrix}
  \end{align*}
\end{Example}

\begin{Definition}
  \label{def:combinatorial-laplacian}
  The operator $\Delta=\del\circ\ddel{}\colon C_0(\mathcal{G}) \to C_0(\mathcal{G})$ is called \emph{combinatorial Laplacian}. 
  Its matrix representation in the standard basis is $\mathbb{L}=\mathbb{BB}\transpose$.
\end{Definition}

\begin{Example}
  \label{exa:combinatorial-laplacian}
  For the graph $\mathcal{G}_c$ we have
  \begin{align*}
    \mathbb{L}=
    \begin{pmatrix}
      3 & -1 & -1 & -1\\
      -1 & 2 & -1 & 0\\
      -1 & -1 & 3 & -1\\
      -1 & 0 & -1 & 2\\
    \end{pmatrix}\,.
  \end{align*}
\end{Example}

\begin{Proposition}
  \label{thm:laplacian-degree-adjacency}
  Let $\mathcal{G}$ be a directed graph without loops and let $\mathbb{A}$ be its adjacency matrix and $\mathbb{D}$ its degree matrix, \cf definitions~\ref{def:adjacency-matrix} and \ref{def:degree-of-vertex}.
  Then $\mathbb{L}=\mathbb{D}-\mathbb{A}$.
  Hence, the Laplacian is invariant under change of orientation of any edge.\,\citep{Cvetkovic1980}
\end{Proposition}

\section{Topological Concepts for Graphs}
\label{sec:topology-for-graphs}

Topology helps to understand the basic structure of a graph.
Moreover, topological considerations allow us to find intuitive bases of the cycle space $Z(\mathcal{G})$ and the cocycle space $Z^{\perp}(\mathcal{G})$.
The omitted proofs of this section can be found in the references \citep{Gross1987,Knauer2011}.

\begin{Definition}
  \label{def:inverse-walk}
  Let $\gamma=\left( e_1,e_2,\dots, e_{n-1}, e_n \right)$ be a semiwalk on a graph.
  The \emph{reverse semi\-walk} is defined to be $\gamma^{-1}\coloneqq\left( e_n^{-}, e_{n-1}^{-}, \dots, e_2^{-}, e_1^{-} \right)$.
  On undirected graphs, there is no orientation and consequently no edgewise orientation inversion is required, only the order is reversed.
\end{Definition}

\begin{Definition}
  \label{def:composition-of-walks}
  Let $\gamma_1=\left( e_{1}, e_{2}, \dots, e_n \right)$ be a semiwalk from $v_0$ to $v_n$.
  Furthermore, let $\gamma_2=\left( e_{n+1}, e_{n+2}, \dots, e_{m} \right)$ be a semiwalk from $v_n$ to $v_m$.
  The semiwalk
  \begin{align*}
    \gamma_2\circ\gamma_1\coloneqq \left( e_1, e_1, \dots, e_n, e_{n+1}, e_{n+2}, \dots, e_m \right)
  \end{align*}
  from $v_0$ to $v_m$ is the \emph{composition} of the two semiwalks.
\end{Definition}

\begin{Definition}
  \label{def:reduction-of-walks}
  Let $\gamma=\left( e_1, e_2, \dots, e_{i-1}, e_{i}, e_{i}^{-}, e_{i+1}, \dots, e_n \right)$ be a semiwalk with one edge occurring multiple times in succession but (on directed graphs) with opposite orientation.
  The semiwalk $\gamma'=\left( e_1, e_2, \dots, e_{i-1}, e_{i+1}, \dots e_n \right)$ is a \emph{reduction} of $\gamma$.
  On the other hand, $\gamma$ is an \emph{expansion} of $\gamma'$.
\end{Definition}

\begin{Remark}
  \label{rmk:reduction-of-walks}
  The composition of a semiwalk with its reverse semiwalk reduces to a trivial path.
\end{Remark}

\begin{Definition}
  \label{def:fundamental-group}
  Let $\mathcal{G}$ be a graph and let $v\in\mathcal{V}(\mathcal{G})$ be a vertex. 
  Let $\otherGamma\!_v$ denote the set of all closed walks starting and ending at $v$.
  Two closed walks $\gamma, \gamma' \in \otherGamma\!_v$ are \emph{homotopic} if they can be transformed into one another by reduction or expansion of an arbitrary amount of edges.
  We denote this equivalence relation by $\gamma \sim \gamma'$.
  The factor space $\otherGamma\!_v/\sim$ forms a group under composition of walks.
  This group is denoted as $\pi_1(\mathcal{G},v)$.
  If $\mathcal{G}$ is connected, then for another vertex $v'$ we have $\pi_1(\mathcal{G},v)\simeq\pi_1(\mathcal{G},v')$.
  This abstract group $\pi_1(\mathcal{G})$ is the \emph{fundamental group} of the connected graph $\mathcal{G}$.
\end{Definition}

\begin{Proposition}
  \label{thm:fundamental-group-of-a-tree}
  Let $\mathcal{T}$ be a tree. 
  Then the fundamental group $\pi_1(\mathcal{T})$ is the trivial group consisting of one element only.
  This one element is the homotopy class of the trivial path.
  The converse is also true: If $\mathcal{G}$ is a connected graph with trivial fundamental group, then it is a tree.
\end{Proposition}

\begin{Example}
  \label{exa:fundamental-group-of-simple-cycle}
  Let $\mathcal{C}^N$ be a circuit.
  Then the fundamental group $\pi_1(\mathcal{C}^N)\simeq \mathbb{Z}$ is given by the integers.
  It counts the winding number of a closed walk in $\mathcal{C}^N$.
\end{Example}

For a graph with non-trivial fundamental group, there is a connection between the fundamental group and special subgraphs, the spanning trees:

\begin{Definition}
  \label{def:spanning-tree}
  Let $\mathcal{G}$ be a graph. Every spanning subgraph that is a forest is called \emph{spanning forest}, a connected spanning forest is a \emph{spanning tree}.
\end{Definition}

\begin{Proposition}
  \label{thm:existence-spanning-tree}
  Every graph has a spanning forest. Every connected graph has a spanning tree.
\end{Proposition}

\begin{Remark}
  \label{rmk:existence-spanning-tree}
  Typically, a connected graph has many different spanning trees.
\end{Remark}

\begin{Definition}
  \label{def:chords}
  Let $\mathcal{G}$ be a connected graph with spanning tree $\mathcal{T}$.
  The elements of $\mathcal{H}(\mathcal{T})\coloneqq\mathcal{E}(\mathcal{G})\setminus\mathcal{E}(\mathcal{T})$ are called \emph{chords} of the spanning tree $\mathcal{T}$.
\end{Definition}

\begin{figure}[htbp]
  \centering
  \begin{subfigure}[b]{0.45\columnwidth}
    \centering
    \includegraphics{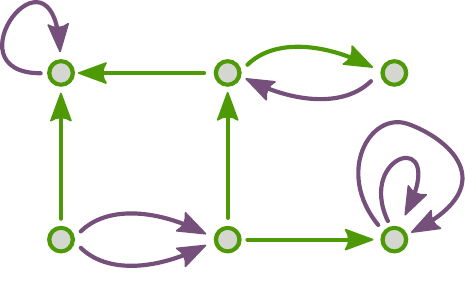}
    \caption{A spanning tree of $\mathcal{G}_a$.}
    \label{fig:directed-spanning-tree-example}
  \end{subfigure}
  \hfill
  \begin{subfigure}[b]{0.45\columnwidth}
    \centering
    \includegraphics{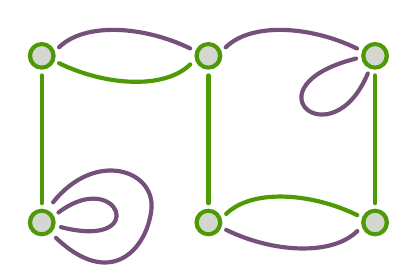}
    \caption{A spanning tree of $\mathcal{G}_b$.}
    \label{fig:undirected-spanning-tree-example}
  \end{subfigure}
  \caption{
  Examples of spanning trees for the graphs from figure~\ref{fig:graph-examples}.
  The spanning trees are depicted in green, their chords are colored violet.
  }
  \label{fig:spanning-trees-examples}
\end{figure}

Spanning trees and chords are tools to construct very intuitive bases for the cycle and the cocycle spaces. We begin with the cycle space $Z(\mathcal{G})$:

\begin{Definition}
  \label{def:fundamental-cycle}
  Adding a chord $\eta\in\mathcal{H}(\mathcal{T})$ to a spanning tree $\mathcal{T}$ creates a semicircuit $\zeta\in C_1(\mathcal{G})$ as a subgraph of $\mathcal{T}+\eta$.
  Aligning the orientation of its edges in the tree to fit the orientation of $\eta$, if necessary, turns $\zeta$ into a circuit $\zeta_\eta\in Z(\mathcal{G})$.
  This circuit $\zeta_\eta$ is called \emph{fundamental cycle} corresponding to $\eta$.
\end{Definition}

\begin{figure}[t]
  \centering
  \begin{subfigure}[b]{0.34\columnwidth}
    \centering
    \includegraphics{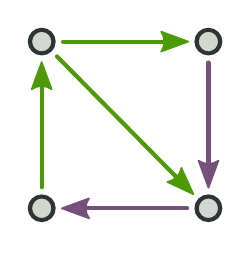}
    \caption{A spanning tree of $\mathcal{G}_c$.}
    \label{fig:simple-spanning-tree-example}
  \end{subfigure}
  \hfill
  \begin{subfigure}[b]{0.64\columnwidth}
    \centering
    \includegraphics{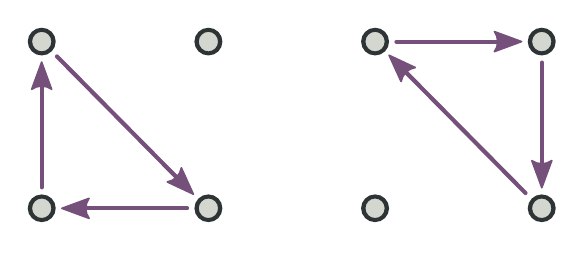}
    \caption{Two fundamental cycles of $\mathcal{G}_c$.}
    \label{fig:fundamental-cycles-example}
  \end{subfigure}
  \caption{
  A spanning tree for the graph $\mathcal{G}_c$ from figure~\ref{fig:labeled-simple-graph-example} directly gives a set of fundamental cycles:
  The spanning tree (\subref{fig:simple-spanning-tree-example}) is depicted in green, its chords are colored violet.
  Every chord provides one fundamental cycle in (\subref{fig:fundamental-cycles-example}).
  The exact construction is given in definition~\ref{def:fundamental-cycle}.
  }
  \label{fig:simple-fundamental-cycles-examples}
\end{figure}

A non-trivial example for fundamental cycles is shown in figure~\ref{fig:simple-fundamental-cycles-examples}.

\begin{Proposition}
  \label{thm:fundamental-cycles-basis}
  Let $\mathcal{T}$ be a spanning tree of a graph $\mathcal{G}$.
  The set $\left\{ \zeta_{\eta} \middle| \eta\in\mathcal{H}(\mathcal{T}) \right\}$ of fundamental cycles is a basis of the cycle space $Z(\mathcal{G})$.
\end{Proposition}

\begin{Remark}
  \label{rmk:fundamental-cycles-basis}
  This basis is not ortho-normalized with respect to the standard scalar product: Let $\eta_i, \eta_j\in\mathcal{H}(\mathcal{T})$ be two chords with fundamental cycles $\zeta_i, \zeta_j$.
  The two fundamental cycles have $\left|\left\langle \zeta_i, \zeta_j \right\rangle\right|$ edges in common.
  The sign of the scalar product indicates whether these common edges are aligned.
  However, by construction we have
  $\left\langle \eta_i, \zeta_j\right\rangle= \zeta_j(\eta_i)=\llbracket i=j \rrbracket$
  so we can write every cycle $z \in Z(\mathcal{G})$ as 
  \begin{align*}
    z= \sum_{\eta\in\mathcal{H}} z(\eta)\, \zeta_{\eta}\,.
  \end{align*}
\end{Remark}

\begin{Proposition}
  \label{thm:fundamental-cycles-fundamental-group}
  The fundamental cycles can be identified with generators of the fundamental group $\pi_1(\mathcal{G})$.
\end{Proposition}

\begin{Definition}
  \label{def:betti-number}
  The number of chords is always $b_1\coloneqq\dim Z(\mathcal{G}) = \norm{\mathcal{G}}-\abs{\mathcal{G}}+1$ and therefore does not depend on the spanning tree.
  This number $b_1$ also counts the number of generators of $\pi_1(\mathcal{G})$ and thus is a topological property of the graph $\mathcal{G}$. It is called \emph{1st Betti number} or \emph{cyclomatic number}.
\end{Definition}

There is also a very nice basis for the cocycle space $Z^{\perp}(\mathcal{G})$ that can be constructed from a given spanning tree $\mathcal{T}$:

\begin{Definition}
  \label{def:fundamental-cocycle}
  The spanning tree $\mathcal{T}$ connects all the vertices of $\mathcal{G}$.
  Let $\tau\in\mathcal{E}(\mathcal{T})$ be any edge in the spanning tree.
  Removing this edge from the spanning tree results in a disconnected graph $\mathcal{T}\setminus\{\tau\} = \mathcal{U}_1 \cup \mathcal{U}_2$ with two connected components $\mathcal{U}_1$ and $\mathcal{U}_2$, where $t(\tau)\in \mathcal{V}(\mathcal{U}_1)$ and $o(\tau)\in\mathcal{V}(\mathcal{U}_2)$.
  Now we define a 0-chain $u_\tau$ via
    $u_\tau(v)\coloneqq i$ for $v \in \mathcal{V}(\mathcal{U}_i)$
  and $i\in\left\{ 1,2 \right\}$.
  The edge-wise difference
  $\zeta^{*}_{\tau}\coloneqq \ddel u_\tau$
  is a 1-chain called \emph{fundamental cocycle} corresponding to the edge $\tau$.
  Note that in the case of an undirected graph, we have $2=1+1=0$ in $\mathbb{F}_2$, so the above definitions still make sense.
\end{Definition}

\begin{Remark}
  \label{rmk:fundamental-cocycles}
  The fundamental cocycle $\zeta^{*}_{\tau}$ is a sum of the edge $\tau$ and some chords:
  The chord set $\mathcal{H}=\mathcal{H}(\mathcal{T})$ decomposes into the chords $\mathcal{H}_i\coloneqq\mathcal{H}\cap\mathcal{E}(\mathcal{U}_i)$ contained in $\mathcal{U}_i$ for $i=\left\{ 1,2 \right\}$ and $\mathcal{H}_0\coloneqq \mathcal{H}\setminus\left( \mathcal{H}_1\cup\mathcal{H}_2 \right)$ connecting $\mathcal{U}_1$ and $\mathcal{U}_2$.
  Orienting the edges $\eta\in\mathcal{H}_0$ such that $t(\eta)\in\mathcal{V}(\mathcal{U}_1)$ and $o(\eta)\in\mathcal{V}(\mathcal{U}_2)$, we have the representation
  \begin{align*}
    \zeta^{*}_\tau = \tau + \sum_{\eta\in\mathcal{H}_0} \eta\,.
  \end{align*}
  In other words: The fundamental cocycle is the sum of all edges of $\mathcal{G}$ connecting $\mathcal{U}_2$ and $\mathcal{U}_1$, possibly reoriented to point from $\mathcal{U}_2$ to $\mathcal{U}_1$.
\end{Remark}

\begin{figure}[t]
  \centering
  \begin{subfigure}[b]{0.30\columnwidth}
    \centering
    \includegraphics{./figures/simple-spanning-tree-example}
    \caption{A spanning tree of $\mathcal{G}_c$.}
    \label{fig:simple-spanning-tree-cocycle-example}
  \end{subfigure}
  \hfill
  \begin{subfigure}[b]{0.68\columnwidth}
    \centering
    \includegraphics{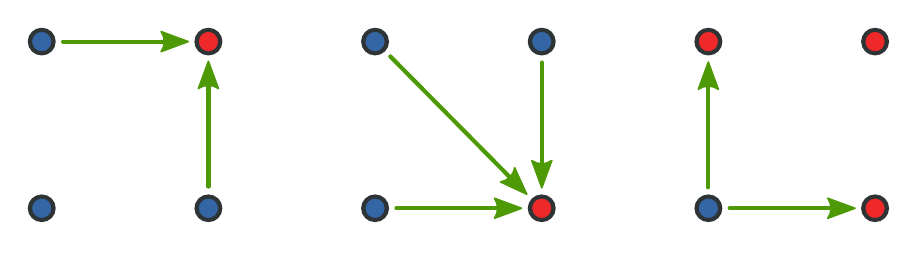}
    \caption{Three fundamental cocycles of $\mathcal{G}_c$.}
    \label{fig:fundamental-cocycles-example}
  \end{subfigure}
  \caption{
  Starting with the spanning tree (\subref{fig:simple-spanning-tree-cocycle-example}) of the graph $\mathcal{G}_c$, we can construct fundamental cocycles (\subref{fig:fundamental-cocycles-example}):
  For every edge of the spanning tree, the graph itself decomposes into two subgraphs, \cf definition~\ref{def:fundamental-cocycle}.
  The vertices of the subgraph $\mathcal{U}_1$ are colored red, the vertices of the subgraph $\mathcal{U}_2$ are colored blue.
  The fundamental cocycle is the sum of all (possibly reoriented) edges that connect $\mathcal{U}_2$ and $\mathcal{U}_1$.
  }
  \label{fig:simple-fundamental-cocycles-examples}
\end{figure}

A non-trivial example for fundamental cocycles is shown in figure~\ref{fig:simple-fundamental-cocycles-examples}.

\begin{Proposition}
  \label{thm:fundamental-cocycles-basis}
  Let $\mathcal{T}$ be a spanning tree of a graph $\mathcal{G}$.
  The set $\left\{ \zeta^{*}_{\tau} \middle| \tau\in\mathcal{E}(\mathcal{T}) \right\}$ of fundamental cocycles is a basis of the cocycle space $Z^{\perp}(\mathcal{G})$.
\end{Proposition}

\begin{Remark}
  \label{rmk:fundamental-cocycles-basis}
  Just as for fundamental cycles, this basis of fundamental cocycles is in general not ortho-normalized with respect to the standard scalar product.
  However, let $\tau_i, \tau_j\in\mathcal{E}(\mathcal{T})$ be two edges of the spanning tree with fundamental cocycles $\zeta^{*}_i, \zeta^{*}_j$.
  By construction we have
  $\left\langle \tau_i, \zeta^{*}_j\right\rangle= \zeta^{*}_j(\tau_i)=\llbracket i=j \rrbracket$
  so we can write every cocycle $y \in Z^{\perp}(\mathcal{G})$ as 
  \begin{align*}
    y= \sum_{\tau\in\mathcal{E}(\mathcal{T})} y(\tau)\, \zeta^{*}_{\tau}\,.
  \end{align*}
\end{Remark}

Another very powerful concept from topology that is tightly related to the fundamental group are the so called covering spaces:

\begin{Definition}
  \label{def:neighborhood-of-a-vertex}
  Given a vertex $v$ in a graph $\mathcal{G}$.
  The subgraph of $\mathcal{G}$ that contains $v$ and all its incident edges and adjacent vertices is the \emph{neighborhood} of the vertex $v$, abbreviated as $\nbh(v)$.
  The set of edges $\out(v)\coloneqq\left\{ e\in\mathcal{E}(\mathcal{G})\middle| o(e)=v \right\}$ originating at $v$ is the \emph{outset} of $v$.
\end{Definition}

\begin{Definition}
  \label{def:covering}
  A \emph{covering} of a directed graph $\mathcal{G}$ is a directed graph $\widetilde{\mathcal{G}}$ together with a graph homomorphism $\theta\colon \widetilde{\mathcal{G}} \to \mathcal{G}$ such that $\theta_\mathcal{V}$ is surjective and for every vertex $\tilde{v}$ in $\widetilde{\mathcal{G}}$, $\theta_\mathcal{E}$ is a bijection of $\out(\tilde{v})$ onto $\out(\theta_\mathcal{V}(\tilde{v}))$.
  In this context, the graph $\mathcal{G}$ is called \emph{base space}, $\widetilde{\mathcal{G}}$ is the \emph{total space} or \emph{covering space} and $\theta$ is the \emph{covering map}.
  For $v\in\mathcal{V}(\mathcal{G})$ the set $\theta_\mathcal{V}^{-1}(v)$ is called \emph{fiber} over $v$.
  Analogously, one defines the fiber over an edge.
  For undirected graphs one first has to choose an orientation of the edges. However, the covering does not depend on the orientations so they can as well be dropped in the end.
\end{Definition}

\begin{Proposition}
  \label{thm:covering-local-isomorphism}
  For simple loop-less graphs, the above definition is equivalent to $\theta$ being a local isomorphism.
  That means $\theta$ maps any neighborhood $\nbh(\tilde{v})$ bijectively onto $\nbh{\theta_{\mathcal{V}}(\tilde{v})}$.
\end{Proposition}

\begin{Definition}
  \label{def:universal-covering}
  Let $\mathcal{G}$ be a graph and let $\theta\colon\widetilde{\mathcal{G}}\to\mathcal{G}$ be a covering of $\mathcal{G}$.
  If $\widetilde{\mathcal{G}}$ is a tree, the covering is called \emph{universal covering}.
\end{Definition}

\begin{Proposition}
  \label{thm:universal-covering}
  For every connected graph $\mathcal{G}$, a universal covering exists.
  Given two universal coverings $\theta_1\colon\widetilde{\mathcal{G}}_1\to\mathcal{G}$ and $\theta_2\colon\widetilde{\mathcal{G}}_2\to\mathcal{G}$.
  Then $\widetilde{\mathcal{G}}_1$ and $\widetilde{\mathcal{G}}_2$ are isomorphic.
\end{Proposition}

\begin{Proposition}
  \label{thm:infinite-covering}
  Let $\mathcal{G}$ be a connected graph with cyclomatic number $b_1>0$.
  Then every universal covering is an infinite graph.
\end{Proposition}

\begin{Example}
  \label{exa:universal-covering-of-a-cycle}
  Let $\mathcal{C}^N$ be an undirected circuit.
  Let $\mathcal{V}(\mathcal{C}^N)=\left\{ 0, \dots, N-1 \right\}$ denote its vertices.
  Let $\mathcal{Z}$ denote the tree with vertex set $\mathbb{Z}$ and undirected edges connecting two successive integers. 
  The operation $\mathrm{mod}\, N\colon \mathbb{Z}\to\{0,\dots,N-1\}$ naturally generalizes to a covering map $\mathrm{mod}\, N\colon \mathcal{Z}\to\mathcal{C}^N$.
  Thus, the integers are a universal covering of any circuit.
\end{Example}

\begin{figure}[htbp]
  \centering
  \includegraphics{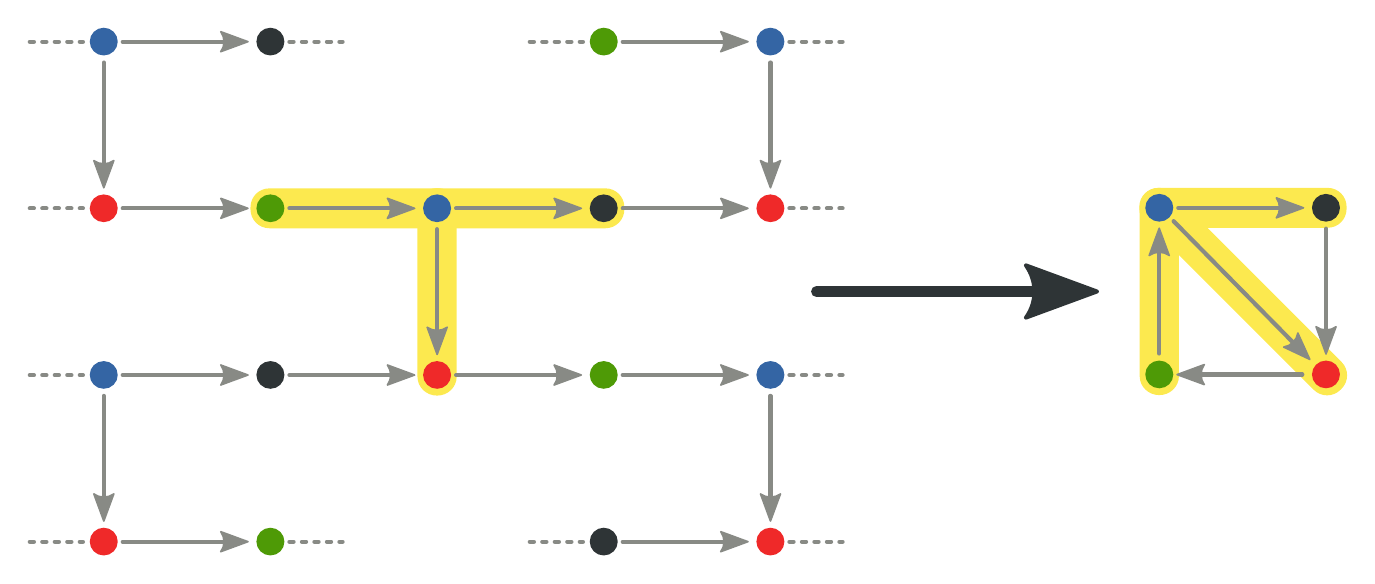}
  \caption{
  A universal covering of the simple directed graph $\mathcal{G}_c$ with cyclomatic number $b_1=2$.
    Vertices with identical color are in the same fibre.
    The neighborhood of a blue vertex in the covering space and its image in the base space are highlighted in yellow.
  }
  \label{fig:universal-covering-example}
\end{figure}

  For a graph with higher cyclomatic number, the universal covering looks more complicated than in example~\ref{exa:universal-covering-of-a-cycle}, \cf figure~\ref{fig:universal-covering-example}.

\begin{Definition}
  \label{def:lift-of-a-walk}
  Let $\theta\colon\widetilde{\mathcal{G}}\to\mathcal{G}$ be a covering and let $\gamma=\left( e_1, e_2, \dots, e_n \right)$ be a walk in $\mathcal{G}$.
  A walk $\widetilde{\gamma}=\left( \tilde{e}_1, \tilde{e}_2, \dots, \tilde{e}_n \right)$ in $\widetilde{\mathcal{G}}$ with $\theta_{\mathcal{E}}(\tilde{e}_i)=e_i$ for $i=1, \dots, n$ is a \emph{lift} of the walk $\gamma$.
\end{Definition}

\begin{Proposition}
  \label{thm:lift-of-a-walk}
  Let $\theta\colon\widetilde{\mathcal{G}}\to\mathcal{G}$ be a covering.
  Let $\gamma$ be a walk in $\mathcal{G}$ from a vertex $v_0$ to a vertex $v_n$.
  Moreover, let $\tilde{v}_0$ be a vertex in the fibre over $v_0$.
  Then there is a unique lift $\widetilde{\gamma}$ starting at $\tilde{v}_0$.
  Its end is a vertex $\tilde{v}_n$ in the fiber over $v_n$.
\end{Proposition}

\begin{Definition}
  \label{def:monodromy-action}
  Given a connected graph $\mathcal{G}$ with covering $\theta\colon\widetilde{\mathcal{G}}\to\mathcal{G}$.
  The \emph{monodromy action} of the fundamental group $\pi_1(\mathcal{G})$ on the fibers over the vertices is defined in the following way:
  Let $\tilde{v}\in\theta_\mathcal{V}^{-1}(v)$ be a vertex in the fiber over $v$.
  Represent the fundamental group as $\pi_1(\mathcal{G},v)$, \ie closed walks starting and ending at $v$.
  The action of $\zeta\in\pi_1(\mathcal{G},v)$ on $\tilde{v}$ is defined as $\zeta \tilde{v}\coloneqq t(\widetilde{\zeta})$, where $\widetilde{\zeta}$ is the unique lift of $\zeta$ starting at $\tilde{v}$.
\end{Definition}

\begin{Example}
  \label{exa:monodromy-action}
  Let us reconsider the (semi)circuit $\mathcal{C}^N$.
  From the examples~\ref{exa:fundamental-group-of-simple-cycle} and \ref{exa:universal-covering-of-a-cycle} we know that both its fundamental group $\pi_1(\mathcal{C}^N)\simeq\mathbb{Z}$
  and its universal covering $\mathrm{mod}\,N\colon \mathcal{Z}\to\mathcal{C}^N$ are given by the integers.
  The monodromy action of $\zeta\in\pi_1(\mathcal{C}^N)\simeq\mathbb{Z}$ on $z\in\mathcal{V}(\mathcal{Z})=\mathbb{Z}$ is given by simple addition: $\zeta z=\zeta+z$.
\end{Example}

\begin{Remark}
  \label{rmk:monodromy-action-non-commutative}
  In general, the fundamental group does not commute.
  So the composition of two different fundamental cycles $\zeta_1, \zeta_2\in\pi_1(\mathcal{G})$ will take you to different points in a fibre, depending on the order: $\zeta_1\zeta_2\tilde{v}\neq\zeta_2\zeta_1\tilde{v}$.
  This can easily be seen in figure~\ref{fig:universal-covering-example}.
\end{Remark}
\section{Spectra of Graphs}

Let us return to algebra.
As seen in the previous sections, graphs can be described by matrices.
Many structural properties of the graph are reflected in the properties of its representing matrices, especially in the characteristic polynomials.\,\citep{Cvetkovic1980}

As noted in the beginning of this chapter, the algebra is nicer on directed graphs. So in the following, $\mathbb{F}$ represents either $\mathbb{R}$ or its algebraic closure, $\mathbb{C}$.

\begin{Definition}
  \label{def:eigenvalue-eigenvector}
  Let $\mathbb{M}\in\mathbb{F}^{N\times N}$ be a square matrix.
  Let $\lambda\in\mathbb{F}$ be a number.
  A non-trivial solution $v\in\mathbb{F}^N$ of the equation
  \begin{align*}
    \mathbb{M} v &= \lambda v
  \end{align*}
  is called \emph{eigenvector} of $\mathbb{M}$.
  The number $\lambda$ is the corresponding \emph{eigenvalue}. 
\end{Definition}

\begin{Definition}
  \label{def:characteristic-polynomial}
  The polynomial $\chi_\mathbb{M}(x)\coloneqq\det\left( \mathbb{M} - x\, \mathbb{U} \right)$ is called \emph{characteristic polynomial} of the square matrix $\mathbb{M}$.
  Here $\mathbb{U}$ denotes the unit (or identity) matrix with the same dimensions as $\mathbb{M}$.
\end{Definition}

\begin{Proposition}
  \label{thm:eigenvalues-is-root-of-charpoly}
  Let $\mathbb{M}\in\mathbb{F}^{N\times N}$ be a matrix and $\chi_\mathbb{M}$ its characteristic polynomial.
  Then for $\lambda\in\mathbb{F}\colon$
  \begin{align*}
    0 &= \chi_\mathbb{M}(\lambda)\quad \Leftrightarrow \quad \text{$\lambda$ is an eigenvalue of $\mathbb{M}$}\,.
  \end{align*}
\end{Proposition}

\begin{Theorem}[Fundamental Theorem of Algebra]
  \label{thm:fundamental-theorem-of-algebra}
  Every polynomial equation over the complex number field $\mathbb{C}$ has a solution.
\end{Theorem}

\begin{Proposition}
  \label{thm:characteristic-polynomial-representation}
  For the characteristic polynomial of a matrix $\mathbb{M}\in\mathbb{C}^{N\times N}$ there is a unique representation 
  \begin{align*}
    \chi_\mathbb{M}(x) = \prod_{i=1}^{\ell} \left( \lambda_i - x \right)^{k_i}\,.
  \end{align*}
  Here $\lambda_i\in\mathbb{C}$ are the eigenvalues of the matrix $\mathbb{M}$ and, moreover, $\sum_{i=1}^\ell k_i = N$.
\end{Proposition}

\begin{Definition}
  \label{def:multiplicity-of-eigenvalue}
  In the above proposition, the number $k_i$ is the \emph{algebraic multiplicity} of $\lambda_i$.
  An eigenvalue with algebraic multiplicity $k=1$ is a \emph{simple} eigenvalue.
  The eigenvalue with the biggest real part is the \emph{dominant eigenvalue}.
\end{Definition}

\begin{Definition}
  \label{def:spectrum-of-matrix}
  The eigenvalues of a matrix $\mathbb{M}$ are also called its \emph{spectrum}, denoted as a tuple $\otherLambda_\mathbb{M}$.
  Typically, an eigenvalue is listed multiple times\thinspace{}---\thinspace{}in accordance to its multiplicity.
  The order of eigenvalues in the spectrum is arbitrary, although they are sometimes semiordered by decreasing real part.
\end{Definition}

\begin{Definition}
  \label{def:spectrum-of-graph}
  Let $\mathcal{G}$ be a directed graph without loops and let $\mathbb{L}=\mathbb{BB}=\mathbb{D-A}$ be the combinatorial Laplacian in the standard basis.
  Then $\otherLambda_{\mathbb{A}}$ is the \emph{ordinary spectrum} and $\otherLambda_{\mathbb{L}}$ is the \emph{Laplacian spectrum} of the graph $\mathcal{G}$.
\end{Definition}

\begin{Proposition}
  \label{thm:laplacian-spectrum}
  The Laplacian spectrum is non-negative and always contains zero.
  If $\mathcal{G}$ is connected, then zero is a simple eigenvalue with eigenvector $\left( 1, \dots, 1 \right)$.\,\citep{Cvetkovic1980}
\end{Proposition}

\begin{Example}
  \label{exa:spectrum-of-graph}
  Let $\mathbb{L}=\mathbb{D-A}$ be the Laplacian matrix of $\mathcal{G}_c$ as it is given in example~\ref{exa:combinatorial-laplacian}.
  Then we have the characteristic polynomials 
  \begin{align*}
    \chi_{\mathbb{A}}(x) &= x^4 - 5 x^2 -4 x\,,  & \chi_{\mathbb{L}}(x) &=x^4 - 10 x^3 + 32 x^2 -32 x
  \end{align*}
  and therefore $\otherLambda_{\mathbb{A}}=\left( \frac{1}{2}\left( 1+\sqrt{17} \right), \frac{1}{2}\left( 1-\sqrt{17} \right), -1, 0 \right)$ and $\otherLambda_{\mathbb{L}}=\left( 4, 4, 2, 0 \right)$.
\end{Example}

\begin{Definition}
  \label{def:circulant-matrix}
  A matrix $\mathbb{M}\in\mathbb{F}^{ N\times N}$ is called \emph{circulant} if the entries satisfy 
  \begin{align*}
    \left( \mathbb{M} \right)^{j}_{i} = \left( \mathbb{M} \right)^{j-i+1}_{1}\,,
  \end{align*}
  \ie row $i$ of $\mathbb{M}$ is identical to row 1 shifted by $i-1$ entries, with indices taken modulo $N$.
\end{Definition}

\begin{Proposition}
  \label{thm:spectrum-for-circulant-matrix}
  Let $\mathbb{M}\in\mathbb{C}^{N\times N}$ be a circulant matrix with first row $\left( m_1, \dots, m_N \right)$.
  Let $\xi=\exp\frac{2\otherpi\I}{N}$ denote the $N$-th root of unity.
  Then the eigenvalues of $\mathbb{M}$ are\,\citep{Knauer2011}
  \begin{align*}
    \lambda_i = \sum_{j=1}^{N} m_j \xi^{\left( j-1 \right)i}\,,\quad i=0, 1, \dots, N-1\,.
  \end{align*}
\end{Proposition}

\begin{Example}
  \label{exa:spectrum-of-simple-cycle}
  Let $\mathcal{C}^N$ be a semicircuit with arbitrary orientation.
  Then the Laplacian matrix is a circulant $N\times N$ matrix with first row $\left( 2, -1, 0, \dots, 0, -1 \right)$.
  The Laplacian spectrum is $\otherLambda_{\mathbb{L}}=\left\{ 2 - 2 \cos \frac{2\otherpi j}{N}\middle| j=0, \dots, N-1\right\}$.
\end{Example}

The Laplacian spectrum is tightly related to the structural properties of the graph via the Matrix--Tree Theorem.
Here, I state the result in a generalized form\,\citep{Chebotarev2006} that will be useful in the later chapters.
We need some more definitions:

\begin{Definition}
  \label{def:rooted-tree}
  Let $\mathcal{T}$ be a directed tree.
  The tree $\mathcal{T}$ with one marked vertex $v\subsc{r}\in\mathcal{V}(\mathcal{T})$ is a \emph{rooted tree} $\left( \mathcal{T}, v\subsc{r} \right)$ with \emph{root} $v\subsc{r}$.
  A \emph{diverging tree} is a rooted tree with all its edges oriented to point away from the root.
  This is possible, since in a tree there is a unique path from the root $v\subsc{r}$ to any other vertex.
  For a set $R$ of roots, a \emph{diverging forest} $\mathcal{F}_R$ is a forest whose connected components are diverging trees $\left( \mathcal{T},v\subsc{r} \right)$, each diverging from a different root $v\subsc{r}\in R$.
\end{Definition}

\begin{Definition}
  \label{def:weight-on-graph}
  Let $\mathcal{G}$ be a directed graph.
  A \emph{weight} $w$ is a real-valued function on the oriented edges.
  It need not be positive for the following considerations\,\citep{Chebotarev2006}.
  In general, $w(e^{+})$ and $w(e^{-})$ are independent of each other, so it is not a 1-chain.
  The weight of a connected subgraph $\mathcal{G}'$ is defined as $w(\mathcal{G}')\coloneqq \prod_{e\in\mathcal{E}(\mathcal{G}')}w(e)$.
  A trivial subgraph $\left\{ v \right\}$ with only one vertex and no edge has the weight $w(\left\{ v \right\})\coloneqq 1$.
  The weight of a set $\mathcal{S}$ of connected subgraphs is the sum over all its connected components: $w(\mathcal{S})=\sum_{\mathcal{G}'\in \mathcal{S}}w(\mathcal{G}')$.
  The weight of the empty set $w(\emptyset)\coloneqq 0$ is zero.
\end{Definition}

\begin{Definition}
  \label{def:weighted-laplacian}
  Let $\mathcal{G}$ be a simple directed graph without loops and let its vertex set be $\mathcal{V}=\left\{ v_1, v_2, \dots, v_N \right\}$.
  Let a weight $w$ on its oriented edges be given and denote the weight of an edge $e^{+}=\left( v_i, v_j \right)$ by $w(e^{+})\eqqcolon w^i_j$.
  The weight of its inverted edge $e^{-}=\left( v_j, v_i \right)$ is $w(e^{-})\eqqcolon w^j_i$.
  The matrix $\mathbb{L}_w$ with entries
  \begin{align*}
    \left( \mathbb{L}_w \right)^i_j = -w^i_j\quad\text{for }i\neq j \qquad\text{and}\qquad \left( \mathbb{L}_w \right)^{i}_{i} = \sum_{\ell=1}^N w^i_j\,.
  \end{align*}
  is the \emph{weighted Laplacian matrix}.
\end{Definition}

\begin{Remark}
  \label{rmk:weighted-laplacian}
  For the constant weight $w(e)\equiv 1$, the weighted and the combinatorial Laplacian matrices coincide.
\end{Remark}

\begin{Proposition}
  \label{thm:laplace-polynomial}
  Let $\mathcal{G}=\left( \mathcal{V}, \mathcal{E}, \iota \right)$ be a directed simple graph without loops.
  Let the order of the graph be $N=\abs{\mathcal{G}}$. 
  Let a weight $w$ on the oriented edges be given.
  Write the characteristic polynomial of the negative weighted Laplacian matrix as $\chi_{-\mathbb{L}_w}(x)=\sum_{i=0}^{N} a_i x^i$.
  Then
  \begin{align*}
    a_i = (-1)^N\sum_{\substack{R\subset\mathcal{V},\\ \abs{R}=i}}w\left( \mathcal{S}_{R} \right)\,,
  \end{align*}
  where $\mathcal{S}_R$ is the set of all diverging forests $\mathcal{F}_R$ with root set $R$.\,\citep{Chebotarev2006}

\end{Proposition}

\begin{Theorem}[Matrix-Tree Theorem]
  \label{thm:matrix-tree-theorem}
  For a simple directed graph without loops and weight $w$ the coefficient $a_1$ in the characteristic polynomial $\chi_{-\mathbb{L}_w}(x)=\sum_{i} a_i x^i$ is given by  \begin{align*}
    a_1 = (-1)^N \sum_{\mathcal{T}} w(\mathcal{T})\,,
  \end{align*}
  where $N$ is the order of the graph and the sum goes over all diverging spanning trees of the graph.The zero order coefficient $a_0$ always vanishes.\,\citep{Chebotarev2006}
\end{Theorem}

\begin{Remark}
  \label{rmk:matrix-tree-theorem}
  With a constant weight $w\equiv 1$ the first order coefficient $a_1$ in the polynomial $\chi_{-\mathbb{L}_w}(x)$ (up to the sign) is identical to the number of diverging spanning trees, the higher order coefficients count the number of diverging spanning forests.
  This shows a direct connection between the structure and the Laplacian spetrum of a graph.
\end{Remark}

\phantomsection
\section*{Summary}

Discrete sets with a neighborhood structure are graphs.
Graph theory provides a language to describe these structures.
Algebraic considerations are especially helpful to deal with functions on the edges of a graph: The edge space decomposes into the cycle space and the orthogonal cocycle space.
Topological arguments allow us to construct very nice and intuitive bases for these two subspaces.
The spectrum of the Laplacian matrix reveals another connection between the algebraic and structural properties of a graph.

\chapter{Probability Theory}
The concept of a probability is very closely related to the mathematical concepts of measure and integration.
In mathematical analysis, the standard measure is the Lebesgue measure on the vector spaces $\mathbb{R}^d$.
I expect the reader is familiar with the general ideas of measure and integration, especially on $\mathbb{R}^d$.
However, the more abstract concepts of probability theory, \eg cumulants and large deviation theory, are likely to be not so well known. 
For a consistent presentation, I also repeat the basic concepts.
Proofs are omitted, however, because they can be found in standard text books\,\citep{VanKampen2011,Capasso2012}\thinspace{}---\thinspace{}unless noted otherwise.

\section{Measure and Probability}
\label{sec:measure-probability}

\begin{Definition}
  \label{def:measurable_space}
  Let $\otherOmega$ be a non-empty set and let $\otherSigma\subset\mathfrak{P}(\otherOmega)$ be a collection of subsets of $\otherOmega$.
  If $\otherSigma$ contains the empty set and is closed under the formation of complements and countable unions, it is a \emph{$\sigma$-algebra} and $(\otherOmega, \otherSigma)$ is a \emph{measurable space}.
\end{Definition}

\begin{Definition}
  \label{def:measurable_mapping}
  Let $\left( \otherOmega_1, \otherSigma_1 \right)$ and $\left( \otherOmega_2, \otherSigma_2 \right)$ be measurable spaces.
  A map $f\colon \otherOmega_1 \to \otherOmega_2$ is called \emph{measurable} if $\forall E \in \otherSigma_2\colon f^{-1}(E) \in \otherSigma_1$.
\end{Definition}

\begin{Definition}
  \label{def:measure}
  Let $(\otherOmega, \otherSigma)$ be a measurable space.
  A function $\mu\colon \otherSigma \to \mathbb{R} $ is called \emph{measure} if it has the following properties:
  \begin{enumerate}
    \item[(M1)] $\forall E\in \otherSigma\colon \mu(E) \geq 0$,
    \item[(M2)] $\mu(\emptyset ) = 0$,
    \item[(M3)] For all $E_{1}, \dots, E_{n} \in \otherSigma$ with $E_{i}\cap E_{j} = \emptyset$:
      $\mu\!\left( \bigcup_{i} E_{i} \right) = \sum_{i} \mu(E_{i})$.
  \end{enumerate}
  A triple $(\otherOmega,\otherSigma,\mu)$ is a \emph{measure space}.
  A measure is called \emph{finite} if $\mu(\otherOmega)<\infty$, it is \emph{$\sigma$-finite} if $\otherOmega$ is a countable union of measurable sets with finite measure.
  Every set $E\in \otherSigma$ with $\mu(E)=0$ is called \emph{$\mu$-null set} or just \emph{null set}, if the measure is given by the context.
\end{Definition}

\begin{Definition}
  \label{def:borel_algebra}
  Let $(\otherOmega,\mathcal{T})$ be a topological space.
  The smallest $\sigma$-algebra $\mathcal{B}$ that contains $\mathcal{T}$ is called the \emph{Borel $\sigma$-algebra}, its elements are \emph{Borel sets}.
\end{Definition}

\begin{Remark}
  \label{rmk:borel_algebra}
  The vector spaces $\mathbb{R}^d$ are understood to be equipped with the Borel $\sigma$-algebra and the usual Lebesgue measure $\mu\subsc{L}$.
\end{Remark}

\begin{Definition}
  \label{def:absolute_continuity}
  Let $\mu_1$ and $\mu_2$ be two measures on a measurable space $\left( \otherOmega, \otherSigma \right)$.
  We call $\mu_1$ \emph{absolutely continuous} with respect to $\mu_2$ if
  \begin{align*}
    \forall E\in \otherSigma \colon\quad \mu_2(E)=0 &\Rightarrow \mu_1(E)=0\,.
  \end{align*}
  A measure that is not absolutely continuous, is called \emph{singular}.
\end{Definition}

\begin{Example}
  \label{exa:delta-distribution-not-absolutely-continuous}
  For any point $x\in \mathbb{R}$, the Dirac $\otherdelta$-distribution defines a finite measure $\mu_{x}$ on $\mathbb{R}$: The measure of a Borel set $E\in\mathcal{B}$ is
  \begin{align*}
    \mu_{x}(E) \coloneqq \int_{E} \otherdelta(y-x)\, \D y= \left\llbracket x\in E \right\rrbracket\,.
  \end{align*}
  This measure is singular with respect to the Lebesgue measure: $\mu_{x}(\left\{ x \right\})=1\neq 0$, but $\left\{ x \right\}$ is a Lebesgue null set.
\end{Example}

\begin{Theorem}[Radon--Nikod\'ym]
  \label{thm:radon-nikodym}
  Let $\mu_1$, $\mu_2$ be two $\sigma$-finite measures on a measurable space $\left( \otherOmega, \otherSigma \right)$ and let $\mu_1$ be absolutely continuous with respect to $\mu_2$.
  Then there is a measurable function $f\colon \otherOmega \to \left[0, \infty\right)$ such that
  \begin{align*}
    \forall E\in \otherSigma\colon\quad \mu_1(E) &= \int_{E} f(x) \mu_2(\D x)\,.
  \end{align*}
\end{Theorem}

\begin{Definition}
  \label{def:radon-nikodym-derivative}
  A function $f$ satisfying the equality in the above theorem is called \emph{Radon--Nikod\'ym derivative} or \emph{density}.
  It is uniquely determined, up to $\mu_2$-null sets.
  The Radon--Nikod\'ym derivative is typically denoted by $\frac{\D \mu_1}{\D \mu_2}$.
\end{Definition}

Measures can also be defined by their densities with respect to other measures, as can be seen in this

\begin{Example}
  \label{exa:normal-distribution-measure}
  The density $\frac{\D \mu}{\D \mu\subsc{L}}= \exp \frac{-x^2}{2}$
  defines a measure on $\mathbb{R}$.
  This measure is finite since $\mu(\mathbb{R})=\sqrt{2\otherpi}$.
\end{Example}

Normalizing a finite measure immediately takes us to the following

\begin{Definition}
  \label{def:prob_space}
  A \emph{probability space} is a measure space $( \otherOmega, \otherSigma, P)$ with normalized measure, \ie $P(\otherOmega) = 1$.
  The set $\otherOmega$ is called \emph{sample space}, the elements of $\otherSigma$ are called \emph{events}, and every element of $\otherOmega$ is called an \emph{elementary event}.
  The measure $P$ is called \emph{probability measure} and for an event $E\in \otherSigma$ the number $P(E)$ is the \emph{probability} of this event.
  Every event $E$ with $P(E)=1$ is said to occur \emph{almost surely} (abbreviation: $\asure$).
\end{Definition}

\begin{Remark}
  \label{rmk:measure_prob_measure}
  A measure space $(\otherOmega,\otherSigma,\mu)$ can be equipped with an additional probability measure that does not need to be connected to $\mu$ in any way: The Dirac $\otherdelta$-dis\-tribution in example~\ref{exa:delta-distribution-not-absolutely-continuous} provides a probability measure on $\mathbb{R}$.
\end{Remark}

\begin{Definition}
  \label{def:uniform_prob_measure}
  Given a measure space $(\otherOmega,\otherSigma,\mu)$ with finite measure. An additional probability measure $P$ on this measure space is called \emph{uniform} if $P=\frac{1}{\mu(\otherOmega)}\mu$.
\end{Definition}

\begin{Example}
  \label{exa:equiprobable}
  Let $(\otherOmega,\mathfrak{P}(\otherOmega),|\,\cdot\,|)$ be a measure space with finite sample space, the power set as $\sigma$-algebra and the cardinality as measure.
  Assuming the space is equipped with the uniform probability measure $P$.
  Then every elementary event $\omega\in \otherOmega$ is \emph{equiprobable} with probability $P(\omega)=\frac{1}{|\otherOmega|}$.
\end{Example}

\begin{Definition}
  \label{def:conditional_prob}
  Let $\left( \otherOmega, \otherSigma, P \right)$ be a probability space and $E, F \in \otherSigma$, $P(F)>0$\@. 
  Then the \emph{probability of $E$ conditional on $F$} is given by
  \begin{align*}
    P\left(E\;\middle| F\right)\coloneqq \frac{P(E\cap F)}{P(F)}\,.
  \end{align*}
\end{Definition}

\begin{Definition}
  \label{def:independent_events}
  Two events $E, F \in \otherSigma$ are called \emph{independent} if
  \begin{align*}
    P(E\cap F) = P(E)P(F)\,.
  \end{align*}  
  With the additional assumption $P(F)> 0$, this is equivalent to $P\left(E\;\middle|F\right)=P(E)$.
\end{Definition}

\section{Random Variables and their Distributions}

\begin{Definition}
  \label{def:random-vector}
  Let $(\otherOmega_1,\otherSigma_1,P)$ be a probability space and let $(\otherOmega_2,\otherSigma_2)$ be a measurable space.
  A measurable mapping $X\colon (\otherOmega_1,\otherSigma_1,P) \to (\otherOmega_2,\otherSigma_2)$ is called \emph{random variable}.
  A random variable taking values in $\mathbb{R}^d$ is referred to as \emph{random vector}.
\end{Definition}

  The sample space $\otherOmega_1$ may be very abstract in nature: A thrown die with its entire molecular configuration and orientation on the surface is a sample space with a very complicated probability measure.
  The random variable “number of pips on the top face” takes values in the set $\left\{ 1, 2, \dots, 6 \right\}$.
  A fair die is characterized by a uniform probability distribution on its values.  
  Consequently, the sample space is not always of interest\thinspace{}---\thinspace{}the probabilities of the values are more important.

\begin{Definition}
  \label{def:probability_distribution}
  The \emph{probability distribution} of a random variable $X$ is given by the push-forward measure $P_X \coloneqq X_{*}P \equiv P\circ X^{-1}$ on $\left(\otherOmega_2, \otherSigma_2\right)$.
  If $X$ is a random vector and $P_X$ is absolutely continuous with respect to the Lebesgue measure $\mu\subsc{L}$, its Radon--Nikod\'ym derivative $\rho=\frac{\D P_X}{\D \mu\subsc{L}}$ is called \emph{probability density}.
\end{Definition}

Sometimes one only considers a probability distribution and does not specify a corresponding random variable.
This is justified with the following

\begin{Proposition}
  \label{thm:existence-of-random-variable}
  Let $P_2$ be a probability measure on a measurable space $\left( \otherOmega_2, \otherSigma_2 \right)$.
  Then there is a probability space $\left( \otherOmega_1, \otherSigma_1, P_1\right)$ and a random variable $X\colon \otherOmega_1\to\otherOmega_2$ such that $P_X=P_2$.
  \begin{Proof}
    Choose $\left( \otherOmega_1,\otherSigma_1, P_1 \right)=\left( \otherOmega_2,\otherSigma_2, P_2 \right)$ and $X=\id$.
  \end{Proof}
\end{Proposition}

  Especially in physics one often writes $\rho(x)\D x$ instead of $P_X(\D x)$ even if $P_X$ is not absolutely continuous.
  In this notation $\rho(x)$ is a generalized function that fully characterizes the distribution of $X$, including the singular parts.
  Therefore these generalized functions are also known under the name \emph{distributions}.
  
  In probability theory, one uses the following

\begin{Definition}
  \label{def:cumulative_distribution}
  Let $X=\left( X_1, X_2, \dots, X_d \right)$ be a random vector.
  The function
  \begin{align*}
    F_X(x) \coloneqq \int_{-\infty}^{x_1}\cdots\int_{-\infty}^{x_d} P_X(\D y)
  \end{align*}
  is called \emph{cumulative distribution function} (CDF).
  It uniquely determines the distribution, even if the latter is singular.
\end{Definition}

We can also integrate functions with respect to the probability measures:

\begin{Proposition}
  \label{thm:expectation-of-funtcion}
  Let $X\colon\otherOmega_1 \to \otherOmega_2$ be a random variable.
  Let $f\colon \otherOmega_2 \to \mathbb{R}^d$ be a measurable function.
  Then $f$ is $P_X$-integrable if and only if $f\circ X$ is $P$-integrable and the integral takes the value
  \begin{align*}
    \left\langle f(X)\right\rangle \coloneqq \int_{\mathbb{R}^d} f(x) P_X(\D x) = \int_{\otherOmega} f\circ X (\omega_1) P(\D \omega)\,.
  \end{align*}
\end{Proposition}

\begin{Definition}
  \label{def:expectation}
  The value $\left\langle f(X)\right\rangle$ of the integral in the above proposition is called the \emph{expectation} of $f$.
\end{Definition}

\begin{Example}
  \label{exa:expectation}
  The expectation for the number of pips in a fair dice cast is $\sum_{i=1}^{6} \frac{i}{6} = 3.5$\,.
\end{Example}

\begin{Definition}
  \label{def:moments}
  Let $X\colon \otherOmega \to \mathbb{R} $ be a random variable.
  For $\nu \in \mathbb{N} $, the expectation $\langle X^\nu\rangle$ is called the \emph{$\nu$-th moment of X}.
  The expectation $\langle (X-\langle X\rangle)^\nu \rangle $ is called the \emph{$\nu$-th central moment of X}.
  Note that, in general, the moments need not be finite.
\end{Definition}

The first moment locates the center of a distribution.
It is not the point with the highest probability(density).
  The second central moment is a quantifier for the width of a probability distribution around that center\thinspace{}---\thinspace{}it is also called \emph{variance}. 

Moments immediately generalize to random vectors via the following

\begin{Definition}
  \label{def:moment_gen_fct}
  Let $X\colon \otherOmega \to \mathbb{R}^d$ be a random vector. The function $G_{X}\colon \mathbb{R}^d\to\mathbb{R}$,
  \begin{align*}
    G_{X}(q)\coloneqq\left\langle \e{q\cdot X}\right\rangle
    = \int_{\mathbb{R}^d} \e{q\cdot x}P_{X}(\D x)
  \end{align*}
  is the \emph{moment-generating function} of $X$.
  The name is justified by the formal identity $\left.\left(\frac{\D}{\D q}\right)^{\nu}\right|_{q=0}G_{X}(q)= \langle X^{\nu}\rangle$ for $d=1$.
  The moment-generating function need not be well defined for all values of $q\in\mathbb{R}^d$, but $G_X(0)=\langle 1\rangle=1$ always holds.
\end{Definition}

\begin{Definition}
  \label{def:joint-moments}
  Let $X=\left( X_1, \dots, X_d \right)\colon\otherOmega\to\mathbb{R}^d$ be a random vector. For any set of indices $1\leq i_1, i_2, \dots, i_\ell \leq d$ the partial derivatives
  \begin{align*}
    \left.\frac{\del}{\del q_{i_1}}\frac{\del}{\del q_{i_2}}\cdots\frac{\del}{\del q_{i_\ell}}\right|_{q=0} G_X(q)
    = \left\langle X_{i_1}X_{i_2}\dots X_{i_\ell}\right\rangle
  \end{align*}
  are the \emph{joint moments} of $X_{i_1}, \dots, X_{i_\ell}$, given that the limits exist.
\end{Definition}

\begin{Definition}
  \label{def:characteristic_fct}
  Let $X\colon \otherOmega \to \mathbb{R}^d$ be a random vector.
  The function
  \begin{align*}
    \varphi_X(q) \coloneqq \left \langle \e{\I q\cdot X} \right \rangle = \int_{\mathbb{R}^d} \e{\I q\cdot x} P_X(\D x) 
  \end{align*}
  is called \emph{characteristic function} of $X$.
\end{Definition}

  Obviously $\varphi_X(q)=G_X(\I q)$, so the preceding definition might seem unnecessary.
  However, the characteristic function has nice mathematical properties:
  \begin{enumerate}
    \item It is well defined for all $q\in\mathbb{R}^d$ because $\left|\left\langle \e{\I q\cdot X} \right\rangle\right| \leq \left\langle\left|\e{\I q\cdot X}\right| \right\rangle = 1$.
    \item It is uniformly continuous. 
    \item It uniquely determines the distribution: $\varphi_{X_1} = \varphi_{X_2} \Leftrightarrow F_{X_1} = F_{X_2}$.
    \item If $P_X$ is absolutely continuous then the density is given by 
      \begin{align*}
	\rho(x) &= \frac{1}{2 \otherpi} \int_{\mathbb{R}^d} \e{-\I q\cdot x } \varphi_X(q) \D q\,.
      \end{align*}
  \end{enumerate}

  A different way to characterize probability distributions are cumulants:

\begin{Definition}
  \label{def:cumulant_gen_fct}
  Let $X\colon \otherOmega \to \mathbb{R}^d$ be a random vector and $G_X(q)$ its moment generating function.
  Then one defines
  \begin{align*}
    g_X(q) \coloneqq \ln G_X(q) = \ln \left\langle \e{q\cdot X} \right\rangle
  \end{align*}
  to be the \emph{cumulant-generating function} (CGF).
  Its derivatives in the origin 
  \begin{align*}
    \kappa(X_{i_1}, X_{i_2}, \dots, X_{i_\ell}) \coloneqq \left.\frac{\del}{\del q_{i_1}}\frac{\del}{\del q_{i_2}}\cdots\frac{\del}{\del q_{i_\ell}} \right|_{q=0} g_X(q)
  \end{align*}
  are called \emph{joint cumulants} of the random variables $X_{i_1}, X_{i_2},\dots, X_{i_\ell}$.
  For $d=1$ the cumulant generating function can be written
  \begin{align*}
    g_X(q) = \sum_{\nu = 1}^{\infty} \frac{q^{\nu}}{\nu !} \kappa_{\nu}(X)\,,
  \end{align*}
  where $\kappa_{\nu}(X)\coloneqq\kappa(X, \dots, X)$ is the \emph{$\nu$-th cumulant of $X$}.
  Note that $g_X(0) = 0$ since $G_X(0) = 1$.
  Just as in the case of moments, cumulants need not exist.
  That is reflected by the cumulant generating function not being smooth.
\end{Definition}

\begin{Proposition}
  \label{thm:convexity-of-cgf}
  The cumulant generating function is always convex.
\end{Proposition}

\begin{Remark}
  \label{rmk:moment_gen_fct}
  Sometimes the cumulant generating function is defined as the logarithm of the characteristic function, rather than the logarithm of the moment generating function.
  The definition \ref{def:cumulant_gen_fct} does not ensure boundedness of the cumulant generating function, but allows a nice connection to Statistical Physics: The partition function in the canonical ensemble is $Z\subsc{c} \sim \int_{\otherGamma} \e{-\beta H} \D^n q \D^n p$, where $H$ is the Hamiltonian defined on the phase space $\otherGamma$.
  This formally has the same structure as a moment generating function.
  The thermodynamic potentials in Statistical Physics formally take the form of cumulant generating functions: They are (basically) logarithms of the partition functions.
  Moreover, their first derivatives are the expectations of the conjugate thermodynamic quantities and their second derivatives quantify the variances of those quantities.
\end{Remark}

\begin{Proposition}
  \label{thm:first_cumulants_and_monents}
  Let $X$ be a real-valued random variable and $\mu_{\nu} = \langle(X-\langle X\rangle)^{\nu}\rangle$ its $\nu$-th central moment.
  Then we have
  \begin{align*}
    \kappa_{1}(X) &= \langle X\rangle & \kappa_{2}(X) &= \mu_{2} = \langle X^{2}\rangle - \langle X\rangle^{2}\\
    \kappa₃(X) &= \mu₃ & \kappa_{4}(X) &= \mu_{4} - 3 \mu _{2}^{2}
  \end{align*}
  as can be seen by direct calculation.
\end{Proposition}

\begin{Example}
  \label{exa:normal-distribution}
  The example~\ref{exa:normal-distribution-measure} gives a very important probability measure.
  Given two numbers $\mu\in\mathbb{R}$ and $\sigma>0$. The density
  \begin{align*}
    \rho(x) \coloneqq \frac{1}{\sqrt{2\otherpi \sigma^2}}\exp\left( -\frac{\left( x - \mu \right)^2}{2\sigma^2} \right)
  \end{align*}
  defines a probability measure called \emph{normal distribution}.
  Its cumulant generating function is $g(q)=\mu q + \frac{1}{2}\sigma^2 q^2$.
  Therefore, its mean is $\mu$ and its variance is $\sigma^2$.
  In fact, the normal distribution is the only absolutely continuous probability distribution on $\mathbb{R}$ with a polynomial as CGF.
  Higher order polynomials contradict positivity of the density.
\end{Example}

\begin{Definition}
  \label{def:covariance-matrix}
  Let $X\colon\otherOmega\to\mathbb{R}^d$ be a random vector.
  The Hessian matrix of its cumulant generating function, \ie the matrix with entries
   $ \kappa\left( X_i, X_j \right)$,
  is the \emph{covariance matrix} of $X$.
  The diagonal elements are the variances of the $X_\ell$, the off-diagonal elements are their \emph{covariances}.
  Whenever the partial derivatives commute, the covariance matrix is symmetric.
\end{Definition}

\begin{Definition}
  \label{def:independen_random_variables}
  Two real-valued random variables $X_{1}$ and $X_{2}$ are called \emph{independent} if their joint probability distribution, \ie the measure of the corresponding vector $X=(X_{1},X_{2})$, factorizes:
   $ P_X(\D x_{1}, \D x_{2}) = P_{X_{1}}(\D x_{1})\cdot P_{X_{2}}(\D x_{2})$.
  They are \emph{uncorrelated} if their joint cumulant $\kappa\left( X_1, X_2 \right)$ vanishes.
\end{Definition}

\begin{Proposition}
  \label{thm:multilinear-joint-cumulants}
  The joint cumulants $\kappa(X_{i_1}, X_{i_2}, \dots, X_{i_\ell})$ with $\ell\geq 2$ are multi-linear in their arguments, while the first cumulants are affine linear.
  This holds irrespective of independence.\,\citep{McCullagh1987}
\end{Proposition}

\begin{Proposition}
  \label{thm:joint-cumulants-and-independence}
  If $X_{i_1}, X_{i_2}, \dots, X_{i_\ell}$ are independent random variables, then all joint cumulants involving only these random variables vanish.
  If a distribution is fully characterized by its cumulants, then the converse is also true: If all mixed cumulants of $X_{i_1}, X_{i_2}, \dots, X_{i_\ell}$ vanish, then they are independent.\,\citep{McCullagh1987}
\end{Proposition}

\begin{Proposition}
  \label{thm:gen_fcts_for_sums_and_linear_transforms}
  Let $X_{1}, X_{2}$ be two independent real-valued random variables and $a,b \in \mathbb{R} $ be constants.
  Then, the cumulants have very easy transformation rules:
  \begin{align*}
    \kappa_{\nu}(X_{1}+X_{2}) &= \kappa_{\nu}(X_{1}) + \kappa_{\nu}(X_{2}) \\
    \kappa_{\nu}(aX_{1} + b) &= a^{\nu} \kappa_{\nu}(X_{1}) + b\cdot\left\llbracket \nu = 1\right\rrbracket
  \end{align*}
\end{Proposition}

Statistically it makes a difference whether the result of one die cast is multiplied by two or the sum of two die casts is taken: The variance in the latter case is half as big as in the former case.

For sums of many random variables we have the following two theorems.

\begin{Theorem}[Law of Large Numbers]
  \label{thm:large-numbers}
  Let $Y_{1}, \dots, Y_{n}$ be $n$ independent and identically distributed real-valued random variables and let $\kappa_1(Y_1)=\mu$ and $\kappa_2(Y_1)=\sigma^2$.
  All higher cumulants are assumed to be bounded, \ie $\exists C<\infty \,\forall \nu \colon \kappa_{\nu}(Y_1) \leq C$.
  Then for the random variable defined as the sample mean $X_n\coloneqq\frac{1}{n}\sum_{i=1}^{n}Y_i$ we have $\kappa_1(X_n)=\mu$, and $\kappa_2(X_n)=\frac{\sigma^2}{n}$.
  Then $\lim_{n\to\infty}X_n=\mu$ almost surely. Therefore, its probability measure is a Dirac $\otherdelta$-distribution with support in $\mu$.
\end{Theorem}

\begin{Theorem}[Central Limit Theorem]
  \label{thm:central_limit}
  Let $Y_{1}, \dots, Y_{n}$ be $n$ independent and identically distributed real-valued random variables and let their cumulants be given by $\kappa_1(Y_1)=0$, $\kappa_2(Y_1) = \sigma^2$.
  All higher cumulants are assumed to be bounded, \ie $\exists C \,\forall \nu \colon \kappa_{\nu}(Y_1) \leq C$.
  Then the density of the random variable $Z_n=\frac{1}{\sqrt{n}} \sum_{i=1}^{n} Y_{i}$ weakly converges for $n\to\infty$ to the normal distribution with density
  \begin{align*}
    \rho_Z(z) = \frac{1}{\sqrt{2\otherpi\sigma^2}} \exp\left(-\frac{z^2}{2\sigma^2}\right)\,.
  \end{align*}
\end{Theorem}

\section{Large Deviation Theory}
\label{sec:largedevtheory}

The mathematical theory of large deviations can be seen as a generalization for the law of large numbers and the central limit theorem.
Here I review the results that I will need later on. 
For a proper introduction to large deviation theory I recommend the book of \citet{Ellis2006}.
A rather brief overview is the review by \citet{Touchette2009}.
The exposition of this section is based on the latter. 

\begin{Definition} \label{def:largedev}
  A sequence of random vectors $X_{n}\colon \otherOmega \to \mathbb{R}^d$ or the corresponding sequence of probability density functions $\rho_n(x)$ is said to fulfill a \emph{large deviation principle} if the following limit exists at least for all $x$ in an open subset of $\mathbb{R}^d$:
\begin{align}
  I(x) &\coloneqq - \lim_{n\to\infty} \frac{1}{n} \ln \rho_{n}(x)\,.   \label{eq:ratefct} 
  \intertext{In this case the pdf can be written as}
  \rho_{n}(x) &\asymp \e{-nI(x)}\qquad \text{for large $n$}\,.\label{eq:largedev_approx}
\end{align}
The symbol $\asymp$ denotes asymptotic equivalence in the sense of equation (\ref{eq:ratefct}).
The function $I(x)$ is called the \emph{rate function} of the sequence $X_n$ of random vectors.
\end{Definition}

\begin{Proposition}
  \label{thm:non-negative-rate-fct}
  The rate function is non-negative.
  Otherwise the limit of $\e{-nI}$ would not be normalized.
\end{Proposition}

\begin{Definition} \label{def:scgf}
  The \emph{scaled cumulant-generating function} (SCGF) of a sequence $X_{n}$ of random vectors is defined as
  \begin{align*}
    λ(q) \coloneqq \lim_{n\to\infty} \frac{1}{n} \ln \left\langle \e{nq\cdot X_{n}} \right\rangle= \lim_{n\to\infty} \frac{1}{n} g_{X_{n}}(nq)\,.
  \end{align*}
\end{Definition}
Just like the CGF, the scaled cumulant-generating function is convex in $q$. The relation of the SCGF to the rate function is given in the following

\begin{Theorem}[Gärtner--Ellis] \label{thm:gaertner-ellis}
  Let $X_{n}$ be a sequence of random vectors.
  Assume the scaled cumulant-generating function $λ(q)$ exists and is differentiable for all $q$. Then the sequence $X_{n}$ satisfies a large deviation principle. Furthermore, the rate function is given by the Legendre transform of the SCGF:
  \begin{align*}
    I(x) = x\cdot q(x) - λ\circ q(x) \,,
  \end{align*}
  where $q(x)$ is the inversion of $x=\nabla λ(q)$.
\end{Theorem}

\begin{Proposition}
  \label{thm:Legendre-transform-convexity}
  The Legendre transform of a convex function is convex as well.
\end{Proposition}

\begin{Remark}
  \label{rmk:exponential-convergence}
  \citet{Ellis2006} also comments on the convergence speed of the large deviation principle:
  Under the conditions of the Gärtner--Ellis theorem, the convergence is exponentially quickly.
  Consequently, the approximation in equation~(\ref{eq:largedev_approx}) is very good already for finite but big $n$.
\end{Remark}

\begin{Definition}
  \label{def:fluctuation-relation}
  A sequence $X_{n}$ of random variables is said to satisfy a \emph{fluctuation relation} with constant $c$ if for large $n$ we have
  \begin{align}
    \frac{\rho_{n}(x)}{\rho_{n}(-x)} \asymp \e{ncx}\,. \label{eq:fluctuation-relation}
  \end{align}
\end{Definition}

\begin{Proposition}
  \label{thm:rate-function-symmetry}
  Let the sequence $X_n$ satisfy a fluctuation relation.
  In case the sequence furthermore satisfies a large deviation principle, this property is equivalent to the following symmetry\,\citep{Touchette2009} of the rate function:
\begin{align}
  I(x) - I(-x) = -cx\,.
\end{align}
\end{Proposition}

The literature always refers to the scaled cumulant-generating function as a whole.
However, its derivatives in the origin have very nice properties that will help in the following chapter. 

\begin{Definition} \label{def:scaled-cumulants}
  Let $λ(q)$ denote a smooth scaled cumulant-generating function of a sequence $X_n=\left( X_n^{(1)}, X_n^{(2)}, \dots, X_n^{(d)} \right)$ of $d$-dimensional random vectors.
  The derivatives at the origin
  \begin{align}
    c\left(X^{(i_1)}, X^{(i_2)}, \dots, X^{(i_\ell)}\right) \coloneqq \left.\frac{\del}{\del q_{i_1}}\frac{\del}{\del q_{i_2}}\cdots\frac{\del}{\del q_{i_\ell}} \right|_{q=0} \lambda(q) \label{eq:scaled-cumulants}
  \end{align}
  are called \emph{joint scaled cumulants}.
  Thus, for $d=1$ we can write
  \begin{align*}
    λ(q) = \sum_{\nu=1}^\infty \frac{c_\nu(X)}{\nu!} q^\nu\,,
  \end{align*}
  where $c_{\nu}(X) = c\left( X, \dots, X \right)$ is the \emph{$\nu$-th scaled cumulant} of the sequence $X_n$.
  Note that we always have $λ(0)=g_{X_n}(0)=0$.
\end{Definition}

\begin{Proposition}
  \label{thm:scaled-cumulants-cumulants}
  Given that the scaled cumulants exist, they have the following connection to the (non-scaled) cumulants:
  \begin{align}
    c_\nu(X)= \lim_{n\to\infty} n^{\nu-1} \kappa_\nu\left(X_n\right)\,.\label{eq:scaling-of-cumulants}
  \end{align}
  So the first two scaled cumulants are
  \begin{align*}
    c_1(X) = \lim_{n\to\infty}\left\langle X_n\right\rangle \quad\text{and}\quad c_2(X)= \lim_{n\to\infty} n\, \kappa_2(X_n)\,.
  \end{align*}
  \begin{Proof}
    This can easily be seen from the definition~\ref{def:scgf} of the SCGF.
  \end{Proof}
\end{Proposition}


\begin{Proposition}
  \label{thm:multilinear-joint-scaled-cumulants}
  The joint scaled cumulants directly inherit the multilinearity from the joint cumulants, \cf proposition~\ref{thm:multilinear-joint-cumulants}.
  \begin{Proof}
    This is just another consequence of the definition~\ref{def:scgf} of the scaled cumu\-lant-generating function.
  \end{Proof}
\end{Proposition}

\begin{Remark}
  \label{rmk:scaled-cumulant-of-rescaled-sum}
  There is a very natural interpretation of scaled cumulants given the special case of rescaled random variables of the form $X_{n}=\frac{1}{n}Y_n$.
  In this case equation~(\ref{eq:scaling-of-cumulants}) becomes
  \begin{align*}
    c_\nu(X) = \lim_{n\to\infty} \frac{1}{n}\, \kappa_\nu\left(n\,X_n\right) = \lim_{n\to\infty} \frac{1}{n}\, \kappa_\nu\left(Y_n\right)\,.
  \end{align*}
\end{Remark}

\begin{Example}
  \label{exa:large-deviations-of-sample-mean}
  Given a sequence $Y_i$ of independent and identically distributed random variables with finite moments.
  We can construct the new sequence $X_n=\frac{1}{n}\sum_{i=1}^{n}Y_i$ of sample means.
  The joint probability distribution factorizes and therefore the SCGF of $X_n$ is identical to the CGF of any of the $Y_i$.
  The latter is differentiable because the moments and therefore the cumulants exist.
  Then according to the Gärtner--Ellis Theorem, the sequence $X_n$ satisfies a large deviation principle.
\end{Example}

\phantomsection
\section*{Summary}

Random variables are characterized by their distributions.
The distributions of random vectors can be described in various ways: densities, moments, and cumulants\thinspace{}---\thinspace{}given those exist.
Sums of real-valued random variables have remarkable properties: IID sample means satisfy the Law of Large Numbers (LLN) and the Central Limit Theorem (CLT).
  The last example shows that the large deviation principle generalizes both the LLN and the CLT: The Law of Large Numbers describes the convergence of the first cumulant, the Central Limit Theorem is a statement about the convergence behavior of the second cumulant.
  Furthermore, it is worth noticing that whenever the prerequisites for those theorems are violated, \eg non-independent $Y_i$, a large deviation principle might still apply, \ie the limit in equation (\ref{eq:ratefct}) might still exist.

\chapter{Markovian Dynamics}
The concepts introduced in the first two chapters help to understand stochastic dynamics.
Here, I only treat Markovian processes on finite state spaces.

The text book by \citet{VanKampen2011} provides a nice introduction.
A mathematically more rigorous treatment can be found in books like that by \citet{Capasso2012}.
The book by \citet{Jiang2004} focuses on the steady state.

\section{Markovian Jump Processes}
\label{sec:markovian-processes}

\begin{Definition}
  \label{def:stochastic-process}
  Let $(\otherOmega_1,\otherSigma_1,P)$ be a probability space and $(\otherOmega_2,\otherSigma_2)$ be a measurable space.
  Let $T\in\mathbb{R}_{>0}\cup\left\{ \infty \right\}$ be a possibly infinite number.
  A family of random variables $X_t\colon\otherOmega_1\to\otherOmega_2$ for $t\in [0, T]$ is called a \emph{stochastic process} on $\otherOmega_1$ valued in $\otherOmega_2$.
  The space $\otherOmega_2$ is called \emph{state space} of the process.
\end{Definition}

\begin{Remark}
  \label{rmk:time-for-stochastic-process}
  I will refer to $t\in [0, T]$ as \emph{time}.
\end{Remark}

Just as in the case of random variables, the sample space $\otherOmega_1$ can be very abstract.
The physically relevant statistics is that on the state space.
Thus, denoting the time-dependent push-forward probability measure also with the symbol $P$ will not lead to confusion.

\begin{Definition}
  \label{def:trajectory-of-process}
  For a given time $t$, the random variable $X_t$ is called \emph{state of the process at time $t$}.
  For a fixed $\omega\in\otherOmega_1$, the mapping $X(\cdot,\omega)\colon t \mapsto X_t(\omega)\in\otherOmega_2$ is the \emph{trajectory of the process corresponding to $\omega$}.
\end{Definition}

\begin{Definition}
  \label{def:jump-process}
  A stochastic process with finite state space $(\mathcal{V},\mathfrak{P}(\mathcal{V}))$ is a \emph{jump process}.
\end{Definition}

\begin{Example}
  \label{exa:ergodic-deterministic-dynamics}
  Consider an ergodic deterministic dynamics $\otherPsi^t$ on a smooth phase space $\otherGamma$.
  Let the ergodic measure on $\otherGamma$ be denoted by $\mu$.
  This ergodic measure is a normalized measure characterized by $\otherPsi^t_{*}\mu \equiv \mu\circ(\otherPsi^t)^{-1}=\mu$.
  Let the finite partition $\mathcal{V}=\left\{ U_1, U_2, \dots, U_N  \right\}$ of phase space satisfy $\mu\left( \bigcup_{U\in\mathcal{V}}U \right)=1$ and let $\otherPi\colon\otherGamma\to\mathcal{V}$ denote the assignment to the phase space cells $U_i\subset\otherGamma$.
  Then the family of random variables $X_t\colon \otherGamma\ni\omega\mapsto \otherPi\circ\otherPsi^t(\omega)\in\mathcal{V}$ is a stochastic jump process.
\end{Example}

\begin{Definition}
  \label{def:jump-times}
  The trajectory of a jump process changes the state on a discrete set of times.
  These times are the \emph{jump times}, $t_i$.
  Their consecutive differences $t_{i}-t_{i-1}$ are the \emph{staying times}.
\end{Definition}

\begin{Definition}
  \label{def:RCLL-process}
  A process is said to be \emph{right continuous with left limits} (RCLL) if every trajectory is almost surely continuous from the right and has a limit from the left, \ie for any time $t$
  \begin{align*}
    \lim_{\tau\to t+0} X_\tau &= X_t \quad\asure \quad\text{and}\quad \lim_{\tau\to t-0} X_\tau\eqqcolon X_{t-}\in\otherOmega_2\quad\text{exists \asure}
  \end{align*}
\end{Definition}

\begin{figure}[htb]
  \centering
  \def\svgwidth{0.85\textwidth}
  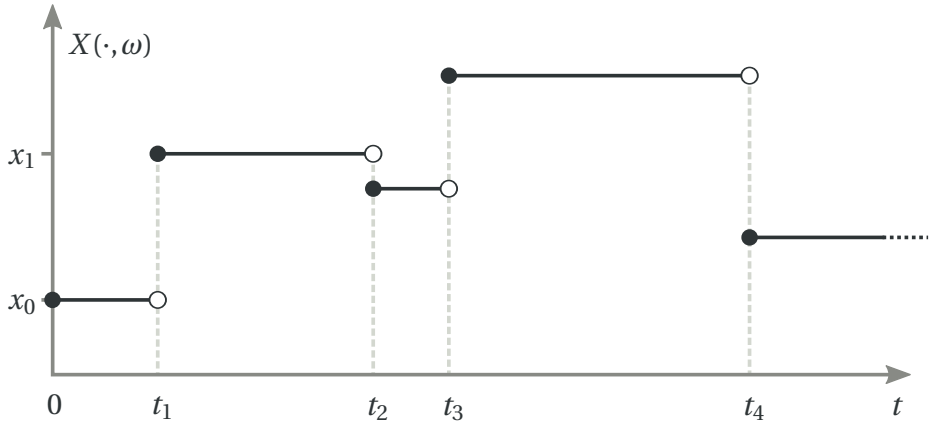
  \caption{Example for a trajectory of an RCLL jump process on a subset of the reals. The jump times $t_1, \dots, t_4$, and the states $x_0$ and $x_1$ are explicitly indicated.}
  \label{fig:random-trajectory-example}
\end{figure}

\begin{Remark}
  \label{rmk:trajectory-of-jump-process}
  A trajectory of an RCLL jump process is fully characterized by an initial state $x_0\in\mathcal{V}$, the number of jumps $n$, a set of jump times $0<t_i<T, i=1,\dots,n$, and the states $x_i\in\mathcal{V}$ the trajectory jumps to at time $t_i$, \cf figure~\ref{fig:random-trajectory-example}.
\end{Remark}

\begin{Definition}
  \label{def:Markov-process}
  A jump process is a \emph{Markovian jump process} if its future only depends on the current state. 
  That means that for all possible trajectories $X(\cdot,\omega)$ we have
  \begin{align*}
    P\left(X_{t_n}=x_n \middle| \left.X(\cdot, \omega)\right|_{[0,t_{n-1})}\right) = P\left( X_{t_n}=x_n \middle| X_{t_{n-1}}=x_{n-1} \right)\eqqcolon p^{x_{n-1}}_{x_n}(t_{n-1},t_n)\,.
  \end{align*}
  A Markovian jump process is \emph{homogeneous} if, in addition, these \emph{transition probabilities} $p^{x_n}_{x_{n-1}}$ only depend on the staying time $t_n-t_{n-1}$.
  These transition probabilities can be gathered in a matrix $\mathbb{P}$. 
\end{Definition}

\begin{Definition}
  \label{def:intensity-matrix}
  Given a finite state space $\mathcal{V}=\{ v_1,\dots, v_N\}$.
  An $N\times N$ square matrix $\mathbb{W}=\left(w^i_j\right)$ with
  \begin{align*}
    w^i_j\geq 0 \quad\text{ for } i\neq j\qquad\text{and}\qquad \sum_j w^i_i= 0
  \end{align*}
  is called \emph{intensity matrix}.
\end{Definition}

\begin{Remark}
  \label{rmk:zero-row-sum-for-intensity-matrix}
  The diagonal elements $w^i_i$ of an intensity matrix $\mathbb{W}$ take negative values with the absolute value $r_i\coloneqq\abs{w^i_i}$ and the vector $(1, 1, \dots, 1)$ always is a right-eigenvector with eigenvalue zero.
\end{Remark}

\begin{Proposition}
  \label{thm:Markov-process-for-intensity-matrix}
  For any $x\in\mathcal{V}$ and given intensity matrix $\mathbb{W}$ there is a unique RCLL Markovian jump process such that $X_0=x$ almost surely.
  Furthermore, the process is homogeneous and $\mathbb{P}(t)=\e{t\mathbb{W}}$ for $t \geq 0$.\,\citep{Capasso2012}
\end{Proposition}

\begin{Proposition}
  \label{thm:jump-time-statistics}
  The staying time in state $v_i$ is exponentially distributed with parameter $r_i$\,\citep{Capasso2012}, \ie
  \begin{align*}
    P\left( \forall \tau \in \left( t_0, t_0+ t \right)\colon X_{\tau}=v_i \middle| X_{t_0}=v_i\right) = \e{-r_i t}\,.
  \end{align*}
\end{Proposition}

\begin{Definition}
  \label{def:transition-escape-rates}
  The two preceding propositions give nice interpretations to the components of the intensity matrix $\mathbb{W}$: 
  The absolute value $r_i$ of its negative diagonal element $w^i_i$ is the \emph{escape rate} out of state $v_i$; its off-diagonal element $w^i_j$ is the \emph{transition rate} from state $v_i$ to state $v_j$.
\end{Definition}

\begin{Remark}
  \label{rmk:transition-rates}
  Sometimes it will be convenient to write $w^{v_i}_{v_j}\equiv w^i_j$ for the transition rate from $v_i$ to $v_j$.
\end{Remark}

\begin{Definition}
  \label{def:reducible-matrix}
  A square matrix $\mathbb{M}$ is \emph{reducible} if there is a simultaneous permutation of the columns and rows such that
  \begin{align*}
    \mathbb{M}=\begin{pmatrix} \mathbb{M}_1 & \mathbb{M}_3\\ 0 & \mathbb{M}_2 \end{pmatrix}
  \end{align*}
  with square matrices $\mathbb{M}_1$ and $\mathbb{M}_2$ of possibly different dimensions.
\end{Definition}

\begin{Definition}
  \label{def:reversible-markov-process}
  Given an intensity matrix $\mathbb{W}$.
  The corresponding Markovian jump process is \emph{dynamically reversible} if $w^i_j>0\Leftrightarrow w^j_i>0$ for $i\neq j$.
\end{Definition}

  A homogeneous and dynamically reversible Markovian jump process can be understood as a random walk on a simple undirected graph $\mathcal{G}$.
  The state space $\mathcal{V}$ is its vertex set.
  For every pair of transition rates $w^i_j,w^j_i>0$ there is an edge $\left\{ v_i, v_j \right\}$ of $\mathcal{G}$.
  Two vertices $v_i$, $v_j$ are not adjacent if $w^i_j=w^j_i=0$.
  If the intensity matrix $\mathbb{W}$ is irreducible, then the graph $\mathcal{G}$ is connected.

  In order to quantify real-valued functions on the edges, it will be convenient to assign an arbitrary orientation to every edge.
  Thus, $\mathcal{G}$ becomes a simple directed graph with no loops.
  Consequently, any trajectory visiting the states $\left( x_0, x_1, \dots, x_n \right)$ defines a walk on $\mathcal{G}$.

\begin{Remark}
  \label{rmk:intensity-matrix-weighted-laplacian}
  An intensity matrix is the negative of a weighted Laplacian matrix with the transition rate $w^i_j$ as weight $w(e)$ of the oriented edge $e=\left( v_i, v_j \right)$, \cf definition~\ref{def:weighted-laplacian}.
\end{Remark}

\begin{Definition}
  \label{def:master-equation}
  For a Markovian jump process one can also discuss the evolution of a random variable on $\mathcal{V}$ as an initial condition, rather than a single initial state $x\in\mathcal{V}$. 
  The random initial condition is given by a discrete probability distribution with probability $p_i$ for state $v_i\in\mathcal{V}$.
  We collect the initial probabilities in a row vector $p(0)=\left( p_1, p_2, \dots, p_N \right)$. Its time evolution along the process is governed by the \emph{master equation}
  \begin{align*}
    \frac{\D}{\D t} p(t) = p(t) \mathbb{W} \quad \Rightarrow \quad p(t) = p(0)\mathbb{P}(t)\,.
  \end{align*}
\end{Definition}

\begin{Remark}
  \label{rmk:ensemble-random-initial-condition}
  Random variables as initial conditions are the mathematical equivalent of what is known as \emph{ensembles} in physics.
\end{Remark}

\begin{Definition}
  \label{def:ergodic-markov-process}
  A Markovian jump process is \emph{ergodic} if $0=\frac{\D}{\D t} p(t)=p(t)\mathbb{W}$ has exactly one solution $p(t)\equiv \pi$.
\end{Definition}

\begin{Proposition}
  \label{thm:Perron-Frobenius}
  Let $\mathbb{W}$ be an irreducible intensity matrix of a dynamically reversible Markovian jump processes.
  Then $\mathbb{W}$ has zero as a simple eigenvalue.
  All other eigenvalues have negative real part.
  Additionally, the left-eigenvector with eigenvalue zero has only positive (or negative) entries.
  \begin{Proof}
    The proof makes use of the Perron--Frobenius Theorem for the non-negative matrix $\mathbb{P}(t)=\e{t\mathbb{W}}$.\,\citep{VanKampen2011}
  \end{Proof}
\end{Proposition}

Thus, dynamic reversibility and irreducibility of the intensity matrix are sufficient for ergodicity:

\begin{Definition}
  \label{def:ergodic-measure}
  The left-eigenvector $\pi$ of $\mathbb{W}$ corresponding to the eigenvalue zero and normalized as $\sum_i \pi_i = 1$, is the \emph{ergodic measure} or \emph{steady state distribution} of the Markov process.
\end{Definition}

  In the following, I always assume a Markovian jump process that is homogeneous, dynamically reversible and takes place on a connected graph.
  The process is characterized by its intensity matrix $\mathbb{W}$.

\begin{Definition}
  \label{def:probability-fluxes}
  The matrix $\otherPhi\coloneqq\diag(\pi)\mathbb{W}$ is the \emph{flux matrix} of the steady state.
  Its entry $\phi^i_j=\pi_i w^i_j$ is the \emph{probability flux} from state $v_i$ to state $v_j$.
  The fluxes can be understood as positive functions on the edges:
    $\phi(e)=\pi_{o(e)}w(e)$.
    Since fluxes are positive on all edges, they are not anti-symmetric. They cannot be understood as elements of the edge space $C_1(\mathcal{G})$, \cf definition~\ref{def:vertex-edge-space}.
\end{Definition}

\begin{Definition}
  \label{def:probability-current}
  The difference $J(e)\coloneqq\phi(e)-\phi(-e)$ is the \emph{probability current} on the edge $e$.
  Hence, the current $J\colon\mathcal{E}\to\mathbb{R}$ is anti-symmetric and an element of $C_1(\mathcal{G})$.
  It can be written as a matrix $\mathbb{J}=\otherPhi-\otherPhi\transpose$ with entries $J^i_j = \phi^i_j - \phi^j_i$.
\end{Definition}

\begin{Remark}
  \label{rmk:probability-current-steady-state}
  Since the steady state distribution is an eigenvector of the intensity matrix, we have
  \begin{align*}
    \forall i\colon 0 = (\pi\mathbb{W})_i
    = \sum_{j=1}^N \left(\pi_j w^j_i - \pi_i w^i_j\right)
    = \sum_{j=1}^N \left( \phi^j_i - \phi^i_j \right)
    = - \sum_{j=1}^{N} J^{i}_j\,.
  \end{align*}
\end{Remark}

\begin{Proposition}
  \label{thm:current-in-cycle-space}
  The probability currents run in cycles, \ie $\del J = 0$ or $J\in Z(\mathcal{G})$, \cf definition~\ref{def:cycle-space}.
  \begin{Proof}
    This proposition is connected to the example~\ref{exa:cycles-cocycles-in-electric-networks} of electric networks. Therefore, I restate the short proof\,\citep{Knauer2011}:
    \begin{align*}
      \del J
      &= \sum_{e\in\mathcal{E}} J(e) \left[ o(e)- t(e) \right]
      = \sum_{v\in\mathcal{V}} v \sum_{e\in\mathcal{E}} J(e)\left(\llbracket o(e)
      =v\rrbracket - \llbracket t(e) = v\rrbracket  \right)\\
      &= \sum_{i=1}^{N} v_{i} \sum_{j\colon j>i}^{N} \left( J^{i}_j - J^j_i \right) 
      = \sum_{i=1}^{N} v_{i} \sum_{j=1}^{N} J^i_j = 0
    \end{align*}
    The last equality is due to remark~\ref{rmk:probability-current-steady-state}.
  \end{Proof}
\end{Proposition}

\begin{Remark}
  \label{rmk:current-in-cycle-representation}
  According to remark~\ref{rmk:fundamental-cycles-basis} the cycle decomposition of the probability current is given by
  \begin{align*}
    J = \sum_{\eta\in\mathcal{H}} J(\eta) \zeta_\eta\,,
  \end{align*}
  where $\mathcal{H}=\mathcal{H}(\mathcal{T})$ is the chord set of a spanning tree $\mathcal{T}$.
\end{Remark}

\begin{Definition}
  \label{def:detailed-balance}
  A Markov process satisfies \emph{detailed balance} if $\otherPhi$ is symmetric or, equivalently, the probability currents on every edge vanish:
  \begin{align*}
    \forall e \colon\phi(e) - \phi(-e) = 0\quad \Leftrightarrow\quad J = 0\,.
  \end{align*}
\end{Definition}

\begin{Proposition}
  \label{thm:conditional-probability-of-states}
  In a Markovian jump process with intensity matrix $\mathbb{W}$, the probability of a state $v_j$ to occur, conditional on another state $v_i$ is\,\citep{Capasso2012}
  \begin{align*}
    P\left( v_j \middle| v_i \right) = \frac{w^{i}_{j}}{r_{i}} = \frac{w^i_j}{\sum_{\ell\neq i}w^i_{\ell}}\,.
  \end{align*}
\end{Proposition}

\begin{Proposition}
  \label{thm:probability-of-walk}
  Let $\mathbb{W}$ be the intensity matrix of an ergodic Markov process with corresponding graph $\mathcal{G}$.
  Any walk $\gamma=\left( e_1, e_2, \dots e_n \right)$, \cf definition~\ref{def:walk-on-a-graph}, on $\mathcal{G}$ has the probability
  \begin{align*}
    p\left[\gamma\right]\coloneqq \pi_{o(e_1)} \prod_{i=1}^{n} \frac{w(e_i)}{r_{o(e_i)}}\,.
  \end{align*}
  to occur in the steady state.
  \begin{Proof}
    A Markovian process has no memory, thus the conditional probabilities of consecutive jumps multiply.
  \end{Proof}
\end{Proposition}

\begin{Remark}
  \label{rmk:probability-of-walk}
  The probability of a walk is basically the probability of a trajectory, ignoring the jump-time statistics that is known from proposition~\ref{thm:jump-time-statistics}.
\end{Remark}

\begin{Proposition}
  \label{thm:average-number-of-jumps-converges}
  The number of jumps $n$ in a Markovian trajectory is a random variable that depends on the running time $T$.
  The average number of jumps for long trajectories is almost surely constant\,\citep{Jiang2004}:
  \begin{align*}
    \lim_{T\to \infty} \frac{n(T)}{T} = \sum_{i} \pi_i r_i \eqqcolon r \quad \asure
  \end{align*}
\end{Proposition}

\section{Observables and Fluctuations}
\label{sec:observables-fluctuations}

In most cases the steady state distribution $\pi$ and the currents $J$ of the steady state are not that interesting.
Rather, one is interested in observables defined on the Markov process.
In the following I will treat \emph{current-like observables}.
These observables are 1-chains $f\in C_1(\mathcal{G})$ satisfying $f(e^{-})=-f(e^{+})$.
Just like in the case of probability currents, for any 1-chain $f\in C_1(\mathcal{G})$ there is an anti-symmetric matrix with entries $f^i_j\coloneqq f(e)$ where $e=\left( v_i, v_j \right)$.

Ergodic processes have a very remarkable property: Very long trajectories sample the state space in a representative way.
That means, asymptotically, time averages and ensemble averages coincide.
Large deviation theory even allows to quantify the convergence of time averages.

\begin{Remark}
  \label{rmk:large-deviations-with-continuous-parameter}
  The definitions and results on large deviations in section~\ref{sec:largedevtheory}, stated for sequences of random variables, immediately apply to stochastic processes with continuous time parameter.
\end{Remark}

\begin{Definition}
  \label{def:ensamble-average}
  Let $f\in C_{1}(\mathcal{G})$ be a 1-chain.
  Then the number
  \begin{align*}
    \left\langle f \right\rangle \coloneqq \left\langle J, f \right\rangle
  \end{align*}
  is its \emph{steady state expectation}.
\end{Definition}

\begin{Definition}
  \label{def:trajectory-average}
  Let $\gamma_{T}=\left( e_1, e_2, \dots, e_{n(T)}\right)$ be the walk given by a trajectory of a Markovian jump process.
  The \emph{trajectory average} of $f\in C_1(\mathcal{G})$ is defined as
  \begin{align*}
    \overline{f}(T)\coloneqq \frac{1}{T}\sum_{j=1}^{n(T)}f(e_j)\,.
  \end{align*}
  Leaving the trajectory unspecified makes $\overline{f}(T)$ a random variable, referred to as \emph{time average}.
\end{Definition}

\begin{Theorem}[Ergodic Theorem]
  \label{thm:ergodic-theorem}
  For any $f\in C_{1}(\mathcal{G})$ the time average satisfies\,\citep{Jiang2004}
  \begin{align*}
    \lim_{T\to\infty}\overline{f}(T) = \left\langle f \right\rangle \quad \asure
  \end{align*}
\end{Theorem}

\begin{Definition}
  \label{def:skewed-intensity-matrix}
  Let $F=\left( f^{(1)}, f^{(2)}, \dots, f^{(d)} \right)$ be a $d$-tuple of 1-chains. One defines its \emph{skewed intensity matrix} $\mathbb{W}(q)$ via 
  \begin{align*}
    \left(\mathbb{W}(q)\right)^i_j \coloneqq 
    \mathbb{W}^i_j \exp\left(\sum_{\ell=1}^d q_\ell {f^{(\ell)}}^i_j\right)\,,\quad\text{ with }q=\left( q_1, q_2, \dots, q_d \right)\in\mathbb{R}^d\,.  \end{align*}
For $q=0$ the intensity matrix $\mathbb{W}$ and the skewed intensity matrix $\mathbb{W}(0)$ are identical.
\end{Definition}

\begin{Proposition}
  \label{thm:skewed-intensity-matrix-eigenvalue}
  The dominant eigenvalue of any skewed intensity matrix is simple, just as in the non-skewed case.\,\citep{Faggionato2011}
\end{Proposition}

\begin{Proposition}
  \label{thm:existence-of-scgf-for-chains}
  Let $E=\left( e_1, e_2, \dots, e_M \right)$ denote the standard basis of $C_1(\mathcal{G})$.
  Let $\overline{E}(T)$ denote its component-wise time average.
  The limit 
  \begin{align*}
    \lambda_{E}(q)=\lim_{T\to\infty}\frac{1}{T} \ln \left\langle \e{ Tq\cdot \overline{E}(T)} \right\rangle\,.
  \end{align*}
 exists for any $q\in\mathbb{R}^M$ and is given by the dominant eigenvalue of the skewed intensity matrix of $E$.\,\citep{Andrieux2007}
\end{Proposition}

The assumption of a finite state space ensures differentiability\,\citep{Touchette2009} of $\lambda_E(q)$.
So the above proposition basically states that the time average of the standard basis has a smooth scaled cumulant-generating function.
Due to the Gärtner--Ellis Theorem~\ref{thm:gaertner-ellis}, the time average of the standard basis satisfies a large deviation principle.
This is not limited to the standard basis:

\begin{Proposition}
  \label{thm:large-deviation-for-currents}
  The time average of any $d$-tuple $F=\left( f^{(1)}, f^{(2)}, \dots, f^{(d)} \right)$ of 1-chains has a smooth SCGF $\lambda_{F}(q)$ that is given by the dominant eigenvalue of its skewed intensity matrix $\mathbb{W}(q)$.
  \begin{Proof}
    There is a matrix $\mathbb{M}$ such that $F=\mathbb{M} E$, hence 
    \begin{align*}
      \lambda_{F}(q)
      =\lim_{T\to\infty}\frac{1}{T}\ln\left\langle \e{Tq\cdot \mathbb{M}\overline{E}(T)} \right\rangle
      =\lim_{T\to\infty}\frac{1}{T}\ln\left\langle \e{T(\mathbb{M}\transpose q)\cdot\overline{E}(T)}\right\rangle
      =\lambda_{E}(\mathbb{M}\transpose q).
    \end{align*}
  \end{Proof}
\end{Proposition}

\begin{Remark}
  \label{rmk:large-deviation-for-0}
  A special case is a tuple of constant zeroes $F_0=\left( 0, 0, 0, \dots, 0 \right)$. 
  This is also a tuple of 1-chains, but its time average is not a random variable\thinspace{}---\thinspace{}it is always constant.
  In this case, obviously, we have $F_0 = \vec{0} E$ with the constant zero matrix $\vec{0}$ of suitable dimension.
  However, the SCGF $\lambda_{F_0}(q)=\lambda_E(\vec{0}\transpose q) = \lambda_E(0)=0$ is a constant and not strictly convex.
  Therefore, all of its derivatives, \ie the scaled cumulants, vanish.
\end{Remark}

\begin{Definition}
  \label{def:fluctuation-spectrum}
  The scaled cumulants $c\left(\overline{f^{(i)}}(T),\dots,\overline{f^{(j)}}(T)\right)$, \ie the partial derivatives of $\lambda_F(q)$, are the \emph{fluctuation spectrum} of the tuple $\overline{F}(T)$ .
  Let $\overline{F_1}(T), \overline{F_2}(T)$ be two tuples with identical spectrum.
  I denote this by $\overline{F_1}(T)\asymp \overline{F_2}(T)$.
\end{Definition}

Since the SCGF of every $d$-tuple of 1-chains is smooth, the scaled cumulants entirely determine the convergence properties and thus the rate function.

In order to avoid an overly cumbersome notation I will write $c\left(f^{(i)},\dots,f^{(j)}\right)$
for the scaled cumulants and $F_1\asymp F_2$ if the fluctuation spectra coincide.

\begin{Remark}
  \label{rmk:finding-root}
  The propositions~\ref{thm:existence-of-scgf-for-chains} and \ref{thm:large-deviation-for-currents} are very nice from a theoretical point of view: They guarantee existence and differentiability for the SCGF.
  Unfortunately, it is a hard problem to find analytical expressions for the eigenvalues of big matrices.
  In order to find the fluctuation spectra, however, this step can be avoided with the help of the following
\end{Remark}

\begin{Theorem}[Implicit Function Theorem]
  \label{thm:implicit-fct-thm}
  Let $h\colon \mathbb{R}^{d}\times\mathbb{R}^k\to\mathbb{R}^k,(q, x)\mapsto h(q, x)$ be a continuously differentiable function.
  Fix a point $(q_0,x_0)$ with $h(q_0,x_0)\eqqcolon h_0$.
  If the matrix $\nabla\!_x h(q_0, x_0)$ is invertible then there is an open set $U\ni q_0$, an open set $V \ni x_0$, and a continuously differentiable function $\lambda\colon U \to V$ with
  \begin{align*}
    \left\{ \left(q, \lambda(q)\right)\middle| q\in U \right\} = \left\{ (q, x)\in U \times V \middle| h(q,x) = h_0 \right\}\,.
  \end{align*}
  Furthermore, the derivative of $\lambda$ at $q_0$ is given by
  \begin{align*}
    \nabla \lambda (q_0) = - \left(\nabla\!_x h(q_0,x_0)\right)^{-1}\nabla\!_{q} h(q_0, x_0)\,.
  \end{align*}
  If $h$ is $\ell$ times continuously differentiable, then so is $\lambda$ and its derivatives can be calculated from those of $h$.
\end{Theorem}

The SCGF $\lambda(q)$ is the unique solution to the eigenvalue problem with $\lambda(0)=0$.
The implicit function theorem allows us to calculate the fluctuation spectrum directly from the characteristic polynomial\thinspace{}---\thinspace{}explicitly finding the roots is not necessary:

Let $\chi(q,x)\coloneqq \chi_{\mathbb{W}(q)}(x)=\sum_{i=0}^{N}a_i(q) x^{i}$ be the characteristic polynomial of $\mathbb{W}(q)$.
The implicit function theorem is applicable in $(0,0)$ since $\mathbb{W}(0)=\mathbb{W}$ and due to the Matrix-Tree Theorem~\ref{thm:matrix-tree-theorem} the coefficient $\frac{\D \chi}{\D x}(0,0)=a_{1}(0)=a_1$ does not vanish.
The scaled cumulants can hence be determined iteratively: 
Take a partial derivative of order $\nu$ with respect to $q$ of the entire equation $0=\chi\left(q,\lambda(q)\right)$ and evaluate at $q=0$, $\lambda(0)=0$.
The partial derivative of $\lambda$ appears only once, so you can solve for it.
This scaled cumulant depends on partial derivatives of $\chi$ of order less or equal to $\nu$. These partial derivatives are the polynomial’s coefficients $a_i(q)$ and their derivatives at $q=0$.

\begin{Example}
  \label{exa:fluctuation-spectrum-of-scalar-observable}
  The first three scaled cumulants of a 1-chain $f\in C_1(\mathcal{G})$ are
  \begin{align*}
    c_1 = &-\frac{a_{0}'}{a_{1}}\,,\qquad\qquad\qquad\qquad c_2 = 2\frac{a_1' a_0'}{a_1^2} -2\frac{a_2\left(a_0'\right)^2}{a_1^3} - \frac{a_0''}{a_1}\,,\\
    c_3 =
    &+\frac{3 a_0' a_1''}{a_1^2}-\frac{6 a_2 a_0' a_0''}{a_1^3}
    +\frac{3a_1' a_0''}{a_1^2}+\frac{6 a_3 \left(a_0'\right)^3}{a_1^4}
    -\frac{12 a_2^2 \left(a_0'\right)^3}{a_1^5}\\
    &-\frac{6 a_2'\left(a_0'\right)^2}{a_1^3}
    +\frac{18 a_2 a_1'\left(a_0'\right)^2}{a_1^4}
    -\frac{6 \left(a_1'\right)^2a_0'}{a_1^3}-\frac{a_0'''}{a_1}\,.
  \end{align*}
  In the above expressions all coefficients $a_i$ and their derivatives have to be evaluated at $q=0$.
\end{Example}

\begin{Proposition}
  \label{thm:cocycles-have-no-fluctuation-spectrum}
  Let $y\in Z^{\perp}(\mathcal{G})$ be a cocycle, \cf definition~\ref{def:cycle-space}.
  Then $y\asymp 0$.
  \begin{Proof}
    The proof goes along the lines of a proof in reference~\citep{Andrieux2007}.
    There, the authors prove a different statement: They only consider a special 1-chain that is neither a pure cycle nor a pure cocycle.

    In the following let $N=\abs{\mathcal{G}}$ be the order of the graph.
    By definition, the entries of the characteristic matrix are
    \begin{align*}
      \mathbb{M}^i_j\coloneqq\left( \mathbb{W}(q)-x\,\mathbb{U} \right)^{i}_{j}=
      \begin{cases}
	-\sum_{\ell=1}^N w^i_\ell - x\,,&\text{if }i=j \\
	w^i_j \e{q y^i_j}\,,&\text{if }i\neq j\,.
      \end{cases}
    \end{align*}
    The determinant of $\mathbb{M}$ can be represented as
    \begin{align*}
      \det(\mathbb{M}) = \chi_{\mathbb{W}(q)}(x)
      = \sum_{\sigma\in\mathfrak{S}_N} \sgn(\sigma)\, \mathbb{M}^1_{\sigma(1)}\mathbb{M}^{2}_{\sigma(2)} \dots \mathbb{M}^{N}_{\sigma(N)}\,,
    \end{align*}
    where $\mathfrak{S}_N$ denotes the permutation group of $N$ symbols and $\sgn(\sigma)$ is the sign of the permutation $\sigma$.
    The permutations can always be decomposed along generators of the group:
    Let $\sigma\in\mathfrak{S}_N$ be a permutation. 
    If there is a vertex $v_i$ for that $v_{\sigma(i)}\notin\nbh(v_i)$, \cf definition~\ref{def:neighborhood-of-a-vertex}, then the corresponding $\mathbb{M}^{i}_{\sigma(i)}$ vanishes and the entire term is absent, hence independent of $q$.
    For any index $\ell$ such that $\sigma(\ell)=\ell$, the matrix entry $\mathbb{M}^\ell_\ell$ does not depend on $q$.
    Obviously, there can be indices with $\sigma^2(j)=j$.
    In this case the product $\mathbb{M}^j_{\sigma(j)}\mathbb{M}^{\sigma(j)}_j$ is independent of $q$ as well, since $y$ is anti-symmetric.
    For the remaining indices $k$, the permutations are cyclic (or anticyclic) permutations along cycles of the graph: They satisfy $\sigma^n(k)=k$ for some minimal $n$ depending on $k$.
    Multiplying the corresponding matrix entries adds the exponents.
    Since $y$ is a cocycle, its sum along any cycle is zero.
    Thus, no term in $\det(\mathbb{M})$ depends on $q$ and all scaled cumulants vanish.
  \end{Proof}
\end{Proposition}

\begin{Proposition}
  \label{rmk:fluctuation-spectrum-cycle-part}
  Let $f, h\in C_1(\mathcal{G})$ be two 1-chains. Then
  $f-h\in Z^{\perp}(\mathcal{G})\Leftrightarrow f\asymp h$.
  \begin{Proof}
    Let $f-h\in Z^{\perp}(\mathcal{G})$. This is equivalent to $\forall z\in Z(\mathcal{G})\colon \langle f, z \rangle = \langle h, z\rangle$.
    As seen in the proof of proposition~\ref{thm:cocycles-have-no-fluctuation-spectrum}, the fluctuation spectra only depend on the cycle sums.
    This can also be seen from multilinearity of the scaled cumulants, \cf proposition~\ref{thm:multilinear-joint-scaled-cumulants}: 
  \end{Proof}
\end{Proposition}

\begin{Definition}
  \label{def:chord-decomposition}
  For any 1-chain $f$ the \emph{chord representation} is given by
  \begin{align*}
    f_{\mathcal{H}}\coloneqq\sum_{\eta\in\mathcal{H}} f_\eta\, \eta\,,
  \end{align*}
  where $f_\eta\coloneqq \left\langle f, \zeta_\eta \right\rangle$ is the \emph{projection} of $f$ onto the fundamental cycle $\zeta_\eta$ corresponding to the chord $\eta$.
\end{Definition}

\begin{Remark}
  \label{rmk:chord-representation}
  The above representation is fundamentally different from the representation of a cycle given in remark~\ref{rmk:fundamental-cycles-basis}: It is defined for all $f\in C_1(\mathcal{G})$ and $f_{\mathcal{H}}$ vanishes on any edge of the spanning tree.
  Therefore, the boundary $\del f_{\mathcal{H}}$ is a linear combination of vertices incident to the chords.
  In general, this boundary does not vanish so $f_{\mathcal{H}}\notin Z(\mathcal{G})$.
\end{Remark}

\begin{Proposition}
  \label{thm:chord-decomposition-of-observable}
  The 1-chain $f$ and its chord representation $f_{\mathcal{H}}$ satisfy $f\asymp f_{\mathcal{H}}$.
  \begin{Proof}
    Let $\zeta$ be a fundamental cycle.
    Then
    \begin{align*}
      \left\langle f - f_{\mathcal{H}}, \zeta\right\rangle
      = \left\langle f, \zeta\right\rangle - \sum_{\eta\in\mathcal{H}} \left\langle f, \zeta_\eta\right\rangle\left\langle \eta, \zeta\right\rangle
      = 0
    \end{align*}
  \end{Proof}
\end{Proposition}

\begin{Proposition}
  \label{thm:fluctuation-spectra-of-1-chains}
  The fluctuation spectrum of a 1-chain $f$ is determined entirely by the joint fluctuation spectra of the chords $\mathcal{H}$ and the projections $f_{\eta}$. The first two scaled cumulants are
  \begin{align*}
    c_1\left(f\right) &= \sum_{\eta\in\mathcal{H}} f_{\eta}\, c_1\left(\eta\right)\,, \\
    c_2\left(f\right) &= \sum_{\eta\in\mathcal{H}} \sum_{\xi\in\mathcal{H}} f_\eta f_\xi\, c\left(  \eta,  \xi  \right)\,.
  \end{align*}
  \begin{Proof}
    This is a direct consequence of proposition~\ref{thm:chord-decomposition-of-observable} and multilinearity of the scaled cumulants .
  \end{Proof}
\end{Proposition}

\begin{Remark}
  \label{rmk:currents-through-chords}
  For any chord $\eta\in\mathcal{H}$, the time average $\bar\eta(T)$ has to be interpreted as an instananeous current through that chord.
  According to the Ergodic Theorem~\ref{thm:ergodic-theorem} its first cumulant is identical to the steady state current $J(\eta)$ on that chord.
  Therefore, the above proposition resembles the cycle representation, as given in remark~\ref{rmk:fundamental-cycles-basis}, and generalizes it to the the entire fluctuation spectrum of any current-like observable.
\end{Remark}

\phantomsection
\section*{Summary}
In this chapter we have seen how to describe Markovian jump processes as random walks on graphs.
Ergodic Markovian jump processes have a unique stationary probability distribution on the vertices.
Observables on ergodic Markov processes satisfy a large deviation principle.
The convergence behavior of their time average is entirely determined by their projection onto the cycle space and the fluctuation spectra of the currents through the chords.

\chapter{Stochastic Thermodynamics}
Now I will use the concepts introduced so far to mathematically model physical systems:
Let a thermodynamical system be given and let it be connected to at least two reservoirs. 
In the language of statistical mechanics the system is assumed to be described by a deterministic, ergodic dynamics on a Hamiltonian phase space, or possibly on a subset thereof.

Physically not all microscopic states in phase space can be measured.
Let us assume that only a finite set $\mathcal{V}$ of \emph{mesoscopic} states can be distinguished by our measurements.
They induce a partition of phase space.
On the set $\mathcal{V}$ of mesoscopic states the ergodic dynamics becomes a stochastic jump process, as seen in example~\ref{exa:ergodic-deterministic-dynamics}.
Let us assume that this process is Markovian.
This assumption is definitely not satisfied for all possible combinations of partitions and ergodic measures on phase space.
However, a separation of time scales between the dynamics within a partition cell and in between different cells results in an approximately Markovian process.\,\citep{Seifert2012}

The connections and interactions with the reservoirs define transition rates between the mesoscopic states in $\mathcal{V}$ and thus an intensity matrix $\mathbb{W}$, \cf definition~\ref{def:intensity-matrix}.
For physical systems it is reasonable to assume dynamical reversibility and connectedness of the corresponding graph $\mathcal{G}$\,\citep{Lebowitz1999,Schnakenberg1976,Seifert2012}.
Consequently, the induced Markovian jump process is ergodic.
Let both an arbitrary orientation on the edges, and spanning tree $\mathcal{T}$ with chord set $\mathcal{H}=\mathcal{H}(\mathcal{T})$ be given.

\section{Thermodynamics of the Steady State}

\begin{Definition}
  \label{def:visible-entropy}
  Let $v_i\mapsto p_i(t)$ be a time dependent probability distribution on the states.
  Then the Shannon entropy of the distribution
  \begin{align*}
    S\subsc{vis}(t) \coloneqq - \sum_{v_i\in\mathcal{V}} p_i(t) \ln p_i(t)
  \end{align*}
  is the \emph{visible entropy} of the system.
  For ergodic Markov processes the visible entropy approaches that of the stationary distribution $\pi$.
\end{Definition}

This definition implicitly assumes that the mesoscopic states $v_i$ have no internal structure and thus no internal entropy. This might not always be the case, but here I assume that all information is known by specifying the mesoscopic state.

\begin{Proposition}
  \label{thm:visible-entropy-derivative}
  The time derivative of the visible entropy can be split into two parts\,\citep{Esposito2010}
  \begin{align}
    \dot{S}\subsc{vis}(t) = \dot{S}\subsc{prod}(t) - \dot{S}\subsc{flow}(t)\,,\label{eq:entropy-splitting}
  \end{align}
  where we use the following 
\end{Proposition}

\begin{Definition}
  \label{def:entropy-prod-flow}
  The two quantities in the above proposition are the \emph{entropy-pro\-duc\-tion rate} within the system, 
  \begin{align*}
    \dot{S}\subsc{prod}(t) &\coloneqq \sum_{i,j} p_i(t) w^i_j \ln \frac{p_i(t)w^i_j}{p_j(t)w^j_i}\,,
  \intertext{and the \emph{entropy-flow rate} out of the system,}
    \dot{S}\subsc{flow}(t) &\coloneqq \sum_{i,j} p_i(t) w^i_j \ln \frac{w^i_j}{w^j_i}\,.
  \end{align*}
\end{Definition}

\begin{Proof}[of proposition~\ref{thm:visible-entropy-derivative}]
  \begin{align*}
    \dot{S}\subsc{vis}(t)
    &=- \frac{\D}{\D t} \sum_{i} p_i(t) \ln\left[p_i(t)\right]
     =-\sum_{i} \dot{p}_i(t) \ln\left[p_i(t)\right] - \underbrace{\sum_{i} \dot{p}_i(t)}_{=0}\\
    &=-\sum_{i} \sum_{j\colon j\neq i} \left( p_j(t) w^j_i - p_i(t) w^i_j \right) \ln\left[ p_{i}(t) \right]
    = \sum_{i}\sum_{j}p_i(t)w^i_j \ln\left[ \frac{p_i(t)}{p_j(t)} \right] \\
    &= \sum_{i}\sum_{j} p_i(t)w^i_j \ln\left[ \frac{p_i(t)w^i_j}{p_j(t)w^j_i} \right]
    - \sum_{i}\sum_{j} p_i(t)w^i_j \ln\left[ \frac{w^i_j}{w^j_i} \right]
  \end{align*}
\end{Proof}

\begin{Proposition}
  \label{thm:positivity-of-entropy-prod}
  The entropy-production rate is non-negative.\,\citep{Esposito2010}
  \begin{Proof}
    \begin{align*}
      \dot{S}\subsc{prod}(t)
      &= \sum_{i}\sum_{j} p_i(t)w^i_j \ln\left[ \frac{p_i(t)w^i_j}{p_j(t)w^j_i} \right]\\
      &= \sum_{i}\sum_{j\colon j>i} \left(p_i(t) w^i_j- p_j(t)w^j_i \right) \ln\left[ \frac{p_i(t)w^i_j}{p_j(t)w^j_i} \right] \geq 0
    \end{align*}
  \end{Proof}
\end{Proposition}
    The above statement justifies the name entropy-production rate.
    So in a sense, equation~\ref{eq:entropy-splitting} generalizes the second law.

In the following I will only consider the steady state with the probability distribution $\pi$ on $\mathcal{V}$.
The fluxes are $\Phi=\diag\left( \pi \right) \mathbb{W}$, the currents are $\mathbb{J}=\Phi-\Phi\transpose$, \cf definitions~\ref{def:probability-fluxes} and \ref{def:probability-current}.
  The entropy-production and entropy-flow rates in the steady state are independent of time and can be expressed by the probability fluxes and currents:
  \begin{align*}
    \dot{S}\subsc{prod} &= \sum_{e\in\mathcal{E}} J(e) \ln \frac{\phi(e)}{\phi(-e)} & &\text{and} &
    \dot{S}\subsc{flow} &= \sum_{e\in\mathcal{E}} J(e) \ln \frac{w(e)}{w(-e)}\,.
  \end{align*}
  Furthermore, $\dot{S}\subsc{prod}=\dot{S}\subsc{flow}$ since $\dot{S}\subsc{vis}=0$. 

\begin{Definition}
  \label{def:equilibrium}
  A Markovian jump process is in \emph{equilibrium} if the entropy-production rate $\dot{S}\subsc{prod}$ in the steady state vanishes.
\end{Definition}

\begin{Definition}
  \label{def:affinity-motance}
  We define $A, B \in C_1(\mathcal{G})$ by
  \begin{align*}
    A(e)& \coloneqq \ln \frac{\phi(e)}{\phi(-e)}=\ln\frac{\pi_{o(e)}w(e)}{\pi_{t(e)}w(-e)}\,, & B(e) &\coloneqq \ln \frac{w(e)}{w(-e)}\,.
  \end{align*}
  The 1-chain $A$ is called \emph{affinity}, $B$ is the \emph{motance}.
\end{Definition}

The name affinity for the quantity $A$ is inspired by chemistry: In chemical reactions one considers concentrations of particles instead of probability distributions and reaction rates rather than transition rates.
These descriptions can be mapped onto each other and then the definition above coincides with the chemical affinity, up to scalar factors involvintemperature and Boltzmann’s constant.\,\citep{Hill1977}

\begin{Proposition}
  \label{thm:affinity-motance-on-cycles}
  The affinity and motance satisfy $A - B \in Z^{\perp}(\mathcal{G})$.
  \begin{Proof}
    For the fundamental cycle $\zeta=\zeta_\eta$ of any chord $\eta\in\mathcal{H}$, we have
    \begin{align*}
      \left\langle A - B , \zeta \right\rangle
      &= \sum_{e\in\mathcal{E}} \left(A(e) -B(e)\right)\zeta(e)
      = \sum_{e\in\zeta} \ln \frac{\pi_{o(e)}}{\pi_{t(e)}} 
      = \sum_{e\in\zeta} \ln \pi_{o(e)} - \sum_{e\in\zeta} \ln\pi_{t(e)}
      = 0
    \end{align*}
    The proposition follows from linear extension.
  \end{Proof}
\end{Proposition}

\begin{Remark}
  \label{rmk:cycle-affinity}
  The above proposition ensures that for any fundamental cycle $\zeta_\eta$ the \emph{cycle affinity} $A_\eta =\left\langle A, \zeta_\eta \right\rangle$ only depends on the transition rates and not on the stationary distribution $\pi$.
\end{Remark}

\begin{Proposition}
  \label{thm:entropy-prod-as-AJ}
  With the cycle representation of $J$ or the chord representation of $A$, \cf remark~\ref{rmk:fundamental-cycles-basis} and definition~\ref{def:chord-decomposition}, we have
  \begin{align*}
    \dot{S}\subsc{prod} = \sum_{e\in\mathcal{E}} J(e) A(e) = \left\langle J, A\right\rangle = \sum_{\eta\in\mathcal{H}} J(\eta) A_{\eta}\,.
  \end{align*}
\end{Proposition}

\begin{Proposition}
  \label{thm:equivalent-formulation-for-equilibrium}
  The following statements are equivalent
  \begin{align*}
    J=0
    \Leftrightarrow A=0
    \Leftrightarrow \dot{S}\subsc{prod}=0
    \Leftrightarrow A\in Z^{\perp}(\mathcal{G})
    \Leftrightarrow B\in Z^{\perp}(\mathcal{G})\,.
  \end{align*}
\end{Proposition}

\begin{Remark}
  \label{rmk:equilibrium-from-intensity-matrix}
  The preceding proposition shows that the concepts of equilibrium and detailed balance are equivalent.
  Moreover, it allows to determine whether the stationary state of an ergodic Markov process is an equilibrium state without the need to actually calculate the stationary distribution $\pi$.
\end{Remark}

\begin{Proposition}
  \label{thm:thermodynamic-potential-in-equilibrium}
  For an equilibrium process, there is a 0-chain $u\in C_0(\mathcal{G})$ such that $\ddel u = B$.
  Writing $u_i \coloneqq u(v_i)$, the detailed balance condition translates to
  \begin{align*}
    \frac{\pi_i}{\pi_j} = \exp\left[ u_{j}-u_{i} \right] \quad\Leftrightarrow\quad \pi_i = \frac{\e{-u_i}}{\sum_{\ell} \e{-u_\ell}}\,.
  \end{align*}
  The 0-chain $u$ is unique up to a constant.
\end{Proposition}

\begin{Definition}
  \label{def:thermodynamic-potential}
  The 0-chain $u$ in the preceding proposition is the \emph{thermodynamic potential} of the equilibrium process.

\end{Definition}

\begin{Example}
  \label{exa:neq-triangle}
  Consider the Markov process on the state space $\mathcal{V} = \left( v_1, v_2, v_3 \right)$ associated to the intensity matrix
  \begin{align*}
    \mathbb{W} = w \begin{pmatrix}
      -1 -x & x & 1 \\
      1 & -2 & 1 \\
      1 & 1 & -2 \\
    \end{pmatrix}\,,
  \end{align*}
  where $w,x>0$.
  The constant $w$ defines the time scale of the process and was introduced only for dimensional purposes.
  The corresponding graph is a circuit $\mathcal{C}^3$.
  A natural choice for the orientations of the edges is $v_1 \to v_2 \to v_3 \to v_1$.
  The graph is its own fundamental cycle $\zeta$, irrespective of the choice of any spanning tree.
  The stationary state is characterized by the distribution $\pi=\left( \frac{1}{2+x}, \frac{2}{3}-\frac{1}{2+x}, \frac{1}{3}  \right)$ and the current $J = \frac{w}{2}\frac{x-1}{x+2}\zeta$.
  The affinity is $A=\left( \ln\frac{3x}{1+2x}, \ln\frac{2+x}{1+2x}, \ln\frac{1+x}{3} \right)$, the motance is $B=\left( \ln x, 0, 0 \right)$.
  The cycle affinity is $\left\langle A, \zeta\right\rangle = \ln x$, so the entropy-production rate is $\dot{S}\subsc{prod}=\frac{w}{2}\frac{x-1}{x+2}\ln{x}$.
  This system is in equilibrium if and only if $x=1$, as is seen most easily in the motance.
  The thermodynamic potential of this equilibrium process is a constant.
  Additionally, under equilibrium conditions the intensity matrix can be expressed by the Laplacian matrix of the graph: $\mathbb{W}=-w\mathbb{L}$.
\end{Example}

\section{Thermodynamics of Random Trajectories}

\begin{Definition}
  \label{def:entropy-change-along-trajectory}
    The time average $\otherSigma_{T}\coloneqq \overline{A}(T)$ of the affinity is called \emph{time-averaged entropy-production rate}.
\end{Definition}

As a result of the ergodic theorem~\ref{thm:ergodic-theorem} and proposition~\ref{thm:entropy-prod-as-AJ} we have
\begin{align*}
  \lim_{T\to\infty}\otherSigma_T = \dot{S}\subsc{prod}\quad\asure
\end{align*}

\begin{Remark}
  \label{rmk:entropy-change-along-trajectory}
  In the literature, the time-averaged entropy-pro\-duction rate is not always defined this way.
  Other definitions, however have the same fluctuation spectrum.
  The definition\,\citep{Lecomte2007} with the most appealing microscopic interpretation is
  \begin{align*}
    \otherSigma_T = \frac{1}{T} \ln\frac{p\left[\gamma\right]}{p\left[\gamma^{-1}\right]}
    = \frac{1}{T}\ln\frac{\pi_{o(\gamma)}r_{t(\gamma)}}{\pi_{t(\gamma)}r_{o(\gamma)}}
    + \frac{1}{T}\sum_{e\in\gamma}\ln\frac{w(e)}{w(-e)}\,,
  \end{align*}
  where $p\left[ \gamma \right]$ is the probability of the walk $\gamma=\left( e_1, e_2, \dots, e_{n(T)} \right)$, \cf proposition~\ref{rmk:probability-of-walk}.
\end{Remark}

\begin{Proposition}
  \label{thm:large-deviation-principle-for-entropy-along-trajectory}
  The scaled cumulant generating function $\lambda(q)$ of the time-averaged entropy-production rate exists and is differentiable.
  It is given by the dominant eigenvalue of the skewed intensity matrix $\mathbb{W}(q)$ with entries
  \begin{align}
    w^i_j(q) \coloneqq w^i_j \left(\frac{w^i_j}{w^j_i}\right)^q\,.\label{eq:tilted-matrix}
  \end{align}
  Consequently, the family $\otherSigma_T$ of random variables satisfies a large deviation principle with rate function $I(s) = s\,q(s) - \lambda\circ q(s)$ where $q(s)$ is the solution to $s=\lambda'(q)$.
\end{Proposition}

\begin{Theorem}[Fluctuation Theoerm]
  \label{thm:fluctuation-theorem}
  The rate function $I(s)$ and the SCGF $\lambda(q)$ of the time-average entropy-production $\otherSigma_T$ have the following symmetries\,\citep{Lebowitz1999,Touchette2009}:
  \begin{align}
    I(s) - I(-s) &= -s\,, & \lambda(-q) &= \lambda(q-1)\,.\label{eq:fluctuation-theorem}
  \end{align}
\end{Theorem}

\section{Displacement, Drift, Diffusion}

Unfortunately, not all observables on the edges are intrinsically current-like, \ie anti-symmetric.
One of the most important examples for such an observable is the displacement from which one can extract a drift coefficient and a diffusion constant.
The latter two scalars are often used to quantify physical motion with stochastic dynamics.

The non-trivial structure of graphs implies a difficulty in defining what the physical displacement of a walk is: The length $n$ of a walk is not a good measure.
A walk hopping between two adjacent states arbitrarily often does not travel far, no matter how distance is defined.

\begin{Definition}
  \label{def:distance}
  Let $\mathcal{G}$ be a graph.
  A \emph{distance} is a symmetric function on the edges, \ie $L\colon \mathcal{E}(\mathcal{G}) \to \mathbb{R}$ with $L(e^{-}) = L(e^{+})$.
  The distance naturally lifts to the universal covering $\theta\colon \widetilde{\mathcal{G}}\to\mathcal{G}$ by $L(\tilde{e})\coloneqq L(\theta_{\mathcal{E}}(\tilde{e}))$ for $\tilde{e}\in\mathcal{E}(\widetilde{\mathcal{G}})$.
\end{Definition}

\begin{Definition}
  \label{def:displacement-of-path}
  Let $\mathcal{G}$ be equipped with a distance $L$.
  Let $\gamma$ be a path in $\mathcal{G}$.
  The \emph{displacement} of the path $\gamma$ is given by $L[\gamma]\coloneqq \sum_{e\in\gamma}L(e)$.
\end{Definition}

  A walk can travel the same edge multiple times and in opposite directions.
  So the above definition will not give the expected value for the displacement, if applied to walks.

\begin{Definition}
  \label{def:displacement-of-walk}
  Let $\gamma$ be a walk in $\mathcal{G}$.
  The lift $\widetilde{\gamma}$ of $\gamma$ is a walk in the universal covering $\widetilde{\mathcal{G}}$.
  Since the universal covering is a tree, there is a unique path $\gamma'$ in $\widetilde{\mathcal{G}}$ from $o(\widetilde{\gamma})$ to $t(\widetilde{\gamma})$.
  The walk $\widetilde{\gamma}$ and the path $\gamma'$ are homotopic.
  The \emph{displacement} of the walk $\gamma$ is given by the displacement of this path $L[\gamma]\coloneqq L[\gamma']$ on the universal covering.
\end{Definition}

This definition also applies to fundamental cycles and thus gives a function $L\colon\pi_1(\mathcal{G},v)\to\mathbb{R}_{\geq 0}$ on the fundamental group.
The exact values of $L$ on $\pi_1(\mathcal{G})$ depend on the representation, \ie the starting and endpoint $v$.
There is no reason to assume that $L$ is invariant under commutation of its arguments, see the following

\begin{Example}
  \label{exa:displacement-of-walk}
  Let us revisit the simple graph $\mathcal{G}_c$.
  Let its vertices be labelled as in figure~\ref{fig:labeled-simple-graph-example}.
  Its universal covering is given in figure~\ref{fig:universal-covering-example}.
  Furthermore, let a constant distance $L(e)\equiv L$ for every edge $e$ be given.
  The neighborhood of vertex $v_1$ gives a spanning tree with the edges $e_2$ and $e_3$ as chords.
  Let $\zeta_2,\zeta_3$ be the corresponding fundamental cycles.
  Obviously, $L[\zeta_2^{-1}\circ\zeta_2]=0$ and $L[\zeta_2^{-1}]=L[\zeta_2]=L[\zeta_3]=3 L$.
  A walk starting and ending at $v_1$ and going along the cycle $\zeta_2^{-1}\circ \zeta_3\circ\zeta_2$ has a displacement of $L\left[ \zeta_2^{-1}\circ\zeta_3\circ\zeta_2 \right]= 7 L\neq 3 L = L[\zeta_3]$.
\end{Example}

Fixing only a running time $T$, the time average $\overline{L}(T)=\frac{1}{T}L[\gamma_T]$ is a random variable, where again $\gamma_T$ is the walk defined by a random trajectory of the Markov process.
Assume $\overline{L}(T)$ has a smooth scaled-cumulant generating function.
Then we have the following
\begin{Definition}
  \label{def:drift-diffusion}
  The first scaled cumulant
  \begin{align*}
    c_1(L) \eqqcolon V
  \end{align*}
  is the \emph{drift coefficient}, the second scaled cumulant
  \begin{align*}
    c_2(L) \eqqcolon 2D
  \end{align*}
  is twice the \emph{diffusion constant}.
\end{Definition}

\begin{Remark}
  \label{rmk:diffusion-constant}
  The above definition implicitly assumes (effective) motion in one dimension.
  In case of independent motion in $d$ different dimensions, the above definiton is used for every dimension independently.
\end{Remark}

The displacement $L$ has a smooth SCGF whenever $L\colon\pi_1(\mathcal{G},v)\to\mathbb{R}_{\geq 0}$ is the absolute value of a homomorphism $\ell\colon\pi_1(\mathcal{G},v)\to\mathbb{R}$: Then the displacement $L[\gamma_T]$ can effectively be expressed as the sum over a 1-chain $\ell\in C_1(\mathcal{G})$.
The anti-symmetry of $\ell$ accounts for reduction and expansion of the lift $\widetilde{\gamma}$ to the corresponding path $\gamma'$ in the universal covering.
The dependence on the reference state $v$ is irrelevant for the fluctuation spectrum.

\begin{Proposition}
  \label{thm:diffusion-constant}
  Utilizing the scaling behavior of scaled cumulants, \cf remark~\ref{rmk:scaled-cumulant-of-rescaled-sum}, we can represent the diffusion constant in its more common form:
  \begin{align*}
    2D = c_2\left(L\right) = \lim_{T\to\infty}\frac{1}{T} \kappa_2\left( T\, \overline{L}(T) \right) = \lim_{T\to\infty}\frac{1}{T} \kappa_2\left( L[\gamma_T] \right)\,.
  \end{align*}
\end{Proposition}

\phantomsection
\section*{Summary}

With the framework presented in this thesis, is is very easy to treat stochastic thermodynamics on finite Markovian jump processes.
The proposition~\ref{thm:fluctuation-spectra-of-1-chains} ensures that the different definitions of the time-averaged entropy-production rate found in the literature in fact are all equivalent in their fluctuations.
Moreover, the concepts also apply in the case of displacement\thinspace{}---\thinspace{}if the motion is assumed to take place on a 1-dimensional track.
Both the drift velocity and the diffusion constant appear in the fluctuation spectrum of the displacement.

\chapter{Coarse Graining}
The fundamental coarse-graining procedure from a microscopic deterministic dynamics to a mesoscopic stochastic dynamics was already described in the preceding chapter.
A priori it is not clear what a “good” partitioning of phase space has to look like.
Obviously, the stochastic dynamics has to be (approximately) Markovian in order to apply the formalism presented in the preceding chapters.
Nonetheless, observers with different experimental setups will require different mesoscopic descriptions.
In order to gain insight into the relation between these different descriptions, it is reasonable to investigate how a finegrained mesoscopic description can be further coarse grainged. 

Esposito~\citep{Esposito2012,Esposito2012a} examines the effects of coarse graining in stochastic thermodynamics, especially on the steady-state expectations.
He accounts for internal structure and thus internal entropy of the mesoscopic states.
Upon merging two adjacent mesoscopic states, the visible entropy is partly hidden in the internal entropy.
However, a separation of time scales for the staying time on different mesoscopic states is a crucial assumption in this approach.

\citet{Altaner2012a} suggest a form of local coarse graining that is inspired by the stochastic dynamics of trajectories: No internal structure of states is assumed.
By construction, it preserves the steady-state currents\thinspace{}---\thinspace{}with preserved cycle affinities also the entropy production $\dot{S}\subsc{prod}$ in the steady state is recovered.
Moreover, the local coarse graining does not require a separation of time scales:


Let $\mathbb{W}$ be the irreducible intensity matrix of a dynamically reversible and, therefore, ergodic Markovian jump process.
Let $\mathcal{G}=\left( \mathcal{V}, \mathcal{E}, \iota \right)$ be the corresponding graph.
As already noted in the preceding chapter, the graph is simple, connected and has no loops.
The aim is to coarse grain the system, \ie construct a new graph $\mathcal{G}'=\left( \mathcal{V}', \mathcal{E}', \iota' \right)$ with reduced number of vertices and new intensity matrix $\mathbb{W}'$.
Moreover, observables have to be modified in order to match their former steady-state expectations.
For observables not directly depending on the transition rates nor on the stationary distribution this is not a problem.
The entropy-production rate plays a special role.
It cannot be changed independently from the transition rates as it explicitly depends on those.

Altogether, \citet{Altaner2012a} suggest the following requirements for the local coarse graining:

\begin{enumerate}
  \item No change in cycle structure of the graph, no change in numerical values of probability currents.
  \item Preserve cycle affinities.
  \item Local changes in structure or transition rates have no influence on steady-state distribution outside the neighborhood.
  \item Entropy change along single trajectories is preserved.
\end{enumerate}

\section{Coarse Graining of Bridges}
\label{sec:bridgegraining}

\begin{figure}[htbp]
  \begin{subfigure}{0.45\textwidth}
    \centering
    \def\svgwidth{0.90\textwidth}
    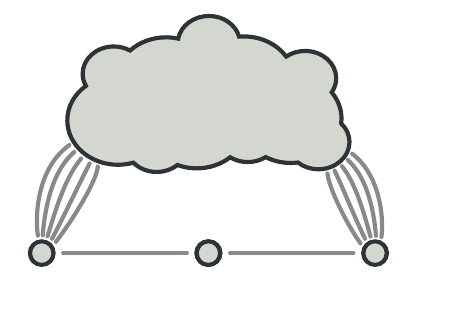
    \caption{Before coarse graining.}
    \label{fig:before-CG}
  \end{subfigure}
  \hfill
  \begin{subfigure}{0.45\textwidth}
    \centering
    \def\svgwidth{0.90\textwidth}
    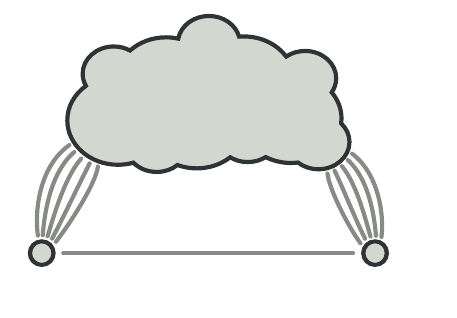
    \caption{After coarse graining.}
    \label{fig:after-CG}
  \end{subfigure}
  \caption{A target for the local coarse graining is a bridge state $v_2$. After coarse graining the bridge is absorbed into its neighbors $v_1$ and $v_3$.}
  \label{fig:bridge-coarse-graining}
\end{figure}

With these requirements a coarse graining of \emph{bridge states} is possible:
Pick a vertex $v_2\in\mathcal{V}$ of degree 2 with its two adjacent vertices $v_1$ and $v_3$ not being adjacent to each other, \cf figure~\ref{fig:bridge-coarse-graining}.
Let the local probability current be $J^1_2=J^2_3\equiv J$.

In the coarse-graining procedure, the bridge state $v_2$ is absorbed into its neighbors $v_1$ and $v_3$, resulting in new vertices $v_1'$, $v_3'$ now connected by a new edge $e'$.
The new transition probabilities on this edge are $w^{1'}_{3'}$ and $w^{3'}_{1'}$.
The structure of the maximal subgraph $\left( \mathcal{V}_0, \mathcal{E}_0, \iota_0 \right)$ not containing any vertex of the neighborhood of $v_2$ is not affected by the coarse graining.

\citet{Altaner2012a} conclude
\begin{align*}
  w^i_{n'} &= w^i_n\phantom{\,p}\quad\text{for } v_i\in\mathcal{V}_0, n\in\{1,3\}\,, &
  w^{1'}_{2'} &= p\, \frac{J+\psi}{\pi_1}\,,\\
  w^{n'}_i &= w^n_i\,p\quad\text{for } v_i\in\mathcal{V}_0, n\in\{1,3\}\,, &
  w^{2'}_{1'} &= p\, \frac{\psi}{\pi_3}\,,
\end{align*}
with the auxiliary parameters
\begin{align*}
  p &= \frac{\pi_1 + \pi_3}{\pi_1 + \pi_2 + \pi_3}\,, &
  \psi &= \frac{\phi^3_2\, \phi^2_1}{J + \phi^3_2 + \phi^2_1}\,.
\end{align*}

The above representation does not suggest a direct physical interpretation.
However, a short calculation results in the following expressions:
\begin{align*}
  w^i_{n'} &= w^i_n\phantom{\,p}\quad\text{for } v_i\in\mathcal{V}_0, n\in\{1,3\}\,, &
  w^{1'}_{2'} &= p\,\frac{w^1_2\, w^2_3}{r_2}\,,\\
  w^{n'}_i &= w^n_i\,p\quad\text{for } v_i\in\mathcal{V}_0, n\in\{1,3\}\,, &
  w^{2'}_{1'} &= p\,\frac{w^3_2 \, w^2_1}{r_2}\,,\\
\end{align*}
with the auxiliary parameter $p = \frac{\pi_1 + \pi_3}{\pi_1 + \pi_2 + \pi_3}$ as before.
The difference is the new parameter $r_2 = w^2_1 + w^2_3$ which has a physical interpretation: It is the escape rate out of the bridge state $v_2$, \cf definition~\ref{def:transition-escape-rates}.
So in total, the coarse graining multiplies the transition rate to the bridge state with the conditional probability $\frac{w^2_n}{r_2}, n\in\left\{ 1,3 \right\}$ of jumping further forward.
The factor $p$ accounts for the new steady-state probability to be conserved within the neighborhood.
Up to this factor $p$ these expressions are in fact identical to former suggestions\,\citep{Hill1977,Puglisi2010} of coarse graining not satisfying the locality requirement.

\section{Coarse Graining of Leaves}
\label{sec:leavegraining}

\begin{figure}[htpb]
  \centering
  \begin{subfigure}{0.45\textwidth}
    \centering
    \def\svgwidth{0.90\textwidth}
    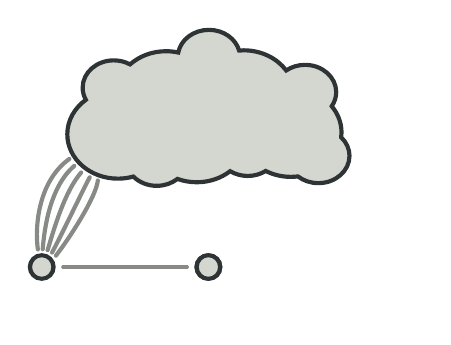
    \caption{Before the coarse graining.}
    \label{fig:cloud-leaf-example}
  \end{subfigure}
  \hfill
  \begin{subfigure}{0.45\textwidth}
    \centering
    \def\svgwidth{0.90\textwidth}
    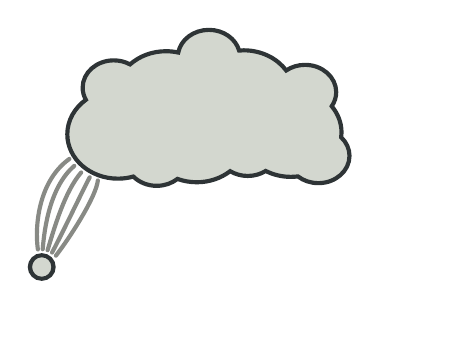
    \caption{After the coarse graining.}
    \label{fig:cloud-leaf-CG-example}
  \end{subfigure}
  \caption{Also leaves can be reduced. After the coarse graining the leaf and its incident edge are absorbed into the neighbor $v_1$.}
  \label{fig:leaf-coarse-graining}
\end{figure}

The local coarse graining is also capable of reducing \emph{leaves}.
A leaf is a vertex $v_2$ with degree 1.
Its single neighbor $v_1$ can have arbitrary degree, \cf figure~\ref{fig:leaf-coarse-graining}.

The coarse graining of a leaf does not have to account for the local current nor for the affinity:
The edge $\left\{ v_1, v_2 \right\}$ connecting the leaf $v_2$ always is in the cocycle space.
The new transition rates $w^{1'}_i = w^1_i\,\frac{\pi_1}{\pi_1 + \pi_2}$ for $v_i\in\mathcal{V}_0$ have to be reduced in order to preserve both the outgoing probability fluxes and the stationary probability in the neighborhood.
All the other transition rates stay untouched.

\phantomsection
\section*{Summary}

The presented coarse-graining procedure addresses single vertices.
Due to the locality of the procedure, it can be applied iteratively to reduce all bridges and leaves in a given system.
The final structure consists only of triangles, \ie cycles of type $\mathcal{C}^3$, at which point no bridges and no leaves are present any more.
By construction, in every single coarse-graining step both the currents and the cycle affinities and thus the entropy-production rate $\dot{S}\subsc{prod}$ are preserved.

\chapter{Model Systems}
The presented framework is able to determine fluctuation spectra of arbitrary observables on Markovian jump processes\thinspace{}---\thinspace{}as long as ergodicity is satisfied.

The simplest model system is a circuit: Its fundamental group is Abelian and there is a natural notion of distance.
There are different ways to drive such a system out of equilibrium.
I will analyze a homogeneously driven circuit that reveals a connection to transport theory.

Another class of model systems that satisfy all of the prerequisites is of biological nature: molecular motors.
These molecular machines are driven by coupling to particle baths, mainly adenosine triphosphate (ATP) and adenosine diphosphate (ADP). 
Transforming ATP into ADP and phosphate releases the energy the molecular machine needs to perform work.
There are different mechanochemical reaction pathway models for different motor proteins: the molecules change their conformation, attach to other molecules, detatch, and react.
Typically only a finite set of states is distinguishable and the dynamics is stochastic.
So these systems can be analyzed with the methods given in this thesis.
Moreover, many motor proteins travel along a one-dimensional structure, irrespective of the topology of the mechanochemical reaction pathway. 
Consequently, there is a notion of distance on the graph.
Drift and diffusion of the molecular motor on its track can be measured experimentally.
The presented formalism allows us to calculate the drift coefficient and the diffusion constant solely from the mechanochemical reaction pathway.
Thus, the quality of a suggested pathway model can be evaluated.
This method has fewer requriments than former attempts\,\citep{Boon2012} and is much simpler: The combinatoric complexity of the expressions involved is hidden in the coefficients of the characteristic polynomial, \cf proposition~\ref{thm:laplace-polynomial} and example~\ref{exa:fluctuation-spectrum-of-scalar-observable}.

\section{Single Circuit}

Let us consider a circuit $\mathcal{C}^N$ with vertex set $\left\{ v_1, v_2, \dots, v_N \right\}$ and order $N\geq 3$.
Let the orientation of its edges be aligned: $v_1\to v_2 \to \dots \to v_N \to v_1$. 
The graph is its own fundamental cycle $\zeta\simeq \mathcal{C}^N$ with $\left\langle \zeta, \zeta\right\rangle = N$.
Then we have the following

\begin{PropositionPlain}
  \label{thm:Label}
  Let $f\in C_1(\mathcal{C}^N)$ be a current-like observable and let $\mathbb{W}$ be an intensity matrix for $\mathcal{C}^N$.
  Then the characteristic polynomial of the skewed intensity matrix $\mathbb{W}(q)$ corresponding to $f$ has the following decomposition:
  \begin{align}
    \chi_{\mathbb{W}(q)}(x) = \chi_{\mathbb{W}}(x)+\det \mathbb{W}(q)\,.\label{eq:circuit-polynomial}
  \end{align}
\begin{Proof}
  We already know that $\mathbb{W}(0)=\mathbb{W}$.
  Furthermore, evaluating the characteristic polynomial of $\mathbb{W}(q)$ for $x=0$ gives the determinant of the skewed intensity matrix while $\chi_{\mathbb{W}}(0)=\det \mathbb{W} = 0$.
  As seen in the proof of proposition~\ref{thm:cocycles-have-no-fluctuation-spectrum}, we write the characteristic polynomial of $\mathbb{W}(q)$ as the determinant of the characteristic matrix with entries
    \begin{align*}
      \mathbb{M}^i_j\coloneqq\left( \mathbb{W}(q)-x\,\mathbb{U} \right)^{i}_{j}=
      \begin{cases}
	-\sum_{\ell=1}^N w^i_\ell - x\,,&\text{if }i=j \\
	w^i_j \e{q y^i_j}\,,&\text{if }i\neq j\,.
      \end{cases}
    \end{align*}
  The dependence of $\det\mathbb{M}$ on $q$ is given by the permutations along cycles of the graph.
  The circuit $\mathcal{C}^N$ consists exactly of its fundamental cycle $\zeta$ and has no other vertices.
  There are only two permutations of the edges along this cycle: the cyclic and the anti-cyclic permutation.
  These two permutations leave no vertex invariant, therefore the corresponding terms in $\det \mathbb{M}$ do not depend on $x$.
  That means the entire dependence on $q$ is contained in the determinant $\det\mathbb{W}(q)$.
\end{Proof}
\end{PropositionPlain}

\phantomsection
\subsection*{Distance and Displacement}
Let $L>0$ be a positive real number.
The constant $\frac{L}{N}$ defines a distance for the edges in $\mathcal{C}^N$.
The fundamental group $\pi_1(\mathcal{C}^N)$ is generated by only one fundamental cycle $\zeta$ and therefore commutes.
The displacement of the generator is $\frac{L}{N}[\zeta]=L$.
The 1-chain $\ell=\frac{L}{N}\zeta \in C_1(\mathcal{G})$ gives a group homomorphism $\pi_1(\mathcal{G})\to\mathbb{R}$.
Its projection onto the fundamental cycle is $\left\langle \ell,\zeta\right\rangle=L$.
Moreover, this 1-chain reproduces the displacement:
For any walk $\gamma$ in $\mathcal{G}$, 
the absolute value $|\sum_{e\in\gamma}\ell(e)|$ is identical to $\frac{L}{N}[\gamma]$ defined via the universal covering, \cf definition~\ref{def:displacement-of-walk}.
In the following, I will refer to $L$ as \emph{system size}.

\phantomsection
\subsection*{Homogeneous Driving}

Let us assume a \emph{homogeneous} driving: Given two constants $w^+,w^->0$, define the transition rate along any edge $e$ as $w(e^{+})\coloneqq w^+$ and $w(e^-)\coloneqq w^{-}$, \ie constant on all edges.
As a consequence, also the escape rate $r\equiv w^+ + w^-$ is constant on all vertices.
The $N\times N$ intensity matrix $\mathbb{W}$ is circulant with first row $\left(-r, \Wr, 0, \dots, 0, w^-  \right)$, \cf definition~\ref{def:circulant-matrix}:
\begin{align*}
  \mathbb{W} = \begin{pmatrix}	-r & \Wr & 0 & \cdots & 0 & \Wl \\
			\Wl & -r & \Wr & 0 & \cdots & 0\\
			0 & \Wl & \ddots & \ddots& \ddots&\vdots\\
			\vdots & 0& \ddots & \ddots&\ddots & 0\\
			0 & \vdots & \ddots & \ddots & -r & \Wr \\
			\Wr & 0 & \cdots & 0 & \Wl & -r
  \end{pmatrix}\,.
\end{align*}
The steady state distribution $\pi_{i} \equiv \frac1N$ is constant,
irrespective of the transition rates $\Wr$ and $\Wl$. 
Without loss of generality, we will assume $\Wr \geq \Wl$ in the following.

\phantomsection
\subsection*{Drift, Diffusion, Current, Affinity}

For the above setup of a homogeneously driven circuit, the drift coefficient $V$ and the diffusion constant $D$ are known\,\citep{Derrida1983}:
\begin{align}
  V &= \frac{L}{N} \left( \Wr - \Wl \right)\label{eq:velocity-sym}\,,\\
  D &= \frac{L^2}{2N^2} \left( \Wr + \Wl \right)\label{eq:diffusion-sym}\,.
\end{align}

For $\Wr=\Wl$ the system satisfies detailed balance, thus the drift coefficient vanishes while the diffusion constant stays finite.
If $\Wr \neq \Wl$ on the other hand, there is a finite probability current $J=\frac{1}{N}\left( \Wr-\Wl \right)\,\zeta$.
Since the steady state distribution $\pi$ is constant, the affinity and the motance are equal: $A = B = \ln\left[\frac{\Wr}{\Wl}\right] \zeta$.
The cycle affinity is not independent of the order $N$ of the graph: $A_\zeta=\left\langle A , \zeta\right\rangle = N \ln \frac{\Wr}{\Wl}$.
The steady-state expectation of the entropy-production, however, is independent of the order $N$:
\begin{align*}
  σ\coloneqq \dot{S}\subsc{prod}=\left\langle J, A\right\rangle = \left( \Wr - \Wl \right)\ln \frac{\Wr}{\Wl} \geq 0\,.
\end{align*}

\phantomsection
\subsection*{Continuum Limit}

It is worth mentioning, that it makes no sense to take the limit $N\to\infty$ while keeping both the transition rates $\Wr,\Wl$ and the system size $L$ constant.
In that case both transport coefficients vanish while the affinity diverges.
So the physical interpretation is no longer justified. 

Now let us assume the exact same physics is described by models with a different number of vertices: We fix the system size $L$ and we choose the cycle affinity $A_\zeta$ and the diffusion constant $D$ to rewrite the transition rates as follows: 
\begin{align}\begin{aligned}
  \Wr&=\frac{DN^2}{L^2} \left( 1+ \tanh \frac{A_\zeta}{2N} \right)\,,\\
  \Wl&=\frac{DN^2}{L^2} \left( 1- \tanh \frac{A_\zeta}{2N} \right)\,. \label{eq:driving-bulk}
\end{aligned}\end{align}
So we have
\begin{align}
  V = L\, J &= \frac{D}{L} 2N \tanh \frac{A_\zeta}{2N}\,,\label{eq:velocity-bulk}\\
  σ = A_\zeta\, J &= \frac{A_\zeta D}{L^2} 2N \tanh \frac{A_\zeta}{2N}\,.\label{eq:average-entropy-production-bulk}
\end{align}
Now it is easy to see that the continuum limit gives well defined results:
\begin{align}
  V &\xrightarrow{N\to\infty}\frac{A_{\zeta}D}{L}\,,\label{eq:velocity-bulk-continuous}\\
  σ &\xrightarrow{N\to\infty}\frac{A_{\zeta}^2D}{L^2}\,.\label{eq:entropy-production-bulk-continuous}
\end{align}

For small $N$ on the other hand, there are problems with a consistent physical interpretation: The drift coefficient and the entropy-production rate change with the number of vertices.
This contradicts the assumption that different $N$ describe the exact same underlying physics.
This observation holds irrespective of which physical quantities are used to parameterize the transition rates.
However, the discrepancy also vanishes at equilibrium and in the \emph{linear response regime}, \ie for $A_\zeta=0$ and $A_\zeta \ll 1$. 

\phantomsection
\subsection*{Rate Function of the Entropy-Production Rate}
\label{sec:singlecycle-statistics}

\begin{figure}[htbp]
  \centering
  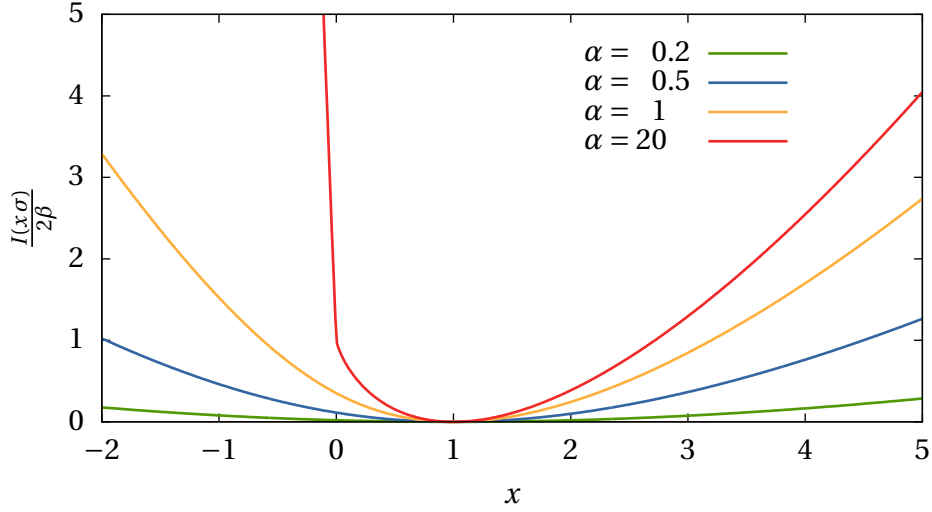
  \caption{The rate function $I$ of the entropy-production rate as a function of $x=\frac{s}{\sigma}$ for different driving parameters $\alpha=\frac{A_\zeta}{2N}$.}
  \label{fig:rate-function}
\end{figure}

The entropy-production rate satisfies a large-deviation principle, as seen in proposition~\ref{thm:large-deviation-principle-for-entropy-along-trajectory}.
The homogeneously driven system can be treated with analytical methods, even for not fixed order $N$.
In order to understand the statistics of the entropy production we need to calculate the dominant eigenvalue of the skewed intensity matrix (\ref{eq:tilted-matrix}) as stated in proposition~\ref{thm:large-deviation-principle-for-entropy-along-trajectory}.
For the homogeneous model this matrix is circulant with first row
\begin{align*}
  \begin{pmatrix} -\Wr -\Wl,& \Wr\left( \frac{\Wr}{\Wl} \right)^q ,& 0\,, &\dots\,,&0\,,& \Wl\left( \frac{\Wl}{\Wr} \right)^q 
  \end{pmatrix}
\end{align*}
Thus, the row sum equals the dominant eigenvalue, \cf proposition~\ref{thm:spectrum-for-circulant-matrix}, giving us the SCGF:
\begin{align}
  λ(q) = \Wr\left( \frac{\Wr}{\Wl} \right)^q + \Wl \left( \frac{\Wr}{\Wl}\right)^{-q} - \Wr -\Wl = 2β\left( \frac{\cosh\left[ \left( 2q+1 \right) α \right]}{\cosh α} - 1 \right)\,,\label{eq:scgf-bulk}
\end{align}
where we use the abbreviations $α \coloneqq \frac{A_\zeta}{2N}$ and $β \coloneqq \frac{DN^2}{L^2}$.
This function is smooth and convex everywhere.
Its derivative is
\begin{align}
  s(q)=λ'(q) &= 4αβ \frac{\sinh\left[ \left( 2q+1 \right) α \right]}{\cosh α}  = σ \frac{\sinh\left[ \left( 2q+1 \right) α \right]}{\sinh α}\,,\label{eq:es}
\end{align}
where we used $σ = 4 α β \tanh α$.
The functional dependence in (\ref{eq:es}) can easily be inverted to determine the function $q(s)$.
This allows us to explicitly calculate the rate function $I(s)=s\, q(s) - λ\circ{}q(s)$.
It is convex as well and has $s = λ'(q)$ as a natural variable.
We find
\begin{align}
  I(s) = - 2β \frac{\cosh\circ\Asinh\left[\frac{s}{σ}\sinh α \right]}{\cosh α} + 2 β +\frac{s}{2α\cosh α}\Asinh\left[\frac{s}{σ}\sinh α \right] - \frac{s}{2}\,.\label{eq:ratefunction-sym}
\end{align}
All of the terms in the rate function are symmetric in $s$, except for the last one, which is linear in $s$. Thus $I(s)$ satisfies the fluctuation relation (\ref{eq:fluctuation-theorem}). 
Moreover, we can rescale the argument of $I$ with the entropy-production rate $\sigma$ to see that the dependence on $β$ is very simple:
\begin{align}
  \frac{I(x\, σ)}{2β} = - \frac{\cosh\circ\Asinh\left[ x \sinh α \right]}{\cosh α}  + 1 + x \tanh[α]\,\big(\Asinh\left[ x \sinh α \right] - α \big) \,.\label{eq:ratefunction-rescaled}
\end{align}
Here it is obvious that the rate function vanishes at $s=σ$, or $x=1$. Since $I$ is both non-negative and convex, this is the global minimum.
In figure~\ref{fig:rate-function}, the rate function is drawn for different driving parameters $\alpha$.

\phantomsection
\subsection*{Approximations of the Rate Function}

In the limit of small $α$, \ie $A\ll N$, we have the expansion
\begin{align}
  I(x\, σ) = α²β\left( x-1 \right)² + \mathcal{O}(α)^4\,. \label{eq:ratefunction-bulk-linresponse}
\end{align}
On the other hand, an expansion of the rate function for values $s$ close to $σ$ yields
\begin{align}
  I(x\, σ) = β &\tanh^2[α]\,\left( x-1 \right)² + \mathcal{O}(x-1)³\,.\label{eq:ratefunction-bulk-minimum-expansion}
\end{align}
Note that the first term in the expansion for itself \emph{cannot} be used to test the fluctuation relation (\ref{eq:fluctuation-theorem}): The only parabola $y(x\, σ)$ with minimum $y(σ)=0$ that satisfies the fluctuation relation is
\begin{align}
  y(x\, σ) = \frac{σ}{4} (x-1)² =α β \tanh [α] \,(x-1)²\,. \label{eq:fluctuation-parabola-bulk}
\end{align}
Solely for small $α$ the approximation (\ref{eq:ratefunction-bulk-minimum-expansion}) and the function (\ref{eq:fluctuation-parabola-bulk}) coincide and take the form (\ref{eq:ratefunction-bulk-linresponse}). The parabolic form of the rate function is equivalent to a Gaussian probability distribution. This case is the so called \emph{linear response regime}.

Far away from the minimum ($|x|\gg 1$) the rate function asymptotically reads
\begin{align}
  \frac{I(\pm |x| σ)}{2β} \approx |x| \tanh[α] \big( \ln\left( 2|x|\sinh α \right)- (1\pm α) \big)\,.\label{eq:large-x}
\end{align}
In contrast to the approximation for the minimum, this asymptotics satisfies (\ref{eq:fluctuation-theorem}), since it is done symmetrically around the origin.
Any approximation ignoring the symmetry around the origin, just as (\ref{eq:ratefunction-bulk-minimum-expansion}) does, will fail to account for the fluctuation relation.

For large $\alpha$, the rate function can be approximated by
\begin{align}
  \frac{I(x\,\sigma)}{2 \beta} \approx 1+ \abs{x} \ln|x| - \abs{x} +\,\alpha \abs{x} (1-\sgn x)
  \begin{cases}= 1 + x \ln(x) -x\,, & \text{if }  x \geq 0\,,\\
    \approx 1 - 2 \alpha \abs{x}\,, & \text{if } x<0\,.
  \end{cases}\label{eq:large-alpha}
\end{align}
In the regime far from equilibrium, the rate function starts to develop a kink in the origin $x=0$.
Negative values for the entropy-production rate become very quickly unlikely as the backward transition rate $w^-$ approaches zero.
For positive $x$ the rate function approaches a scaling form far away from equilibrium.

\phantomsection
\subsection*{Fluctuation Spectra}

The analysis of the statistical properties of the entropy-production rate so far depended on the fact, that both the dominant eigenvalue, equation (\ref{eq:scgf-bulk}), of the skewed intensity matrix is analytically accessible and the relation (\ref{eq:es}) is invertible.
In a more general setting where the transition rates explicitly depend on position, this might not be the case.
Fortunately, from (\ref{eq:scgf-bulk}) we can directly calculate the scaled cumulants, without trying to determine the rate function:
\begin{subequations}
\begin{align}
  c_\nu &= 2 β \left( 2α \right)^\nu \cdot
	  			\begin{cases} \tanh α\,,&\text{for odd } \nu \\
    						1\,,&\text{for even }\nu
				\end{cases} \\
      &= 2D \frac{N^2}{L^2} \left( \frac{A_\zeta}{N} \right)^\nu \cdot
				\begin{cases} \tanh\frac{A_\zeta}{2N}\,,&\text{for odd }\nu \\
    						1\,,&\text{for even }\nu 
					      \end{cases}\label{eq:fluctuation-spectrum-bulk}
\end{align}
\end{subequations}
The first two scaled cumulants of the entropy-production rate equal the transport coefficients, up to multilinear factors:
\begin{align*}
  c_1 &= V \frac{A_\zeta}{L}\,, &  c_2 &= 2D\left(\frac{A_\zeta}{L}\right)^2\,.
\end{align*}
This is a consequence of proposition~\ref{thm:fluctuation-spectra-of-1-chains}: In a system with only one fundamental cycle, all current-like observables have the same fluctuation spectrum\thinspace{}--\thinspace{}up to multilinear factors.

Keeping $D$, $L$ and $A_\zeta$ constant again, we can discuss the continuum limit for the entropy-production rate: In equation (\ref{eq:fluctuation-spectrum-bulk}) it is easy to see that all higher cumulants ($\nu>2$) vanish for $N\to\infty$\,.
Mathematically, the continuum limit is equivalent to the linear response regime, \cf equation (\ref{eq:ratefunction-bulk-linresponse}).
That means the distribution of any current-like observable is Gaussian and therefore fully described by the drift coefficient $V$ and the diffusion constant $D$.
This is in accordance with the typical approximations in transport theory\,\citep{Groot1984}.
In addition, the fluctuation relation for the entropy-production rate demands $2 c_1 = c_2$ in the Gaussian case, which resembles the relation (\ref{eq:velocity-bulk-continuous}).

\phantomsection
\subsection*{Coarse Graining}

\begin{figure}[htbp]
  \centering
  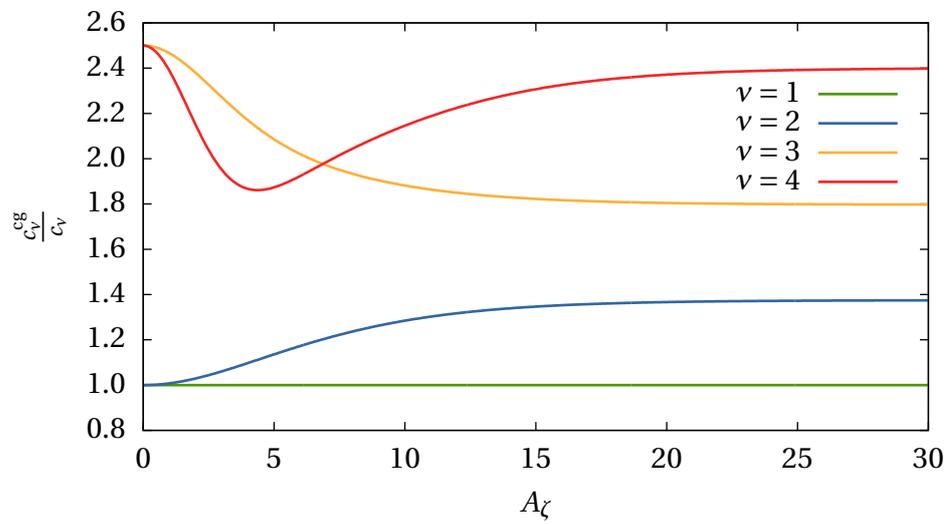
  \caption{The effect of coarse graining: The ratios $\frac{c_{\nu}\supsc{cg}}{c_{\nu}}$ are given as functions of the cycle affinity $A_\zeta$ that drives the system out of equilibrium.}
  \label{fig:cumulants-bulk}
\end{figure}

In the continuum limit, the homogeneously driven circuit exhibits very simple fluctuation spectra. Hence we conclude: The effects of coarse graining the circuit $\mathcal{C}^N$ with homogeneous driving will be highest for small $N$.
In the following, I consider a circuit of order $N=4$.
It is no problem to apply the coarse-graining procedure described in section~\ref{sec:bridgegraining} to an arbitrary edge in $\mathcal{C}^4$ to obtain a new intensity matrix on a reduced circuit of type $\mathcal{C}^3$.
In this case already, the SCGF for the entropy-production rate cannot be conveniently calculated as the dominant eigenvalue of the skewed intensity matrix.
As described in section~\ref{sec:observables-fluctuations}, the fluctuation spectrum can nonetheless be determined via the Implicit Function Theorem.
The exact expressions are rather involved.
In figure~\ref{fig:cumulants-bulk} the ratios of the first four scaled cumulants are given as functions of the cycle affinity $A_\zeta$.
Note that due to multilinearity and preserved cycle affinity $A_\zeta$ these ratios are actually valid for all current-like observables:

The first cumulant does not change, by construction.
The coarse graining leads to changes in the other scaled cumulants, as is clearly visible in the plot.
For the first four scaled cumulants the changes are factors of order 1.
It is very remarkable that the second scaled cumulant $c_2$ actually agrees at equilibrium, \ie $A_\zeta=0$.
Further away from equilibrium, however, the ratio rises to a value of $1.4$\thinspace{}---\thinspace{}a change of about $\SI{40}{\percent}$.
As the second scaled cumulant of the displacement is the diffusion constant, this is a significant change.
The third and fourth cumulants quantify the skewness and kurtosis of a distribution, \cf proposition~\ref{thm:first_cumulants_and_monents}.
In this case, the coarse graining makes the asymptotic distribution even more skew and causes a sharper peak with fatter tails, relative to the asymptotic distribution before the coarse graining.

\phantomsection
\section*{Summary}

A homogeneously driven circuit is analytically fully tractable: The rate function can be calculated directly.
Applying the coarse graining on any one vertex results in a non-homogeneously driven circuit.
With the tools given in this thesis, the fluctuation spectrum of any current-like observable can be determined analytically\thinspace{}---\thinspace{}even though neither the scaled cumulant-generating function, nor the rate function are analytically accessible.

\chapter{Conclusion}
Markovian jump processes naturally arise as models for many systems in physics, chemistry and biology for which the exact microscopic state of the system is experimentally not accessible.
These systems, especially in biology, need not be in equilibrium or even close to equilibrium.
As hallmarks of non-equilibrium situations, currents are of special interest.
In small systems, fluctuations of the currents are not negligible and empirically accessible in modern experiments.\,\cite{Seifert2012}

In this thesis, I presented a consistent framework to treat Markovian jump processes on finite state spaces and to analytically quantify the fluctuations of current-like observables in forms of scaled cumulants.
I used the mathematical theories of graphs and large deviations to derive the main result, proposition~\ref{thm:fluctuation-spectra-of-1-chains}: The fluctuation spectrum of any current-like observable is (up to multilinear factors) identical to the fluctuation spectrum of the probability currents on only a few edges\thinspace{}---\thinspace{}the chords of a suitable spanning tree.

A direct application for the results is Stochastic Thermodynamics: The entropy-production rate is a current-like observable.
Its time-average along trajectories is not defined consistently throughout the literature.
However, the main result shows that all definitions in use are asymptotically equivalent.
For some systems, also the displacement can be described as a current-like observable.
Consequently, the drift coefficient and the diffusion constant can be determined analytically.

A systematic coarse graining is a possibility to construct a new model for a given system.
The presented methods quantify the differences in the predicted fluctuations. 
This is in fact not restricted to a special coarse graining: Any two models for the same physical phenomenon can be compared analytically.
This comparison can then be the basis for deciding what model to use in which context or even to rule out suggested models if they significantly deviate from experimental findings.

The simplest model with finite currents, a single circuit, was analyzed in detail.
As long as the driving is entirely homogeneous, all of the large deviation functions are analytically accessible.
After coarse graining this system once, this convenience is no longer given.
However, it is still possible to calculate the statistics in form of the scaled cumulants.

\phantomsection
\section*{Outlook}

So far, only finite state spaces have been addressed.
An obvious extension of the presented tool-set are countably infinite state spaces.
In these cases, however, the scaled cumulant-generating function does not need to be differentiable\,\citep{Rakos2008}.

Up to now, another important prerequisite is dynamical reversibility, \cf definition~\ref{def:reversible-markov-process}.
In the homogeneously driven circuit, the limit far from equilibrium reveals a rate function that is not differentiable anymore.
This is due to vanishing dynamical reversibility.
Nonetheless, the system is ergodic in this limit.
So, in principle a generalization of the presented framework to merely ergodic processes should be possible.

The homogeneously driven circuit model (section \ref{sec:singlecycle-statistics}) is analytically solvable, even with arbitrary order $N$.
In general, the methods at hand allow us to treat such parametric state-space descriptions only on a case-by-case basis. They require an exact knowledge of the state-space structure.

Further, a systematic understanding of the errors introduced by the local coarse-graining procedure is desirable:
In the driven circuit model, the coarse graining of a bridge (section~\ref{sec:bridgegraining}) caused the diffusion constant to rise.
On the contrary, the elimination of leaves (section~\ref{sec:leavegraining}) attached to a circuit results in a lower diffusion constant (not presented in this work).
My hypothesis regarding this observation is the following: Coarse graining of an edge in the cycle space raises the diffusion constant, coarse graining an edge in the cocycle space, on the other hand, will always lower the diffusion constant.

\appendix

\cleardoublepage
\printbibliography

\begin{otherlanguage}{ngerman}
\Declaration
\end{otherlanguage}
\end{document}